\documentclass[a4paper,oneside,10pt]{amsart}

\usepackage[T1]{fontenc}
\usepackage{amsmath,amssymb,epsf}
\usepackage{amsfonts,amsbsy,amscd,stmaryrd}
\usepackage{latexsym,amsopn}
\usepackage{amsthm}
\usepackage{esint}
\usepackage{mathrsfs}

\usepackage{geometry}
\usepackage{setspace}
\usepackage{graphicx}
\usepackage{color}
\definecolor{Blue}{rgb}{0.,0.,1.}
\definecolor{Red}{rgb}{1.,0.,0.}

\newcounter{smallarabics}
\newenvironment{arabicenumerate}
{\begin{list}{{\normalfont\textrm{(\arabic{smallarabics})}}}
  {\usecounter{smallarabics}\setlength{\itemindent}{0cm}
   \setlength{\leftmargin}{5ex}\setlength{\labelwidth}{4ex}
   \setlength{\topsep}{0.75\parsep}\setlength{\partopsep}{0ex}
   \setlength{\itemsep}{0ex}}}
{\end{list}}

\newcounter{smallroman}

\newenvironment{notations}
{\begin{list}{{\normalfont\textrm{-}}}
  {\setlength{\itemindent}{0cm}
   \setlength{\leftmargin}{2ex}\setlength{\labelwidth}{4ex}
   \setlength{\topsep}{0.75\parsep}\setlength{\partopsep}{1ex}
   \setlength{\itemsep}{1ex}}
}
{\end{list}}

\let\origmaketitle\maketitle
\def\maketitle{
  \begingroup
  \def\uppercasenonmath##1{} 
  \let\MakeUppercase\relax 
	\origmaketitle
  \endgroup
}

\newcommand{\ben}{\begin{arabicenumerate}}  
\newcommand{\een}{\end{arabicenumerate}}

\def\init{\setcounter{equation}{0}}

\newtheorem{theorem}{Theorem}[section]

\newtheorem{proposition}[theorem]{Proposition}
\newtheorem{lemma}[theorem]{Lemma}
\newtheorem{definition}[theorem]{Definition}

\newtheorem{remark}[theorem]{Remark}
\newtheorem{example}[theorem]{Example}
\newcommand{\beq}{\begin{equation}}
\newcommand{\eeq}{\end{equation}}

\newcommand{\bex}{\begin{example}}
\newcommand{\eex}{\end{example}}
\def\bel{\begin{lemma}}
\def\eel{\end{lemma}}
\def\bet{\begin{theorem}}
\def\eet{\end{theorem}}
\def\bed{\begin{definition}}
\def\eed{\end{definition}}
\def\ber{\begin{remark}}
\def\eer{\end{remark}}

\def\rr{{\mathbb R}}
\def\zz{{\mathbb Z}}
\def\cc{{\mathbb C}}
\def\nn{{\mathbb N}}

\def\slim{{\rm s-}\lim}
\def\wlim{{\rm w-}\lim}

\def\bar{\overline}

\def\cinf{C^\infty}
\def\c0inf{C_0^\infty}

\def\proof{
\noindent{\bf Proof.}\ \ }

\usepackage{calrsfs}
\DeclareMathAlphabet{\pazocal}{OMS}{zplm}{m}{n}
\DeclareMathAlphabet{\mathsfsl}{OMS}{cmss}{m}{n}

\DeclareSymbolFont{altletters}  {OML}{zplm}{m}{it}
\DeclareMathSymbol{\altdelta}{\mathalpha}{altletters}{"0E}
\DeclareMathSymbol{\alteta}{\mathalpha}{altletters}{"11}

\def\cY{{\pazocal Y}}
\def\cV{{\mathcal V}}

\def\cD{{\pazocal D}}
\def\cU{{\mathcal U}}

\def\cN{{\pazocal N}}

\def\cW{{\pazocal W}}

\def\wf{{\rm WF}}

\def\free{{{\rm free}}}

\def\i{{\rm i}}

\DeclareMathOperator{\Dom}{Dom}

\def\vac{{\rm vac}}

\def\qed{$\Box$\medskip}
\newcommand{\qeds}{\qed}

\DeclareMathOperator{\ind}{ind}

\DeclareMathOperator{\Ker}{Ker}
\DeclareMathOperator{\Ran}{Ran}
\DeclareMathOperator{\coKer}{coKer}

\def \p{ \partial}
\def\12{\frac{1}{2}}
\def\14{\frac{1}{4}}
\def\x{\langle x \rangle}

\def\supp{{\rm supp}}
\newcommand{\one}{\boldsymbol{1}}
\def\cH{{\pazocal H}}

\def\coinf{C_{\rm c}^\infty}

\def\c{{\pazocal }}

\def\cF{{\pazocal F}}

\def\cX{{\pazocal X}}

\def\cK{{\pazocal K}}

\def\12{\frac{1}{2}}
\def\x{\langle x \rangle}

\def\supp{{\rm supp}}

\def\Diff{{\rm Diff}}
\def\rx{{\rm x}}
\def\bx{{\rm x}}

\def\bep{\begin{proposition}}
\def\eep{\end{proposition}}

\def\Op{{\rm Op}^{\rm w}}

\newcommand{\mat}[4]{\begin{pmatrix}#1 &#2  \\ #3 &#4 \end{pmatrix}}
\newcommand{\col}[2]{\begin{pmatrix}#1 \\#2\end{pmatrix}}

\def\CARal{{\rm C\hskip 0.25 em \hbox{\raise 1.72 ex 
\hbox{$\scriptscriptstyle\rm al$}\kern -0.57 em A}R}}

\def\otimesal{\mathop{\hbox{\raise 1.5 ex
  \hbox{$\scriptscriptstyle\rm al$}
\kern -0.92 em \hbox{$\otimes$}}}}
\def\oplusal{\mathop{\hbox{\raise 1.5 ex
  \hbox{$\scriptscriptstyle\rm al$}
\kern -0.92 em \hbox{$\oplus$}}}}
\def\Gammal{\hbox{\raise 1.68 ex 
\hbox{$\scriptscriptstyle\rm al$}\kern -0.50 em $\Gamma$}}
\def\Bal{\hbox{\raise 1.68 ex 
\hbox{$\scriptscriptstyle\rm  al$}\kern -0.50 em $B$}}
\def\CARal{{\rm C\hskip 0.25 em \hbox{\raise 1.72 ex 
\hbox{$\scriptscriptstyle\rm al$}\kern -0.57 em A}R}}

\def\cE{\pazocal{E}}

\DeclareMathAlphabet{\mathpzc}{OT1}{pzc}{m}{it}
\newcommand{\bra}{\langle} 
\newcommand{\ket}{\rangle}

\DeclareSymbolFont{boldoperators}{OT1}{cmr}{bx}{n}
\SetSymbolFont{boldoperators}{bold}{OT1}{cmr}{bx}{n}

\makeatletter
\newcommand*{\defeq}{\mathrel{\rlap{%
                     \raisebox{0.3ex}{$\m@th\cdot$}}%
                     \raisebox{-0.3ex}{$\m@th\cdot$}}%
                     =}
                     
\makeatother
\makeatletter
\newcommand*{\eqdef}{=\mathrel{\rlap{%
                     \raisebox{0.3ex}{$\m@th\cdot$}}%
                     \raisebox{-0.3ex}{$\m@th\cdot$}}%
                     }
\makeatother
\DeclareMathAlphabet{\mathpzc}{OT1}{pzc}{m}{it}

\def\Op{{\rm Op}}
\def\WF{{\rm WF}}

\def\altb{{\rm\textit{b}}}
\def\altc{{\rm\textit{c}}}
\def\altch{\skew3\hat{\rm\textit{c}}}
\def\altgh{\skew3\hat{\rm\textit{g}}}
\def\altgt{\skew3\tilde{\rm\textit{g}}}
\def\altk{{\rm\textit{k}}}
\def\altg{{\rm\textit{g}}}
\def\alth{{\rm\textit{h}}}
\def\altV{{\rm\textit{V}}}
\def\altW{{\rm\textit{W}}}
\def\altm{{\rm\textit{m}}}
\def\altVh{\skew5\hat{\rm\textit{V}}}
\def\altVt{\skew5\tilde{\rm\textit{V}}}
\def\altR{{\rm\textit{R}}}

\newcommand{\bea}{\begin{aligned}}
\newcommand{\beal}{\begin{array}{l}}
\newcommand{\eeal}{\end{array}}
\newcommand{\eea}{\end{aligned}}

\def\cf{C^\infty}
\def\cof{C_{\rm c}^\infty}

\def\td{{\rm td}}
\def\std{{\rm std}}
\def\Htd{{\rm td}}
\def\Hstd{{\rm std}}
\def\aM{{\rm aM}}
\def\scc{{\rm sd}}
\def\bg{{\rm bg}}
\def\nt{{\rm nt}}
\def\ast{\rm ast}
\def\pos{{\rm pos}}
\def\adg{{\rm ad}}
\def\dg{{\rm d}}

\def\rf{{\rm ref}}

\def\sca{{\rm out/in}}

\def\inout{{\rm in/out}}
\def\inn{{\rm in}}
\def\out{{\rm out}}
\def\F{{\rm F}}
\def\aF{{\rm \overline{F}}}

\def\spexi{{k}}

\newcommand{\traa}[1]{\mskip-6mu\upharpoonright_{#1}}
\def\pe{\overline{\p}}

\def\zero{{\mskip-4mu{\rm\textit{o}}}}
\def\diag{{\rm diag}}
\def\ry{{\rm y}}
\def\cinfb{\cinf_{\rm b}}
\def\BT{{\rm BT}}

 \def\gdia{G^{\dg}}
\def\outin{{\rm out/in}}
\def\varo{\varrho}
\def\varT{t_{+}}
\def\sd{{\rm sd}}
\def\sobo{{m}}
\def\AH{H}

\begin{document}
\title[Feynman propagators and  Hadamard states from scattering data on asymptotically Minkowski spacetimes]{\large Feynman propagators and Hadamard states from scattering data\\ for the Klein-Gordon equation on asymptotically Minkowski spacetimes}
\author{}
\address{Universit\'e Paris-Sud XI, D\'epartement de Math\'ematiques, 91405 Orsay Cedex, France}
\email{christian.gerard@math.u-psud.fr}
\author{\normalsize Christian \textsc{G\'erard} \& Micha{\l} \textsc{Wrochna}}
\address{Universit\'e Grenoble Alpes, Institut Fourier, UMR 5582 CNRS, CS 40700, 38058 Grenoble \textsc{Cedex} 09, France}
\email{michal.wrochna@ujf-grenoble.fr}
\begin{abstract}We consider the massive Klein-Gordon equation on a class of asymptotically static spacetimes. We prove the existence and Hadamard property of the \emph{in} and \emph{out} states constructed by scattering theory methods. Assuming in addition that the metric approaches that of Minkowski space at infinity in a short-range way, jointly in time and space variables, we define Feynman scattering data and prove the Fredholm property of the Klein-Gordon operator with the associated Atiyah-Patodi-Singer boundary conditions. We then construct a parametrix (with compact remainder terms) for the Fredholm problem and prove that it is also a Feynman parametrix in the sense of Duistermaat and H\"ormander.
\end{abstract}

\maketitle

\section{Introduction \& summary}

\subsection{Hadamard property of $\rm in/out$ states}
The construction of quantum states from scattering data is a subject that has been studied extensively in various contexts in Quantum Field Theory, including the case of the wave and Klein-Gordon equation --- set either on Minkowski space, in external electromagnetic potentials \cite{isozaki,lundberg,rui,seiler}, or on curved spacetimes with special a\-symp\-totic symmetries, to mention only the works \cite{wald0,DK0,DK1,DK2,Mo1}. On the physics side, the primary motivation is to give meaning to the notion of particles and anti-particles and to describe quantum scattering phenomena. From the mathematical point of view, the problems often discussed in this context in the literature involve existence of scattering and M{\o}ller operators, the question of asymptotic completeness, as well as specific properties of states such as the ground state or thermal condition with respect to an asymptotic dynamics, see e.g. \cite{drouot,DD,DRS,GGH,nicolas} for recent developments on curved backgrounds.

In the present paper we address the question of whether the so-called \textit{in} and \textit{out} states on asymptotically static spacetimes satisfy the \emph{Hadamard condition} \cite{KW}. Nowadays regarded as an indispensable ingredient in the perturbative construction of interacting fields (see e.g. recent reviews \cite{HW,KM,FV2}), this property accounts for the correct short-distance behaviour of expectation values of fields. It can be conveniently formulated as a condition on the wave front set of the state's two-point functions \cite{radzikowski} --- a terminology that we explain in the paragraphs below. It is known that in the special case of the conformal wave equation, one can study the wave front set of the two-point functions quite directly in the geometrical setup of conformal scattering on asymptotically flat spacetimes \cite{Mo2,characteristic} (cf. \cite{DMP1,DMP2,BJ} for generalizations on the allowed classes of spacetimes). Furthermore, propagation estimates in ${\rm b}$-Sobolev spaces of variable order were used recently to show a similar result in the case of the wave equation on asymptotically Minkowski spacetimes \cite{VW}. The two  methods being however currently limited to a special value of the mass parameter, our focus here is instead on the proof of the Hadamard property of the \textit{in} and \textit{out} state for the Klein-Gordon operator $P=-\Box_\altg+\altm^2$ for any positive mass $\altm$, or more generally for $P=-\Box_\altg+\altV$ with a real-valued potential $\altV\in\cf(M)$ satisfying an asymptotic  positivity condition. \medskip

Specifically, we first consider the special case of a $1+d$-dimensional globally hyperbolic spacetime $(M,\altg)$ with Cauchy surface $\Sigma$ and metric of the form $\altg=-dt^2 + \alth_t$, with $\alth_t$ a Riemannian metric smoothly depending on $t$. The Klein-Gordon operator can be written in the form
\beq\label{Pintro}
P=\p_t^2 + r(t)\p_t + a(t,\bx,\p_\bx),
\eeq
where $r(t)$ is the multiplication operator $|\alth_t|^{-\12}\p_t |\alth_t|^{\12}$ and $a(t,\bx,D_\bx)\in\Diff^2(\Sigma)$ has principal symbol  $\spexi \cdot \alth_t^{-1}(\bx)\spexi$ (where $\xi=(\tau,k)$ is the dual variable of $x=(t,\bx)$) and is bounded from below. Now, supposing $\Sigma$ is a \emph{manifold of bounded geometry} (see Subsect. \ref{ss:mbg}), there exist uniform pseudo\-differential operator classes $\Psi^m(\Sigma)$ due to Kordyukov and Shubin \cite{Ko,Sh2} that generalize the well-known pseudo\-differential calculus of H\"ormander on $\rr^d$ and closed manifolds. Here in addition, in order to control decay in time, we introduce $t$-dependent pseudo\-differential operators $\Psi^{m,\delta}_\td(\rr;\Sigma)$ as quantizations of $t$-dependent symbols $a(t, \rx, \spexi)$ that satisfy
\[
|\p^{\alpha}_{t}\p_{\rx}^{\beta}\p_{\spexi}^{\gamma}a(t, \rx, \spexi)|\leq C_{\alpha\beta\gamma}\langle t\rangle^{\delta- \alpha}\langle \spexi\rangle^{m- |\gamma|}, \ \ \alpha\in \nn, \ \beta, \gamma\in \nn^{d},
\]
where $\bra t\ket=(1+t^2)^{\12}$, $\bra \spexi\ket=(1+|\spexi|^2)^{\12}$, and the constants $C_{\alpha\beta\gamma}$ are uniform in an appropriate sense. This allows us to state a hypothesis that accounts for asymptotic ultra-staticity of $(M,\altg)$ at future and past infinity. Namely, we assume that there exists $a_\out,a_\inn\in\Psi^2(\Sigma)$ elliptic and bounded from below (by a positive constant), such that on $\rr_{\pm}\times\Sigma$,
\[
(\Htd) \  \ \ \beal
a(t, \rx, D_{\rx})= a_{\outin}(\rx, D_{\rx})+ \Psi_{\td}^{2, -\delta}(\rr; \Sigma) ,\ \delta>0,\\[2mm]
r(t)\in \Psi_{\td}^{0, -1-\delta}(\rr; \Sigma).
\eeal  
\]
In practice, in our main cases of interest $a_{\outin}(\rx, D_{\rx})$ will simply be the Laplace-Beltrami operator of some asymptotic metric $\alth_\outin$ plus the mass or potential term. 

Let now $\cU(t,s)$ be the Cauchy evolution of $P$, i.e. the operator that maps Cauchy data of $P$ at time $s$ to Cauchy data at time $t$. 
In this setup, what we call \emph{time-$t$ covariances of the $\out$ state} are the pair of operators defined by
\beq\label{eq:thelimit}
c_{\out}^\pm(t)\defeq \lim_{t_+\to\infty} \cU(t,t_+) c_\out^{\pm,\vac}  \cU(t_+,t)
\eeq
whenever the limit exists (in a sense made precise later on), where $c_\out^{\pm,\vac}$ equals
\[
c_\out^{\pm,\vac}= \12 \begin{pmatrix}\one & \pm a_\out^{\12} \\ \pm a_\out^{-\12} & \one\end{pmatrix}.
\]
To elucidate the interpretation of $c_\out^{\pm,\vac}$ let us point out that $c_\out^{\pm,\vac}$ is the spectral projection on $\rr^{\pm}$ of the generator\footnote{This generator is selfadjoint for the energy scalar product.} of the Cauchy evolution $\cU_\out(t,s)$ corresponding to the asymptotic Klein-Gordon operator $P_\out\defeq\p_t^2+a_\out$. On the other hand, to $c_\out^{\pm,\vac}$, $c_\out^\pm$ we can associate pairs of operators $\Lambda_\out^{\pm,\vac}$, $\Lambda_\out^\pm: \cf_{\rm c}(M)\to \cf(M)$ by
\[
\bea
\Lambda_\out^{\pm,\vac}(t,s)&\defeq \mp\pi_0\cU_\out(t,0)c_{\out}^{\pm,\vac}\cU_\out(0,s)\pi_1^*, \\ \Lambda_\out^\pm(t,s)&\defeq \mp\pi_0\cU(t,0)c_{\out}^\pm(0)\cU(0,s)\pi_1^*,
\eea
\]
where we wrote $\Lambda_\out^{\pm,\vac}$, $\Lambda_\out^\pm$ as operator-valued Schwartz kernels in the time variable and $\pi_0$, $\pi_1$ are the respective projections to the two pieces of Cauchy data. In QFT terms (strictly speaking, using the terminology for charged fields), the operators $\Lambda_\out^{\pm,\vac}$, $\Lambda_\out^\pm$ are \emph{two-point functions}, i.e. they satisfy
\[
\bea
P_\out \Lambda^{\pm,\vac}_{\out}=\Lambda^{\pm,\vac}_\out P_\out = 0,& \ \ \Lambda^{+,\vac}_\out-\Lambda^{-,\vac}_\out=\i G_\out, \ \ \Lambda^{\pm,\vac}_{\out}\geq 0, \\
P \Lambda^\pm_{\out}=\Lambda^\pm_{\out} P = 0, \quad & \ \ \Lambda^+_{\out}-\Lambda^-_{\out}=\i G, \quad \Lambda^\pm_{\out}\geq 0,
\eea
\] 
where $G_\out$, $G$ are the \emph{causal propagators} for respectively $P_\out$, $P$, i.e. 
\[
G_\out(t,s)=\i \pi_0\cU_\out(t,s)\pi_1^*, \ \ G(t,s)=\i \pi_0\cU(t,s)\pi_1^*.
\] 
As a consequence, using the standard apparatus of algebraic QFT one can associate states $\omega_\out^\vac$, $\omega_\out$ on the corresponding ${\rm CCR}$ $C^*$-algebras: $\omega_\out^\vac$ is then the very well studied ground state associated with $P_\out$ and $\omega_\out$ is the \emph{out} state that we study.

Our first result can be expressed as follows in terms of the two-point functions $\Lambda^\pm_\out$. 

\begin{theorem}\label{thm:main1} Assume $(\Htd)$. Then the limit \eqref{eq:thelimit} exists and $\omega_\out$ is a Hadamard state, i.e.
the two-point functions $\Lambda^\pm_\out$ satisfy the {Hadamard condition}:
\beq\label{eq:had1}
\wf'(\Lambda^\pm_\out)=  \textstyle\bigcup_{t\in\rr}(\Phi_t(\diag_{T^*M})\cap \pi^{-1}\cN^\pm),
\eeq
where $\cN^+$, $\cN^-$ are the two connected components of the characteristic set $\cN\subset T^*M\setminus\zero$ of $P$, $\Phi_t$ is the bicharacteristic flow acting on the left component of $\diag_{T^*M}$ (the diagonal in $(T^*M\times T^*M)\setminus\zero$), and $\pi:\cN\times\cN\to\cN$ is the projection to the left component.
\end{theorem}

Above, $\wf'(\Lambda_\out^\pm)$ stands for the primed wave front set of $\Lambda_\out^\pm$, i.e. it is the image of the wave front set of the (full) Schwartz kernel of $\Lambda_\out^\pm$ by the map $(x,\xi,x',\xi')\mapsto (x,\xi,x',-\xi')$. We refer to \cite{hoermander} for the definition and the basic properties of the wave front set of a distribution, cf. \cite{BDH} for a concise introduction. The \emph{bicharacteristic flow} $\Phi_t$ is the Hamilton flow of $p(x,\xi)=\xi\cdot \altg^{-1}(x)\xi$ restricted to $\cN=p^{-1}(\{0\})$ understood as a subset of $T^*M\setminus\zero$ (where $\,\zero$ is the zero section of the cotangent bundle), see \cite{hoermander}.  


The essential feature of the Hadamard condition \eqref{eq:had1} is that it constraints $\wf'(\Lambda^\pm_\out)$ to the positive/negative frequency components $\cN^\pm\times\cN^\pm$. Thus, on a very heuristic level, the plausibility of this statement can be explained as follows. In a static situation, $c^{\pm,\vac}_\out$ can be interpreted as projections that single out Cauchy data that propagate as superpositions of plane waves with positive/negative frequency, and thus with wave front set in $\cN^\pm$. On a generic asymptotically flat spacetime it is not immediately clear what the analogous decomposition at finite times is, but instead one can try to use the decomposition given by $c^{\pm,\vac}_\out$ at \emph{infinite times}: this is what indeed motivates the definition of $\Lambda^\pm_{\out}$. The difficulty is however to control the wave front set of the infinite time limit \eqref{eq:thelimit}. \medskip

In addition to the statement of Thm. \ref{thm:main1}, we get in a similar vein a Hadamard state $\omega_\inn$ by taking the analogous limit with $t_- \to -\infty$ instead of $t_+\to + \infty$; this is the so-called \emph{in} state. 

Furthermore, our results extend to a more general class of asymptotically static spacetimes $M=\rr\times\Sigma$ with metric of the form
\[
\altg= - \altc^{2}(x)dt^{2}+ (d\rx^{i}+ \altb^{i}(x)dt)\alth_{ij}(x)(d\rx^{j}+ \altb^{j}(x)dt),
\]
where $(\Sigma,\alth)$ is a manifold of bounded geometry and $\altc,\alth,\altb$ as well as their inverses are bounded with all derivatives (with respect to the norm defined using a reference Riemannian metric). By \emph{asymptotically static} we mean that there exist Riemannian metrics  $\alth_\outin$ and smooth functions $\altc_{\outin}$ on $\Sigma$, such that on $\rr_\pm\times\Sigma$,
\[
(\ast)\ \ \beal
\alth(x)- \alth_{\outin}(\rx)\in S^{-\mu},\\[2mm]
\altb(x)\in S^{-\mu'},  \mbox{ and } \altc(x)- \altc_{\outin}(\rx)\in S^{-\mu}
\eeal
\]
for some $\mu>0$, $\mu'>1$; in a similar vein the potential $\altV$ is required to satisfy $\altV(x)- \altV_{\outin}(\rx)\in S^{-\mu}$ for some smooth $\altV_{\outin}$. Above, the notation $f\in S^{-\mu}$ means symbolic decay in time, i.e. $\p^{\alpha}_{t}f\in O(\langle t\rangle^{-\mu- |\alpha|})$ for all $\alpha\in \nn^{1+d}$; we refer to Subsect. \ref{ss:asast} for the precise formulation.

In this more general situation, the Klein-Gordon operator is not necessarily of the form \eqref{Pintro} considered so far. However, under a positivity assumption $(\pos)$ on $\altV_{\outin}$, it turns out that there are natural coordinates in terms of which the Klein-Gordon operator is very closely related to an operator \eqref{Pintro} satisfying $(\Htd)$, i.e. one is obtained from the other by conjugation with some multiplication operators. This allows us to give a very similar definition of the \textit{out}/\textit{in} state $\omega_\outin$ and to prove a direct analogue of Thm. \ref{thm:main1}.  

\subsection{Fredholm problem for the Klein-Gordon equation on asymptotically Minkowski spacetimes}
Our second main result makes use  of asymptotic data at  future and past infinity and at the same time relies on good control of what happens  at spatial infinity, and thus requires a more refined setup. 

To formulate the problem, let us first recall that in Quantum Field Theory, expectation values of time-ordered products of fields involve a \emph{Feynman propagator}, which in the present setup is an operator of the form
\[
G_{\F,\omega}=\i^{-1}\Lambda^+ +  G_- = \i^{-1}\Lambda^+ +  G_+,
\]
where $\Lambda^\pm$ are two-point functions of a state $\omega$ and $G_\pm$ the retarded/advanced propagator\footnote{By retarded/advanced propagator $G_\pm$ one means the inverse of $P$ that solves the inhomogeneous problem $P u = f$ for $f$ vanishing
at respectively past/future infinity.} of $P$. We call any such operator $G_{\F,\omega}$ a \emph{time-ordered Feynman propagator} to distinguish it from other related (approximate) inverses of $P$. Apart from playing an essential role in perturbative computations in interacting QFT, time-ordered Feynman propagators on curved spacetimes provide the link between the Hadamard condition \eqref{eq:had1} and the theory of distinguished parametrices of Duistermaat and H\"ormander \cite{DH}. In fact, Radzikowski's theorem \cite{radzikowski} asserts that $\omega$ is Hadamard if and only if the primed wave front set of $G_{\F,\omega}$ is the same as that of Duistermaat and H\"ormander's `Feynman parametrix', i.e. if
\beq\label{eq:fewf}
\WF'(G_{\F,\omega})= (\diag_{T^*M})\cup\textstyle\bigcup_{t\leq 0}(\Phi_t(\diag_{T^*M})\cap \pi^{-1}\cN),
\eeq
This plays a crucial role in applications as it implies that two-point functions of Hadamard states are unique modulo smooth terms.

Recently, a very different point of view was proposed by Gell-Redman, Haber and Vasy \cite{GHV,positive}, basing on earlier developments \cite{kerrds,BVW,semilinear}, who proved that the wave operator on asymptotically Minkowski spacetimes is Fredholm when interpreted as an operator acting on carefully chosen Hilbert spaces of distributions. A remarkable consequence is that it has a generalized inverse $G_\F$ such that its primed wave front set $\WF'(G_\F)$ is given precisely by \eqref{eq:fewf} \cite{GHV,VW}. It is however not expected to be equal $G_{\F,\omega}$ for some Hadamard state $\omega$, even in cases when $G_{\F}$ is an exact inverse of $P$: although one could define some operators $\Lambda^\pm$ by setting $G_\F\eqdef\i^{-1}\Lambda^+ +  G_- = \i^{-1}\Lambda^+ +  G_+$, they will generically not satisfy the positivity condition $\Lambda^\pm\geq 0$ and thus they will not be two-point functions. On the other hand, one can argue that $G_\F$ is a canonical object (modulo finite-dimensional choices, unless some geometrical assumptions are made), and that it bears much more resemblance  to elliptic inverses than the retarded and advanced propagators do. Moreover, a recent work of B\"ar and Strohmaier that treats the Dirac equation on a finite Lorentzian cylinder \cite{BS} achieves to set up a Fredholm problem which is in many ways similar to that of Gell-Redman, Haber and Vasy. Interestingly, they prove a Lorentzian analogue of the Atiyah-Patodi-Singer theorem \cite{APS1,APS2} and relate the index to quantities of direct physical interest, in the so-called chiral anomaly \cite{BS2}.

Our aim is to set up a Fredholm problem on a class of spacetimes similar to ones considered in \cite{GHV}, but for the massive Klein-Gordon equation instead of the wave equation. On the other hand, we use an approach that is more closely related to the method of \cite{BS} and that in fact can be thought of as its non-compact generalization, at least if one disregards distinct features of the Dirac and Klein-Gordon equations.

We are primarily interested in the class of \emph{asymptotically Minkowski spacetimes}, in the sense that $(M,\altg)$ is a Lorentzian manifold (without boundary) such that $M=\rr^{1+d}$ and:
\[
(\aM)\ \begin{array}{rl}
&\altg_{\mu\nu}(x)- \alteta_{\mu\nu} \in S^{-\delta}_{\std}(\rr^{1+d}), \ \delta>1,\\[2mm]
&(\rr^{1+d}, \altg)\hbox{ is globally hyperbolic},\\[2mm]
&(\rr^{1+d}, \altg) \hbox{ has a  time function }\tilde{t}\hbox{ such that }\tilde{t}-t\in S^{1-\epsilon}_{\std}(\rr^{1+d}),\ \epsilon>0,\\[2mm]
\end{array}
\]
where $\alteta_{\mu\nu}$ is the Minkowski metric and $S_{\std}^{\delta}(\rr^{1+d})$ stands for the class of smooth functions $f$ such that 
\[
\p^{\alpha}_{x}f\in O(\langle x\rangle^{\delta- |\alpha|}), \ \alpha\in \nn^{1+d}.
\] 
This way, $\altg$ decays to the flat Minkowski metric simultaneously in time and in the spatial directions in a short-range\footnote{This corresponds to the assumption $\delta>1$.} way. In a similar vein the potential is required to satisfy $\altV(y)- \altm^{2}\in S^{-\delta}_{\std}(\rr^{1+d})$, $\altm>0$. Note that the definition $(\aM)$ covers a similar class of spacetimes as those considered in \cite{BVW,GHV} (the latter are also called asymptotically Minkowski spacetimes therein), but strictly speaking they are not exactly the same. In our setup one has for instance $(M,\altg)$ is globally hyperbolic, which is not clear from the outset in \cite{BVW,GHV}. 

 \medskip

The main idea in the formulation of the Fredholm problem is to consider `boundary conditions' that select asymptotic data which account for propagation of singularities within only one of the two connected components $\cN^\pm$ of the characteristic set of $P$. While in \cite{BS} there is indeed a boundary at finite times, here we need to consider infinite times instead, so boundary conditions are not to be understood literally as they are rather specified at the level of scattering data. 

In order to define scattering data in the setting of asymptotically Minkowski spaces we first make a change of variables by means of a diffeomorphism $\chi$ (see Subsect. \ref{s11.1}), which allows to put the metric in the form
\[
\chi^* \altg = -  \altch^{2}(t, \rx)dt^{2}+ \hat  \alth(t, \rx)d\rx^{2},
\]
where $\altch$ tends to $1$ for large $|x|$, while $\hat\alth$ tends to some asymptotic metrics $\hat\alth_{\inout}$ depending on the sign of $t$. In these coordinates, a convenient choice of Cauchy data is $\varrho_s u\defeq (u,-\i\altch^{-1}\p_{t}u)\traa{t=s}$. On the other hand, the natural reference dynamics in this problem (at both future and past infinity) is that of the free Klein-Gordon operator 
\[
P_\free=-\p_{t}^{2}- \Delta_{\rx}+ \altm^{2}. 
\]
Let now $\cU(t,s)$, $\cU_\free(t,s)$ be the respective Cauchy evolutions for $P$ and $P_\free$, and let us fix as reference time $t=0$. We define the \emph{Feynman} and \emph{anti-Feynman scattering data maps}: 
\[
\bea
\varrho_{\F}&\defeq \slim_{t_\pm\to\pm\infty} \left(c_{\free}^{+,\vac} \cU_{\free}(0,t_+)\varrho_{t_+} + c_{\free}^{-,\vac} \cU_{\free}(0,{t_-})\varrho_{t_-}\right), \\
\varrho_{\rm \overline{F}}&\defeq \slim_{t_\pm\to\pm\infty}\left( c_{\free}^{+,\vac} \cU_{\free}(0,{t_-})\varrho_{{t_-}} + c_{\free}^{-,\vac} \cU_{\free}(0,t_+)\varrho_{t_+}\right),
\eea
\]
as appropriate strong operator limits, where $c_{\free}^{\pm,\vac}$ is defined as $c_{\out}^{\pm,\vac}$, but with $- \Delta_{\rx}+ \altm^{2}$ in the place of $a_\out$.
We abbreviate  the Sobolev spaces $H^\sobo(\rr^d)$ by $H^\sobo$.  Our main result can be stated as follows. 

\begin{theorem}\label{thm:main2}Assume $(\aM)$ and let $m\in\rr$. Consider the Hilbert space 
\beq
\cX_\F^m \defeq  \big\{ u \in (\chi^{-1})^{*} \big(C^{1}(\rr; H^{\sobo+1})\cap C^{0}(\rr; H^{\sobo})\big): \ P u \in \cY^m, \ \varrho_{\aF} u=0\big\},
\eeq
where $\cY^m\defeq(\chi^{-1})^{*}\left(\bra t \ket^{-\gamma} L^2(\rr;H^m)\right)$ and $\textstyle\12<\gamma<\textstyle\12+\delta$. Then $P:\cX_\F^m\to \cY^m$ is Fredholm of index 
\beq\label{eq:theindex2}
\ind P|_{\cX_\F^m\to \cY^m}= \ind (c^{-,\vac}_{\free}W_{\out}^{-1}+ c^{+, \vac}_{\free}W_{\inn}^{-1}),
\eeq
where $W_\outin^{-1}=\lim_{t_\pm\to \pm\infty}\cU_{\free}(0, t_\pm)\cU(t_\pm, 0)$. In particular  the index  is independent on $m$. Furthermore, there exists $G_\F:\cY^m\to\cX_\F^m$ with $\wf'(G_\F)$ as in \eqref{eq:fewf} and such that $\one-PG_\F$ and $\one-G_\F P$ are compact and have smooth Schwartz kernels. 
\end{theorem}

Note that  the space  $\cX_{\F}^{m}$ is  a closed subspace of the Hilbert space 
\[
\cX^{m}= \{ u \in (\chi^{-1})^{*} \big(C^{1}(\rr; H^{\sobo+1})\cap C^{0}(\rr; H^{\sobo})\big): \ P u \in \cY^m\}
\]
equipped with the norm    $\| u\|^{2}_{\cX^{m}}= \| \varrho_{0}(\chi^{-1})^{*}u\|_{\cE^{m}}+ \|Pu\|^{2}_{\cY^{m}}$,  where $\cE^{m}$ is the energy space, see 
Def. \ref{defowit}.

As pointed out in \cite{BS}, the condition $\varrho_{\aF}u=0$ can be seen as an analogue of the Atiyah-Patodi-Singer boundary condition (even though this is less evident here as we do not consider the Dirac equation). Furthermore, one could equally well consider the \emph{anti-APS} boundary condition $\varrho_\F u=0$, which leads to an `anti-Feynman' counterpart of Theorem \ref{thm:main2} --- interestingly, just as in \cite{BS}, this differs from the Riemannian case where one boundary condition is preferred over the other. On the other hand, the kernel of $P:\cX_\F^m\to \cY^m$ consists of smooth functions and $G_\F$ satisfies a positivity condition $\i^{-1}(G_\F-G_\F^*)\geq0$ reminiscent of the limiting absorption principle. As pioneered in \cite{BS} and \cite{BVW,positive}, this shows a striking similarity to the elliptic case.

\subsection{Outline of proofs}

\subsubsection{Proof of Hadamard property}

The main technical ingredient that we use in the proof of both theorems is an approximate diagonalization\footnote{On a side note, let us mention that a different diagonalization procedure was proposed by Ruzhansky and Wirth in the context of dispersive estimates \cite{RW,wirth}; in their method it is the (full) symbol of the generator of the Cauchy evolution that is diagonalized (rather than the Cauchy evolution itself).} of the Cauchy evolution by means of elliptic pseudo\-differential operators, derived in detail in \cite{bounded} and based on the strategy developed successively in the papers \cite{junker,JS,GW,GW2}. Specifically, its outcome is that the Cauchy evolution of $P$ can be written as
\beq\label{eq:modsmoj}
\cU(s,t)=T(t) \cU^{\adg}(t,s) T(s)^{-1}
\eeq
where $T(t)$ is a $2\times 2$ matrix of pseudo\-differential operators (smoothly depending on $t$). The superscript $\adg$ stands for `almost diagonal' and  indeed $\cU^{\adg}(s,t)$ is the Cauchy evolution of a time-dependent operator of the form $\i\p_t+H^{\adg}(t)$, where
\[
H^{\adg}(t)=\mat{\epsilon^{+}(t)}{0}{0}{\epsilon^-(t)},
\]
modulo smooth terms (more precisely, modulo terms in $C^\infty(\rr^2,\cW^{-\infty}(\Sigma)\otimes \cc^2)$, where $\cW^{-\infty}(\Sigma)$ are the operators that map $H^{-m}(\Sigma)$ to $H^{m}(\Sigma)$ for each $m\in\nn$), and $\epsilon^\pm(t)$ are elliptic pseudodifferential operators of order $1$ with principal symbol $\pm(k\cdot \alth_t^{-1} k)^{\12}$. Now, because of this particular form of the principal symbol, solutions of $(\i^{-1}\p_t +\epsilon^\pm(t))$ propagate with wave front set in $\cN^\pm$. This serves one to prove that if we fix some $t_0\in\rr$ and set
\[
c^\pm_{\rm ref}(t_0)\defeq T(t_0)\pi^\pm  T^{-1}(t_0), \ \mbox{ where \ } \pi^+=\mat{\one}{0}{0}{0}, \ \ \pi^-=\mat{0}{0}{0}{\one},
\]
then $\Lambda_{\rm ref}^\pm(t,s)\defeq \mp\pi_0\cU(t,t_0)c_{\rm ref}^\pm(t_0)\cU(t_0,s)\pi_1^*$ have wave front set only in $\cN^\pm\times\cN^\pm$ and therefore satisfy the Hadamard condition \eqref{eq:had1}. As a consequence, to prove the Hadamard condition for $\Lambda^\pm_{\inout}$ it suffices to show that 
\[
c^\pm_\inout - c_{\rm ref}^\pm \in \cW^{-\infty}(\Sigma)\otimes B(\cc^2).
\]
To demonstrate that this is the case, we use assumption $(\Htd)$ to control the decay in time of various remainders in identities `modulo smooth'. The most crucial estimate here is
\beq\label{eq:tima}
H^{\adg}(t)-\begin{pmatrix}a(t)^{\12} & 0 \\ 0 & -a(t)^{\12} \end{pmatrix}\in \Psi_\td^{0,-1-\delta}(\rr;\Sigma)\otimes B(\cc^2),
\eeq
which then yields time-decay of various commutators that appear in the proofs. We obtain \eqref{eq:tima} by revisiting the approximate diagonalization \eqref{eq:modsmoj} using poly-homogeneous expansions of pseudodifferential operators in $\Psi_\td^{m,-\delta}(\rr;\Sigma)$ in both $m$ and $\delta$; more details are given in Sect. \ref{secscat}. 

\subsubsection{Proof of Fredholm statement} The proof of our second result, Theorem \ref{thm:main2}, is based on a refinement of the above strategy. First, we show that the original problem on asymptotically Minkowski spacetimes can always be reduced to a special case of assumption $(\Htd)$, with $\Sigma=\rr^d$ and the $\Psi^{m,\delta}_\td$ remainders replaced by a subclass $\Psi^{m,\delta}_\std$ that accounts for decay in both $t$ and $\rx$ (rather than just in $t$). This allows us to derive a better estimate for $c^\pm_\inout - c_{\rm ref}^\pm$, in particular with decay in $\x$ that is sufficient to get that $c^\pm_\inout - c_{\rm ref}^\pm$ is a compact operator. This way, we conclude that
\beq\label{eq:compa}
c^+_\inn + c^-_\out = \one + \hbox{a compact, smoothing term},
\eeq
so in particular $c^+_\inn + c^-_\out$ is Fredholm. From this point on we can use standard arguments from Fredholm theory, to a large extent drawing from \cite{BS}. To give only a rough intuition, let us point out that if we took instead of $\varrho_\aF$ standard scattering data at future or past infinity, then the associated boundary conditions would simply give rise to the forward or backward inhomogeneous problem, which is invertible. If for the sake of the argument, $c^\pm_\inn=c^\pm_\out$ then the same can be said about the Feynman problem. Although generically,  $c^\pm_\inn$ does not equal $c^\pm_\out$, \eqref{eq:compa} ensures that we are in the `next best possible' case, where the obstruction to invertibility of $c^+_\inn + c^-_\out$ is finite dimensional. 

The construction of the Feynman parametrix $G_{\F}$ is then based on a formula that makes use of the approximate diagonalization again. It is interesting to note that although our techniques differ a lot from that of \cite{VW}, the final formulas are quite similar. This provides further evidence that our result can be seen as an analogue of that of \cite{GHV} in the case of the massive Klein-Gordon equation.


It is also worth mentioning that compactness of the remainder term in \eqref{eq:compa} was already studied in an analogous problem for the Dirac operator on Minkowski space with external potentials \cite{matsui1,matsui2,BH}, where index formulas have also been derived and the interpretation of the index in terms of particle creation was discussed (see also \cite{BS,BS2}). An interesting topic of further research would thus be to find a short-hand index formula in our setting.

\subsection{Plan of the paper} The paper is structured as follows.

In Sect. \ref{sec1} we fix some basic terminology and recall the definition of two-point functions and covariances of states in the context of non-interacting Quantum Field Theory. 

Sect. \ref{secbounded} contains a brief overview of the pseudo\-differential calculus on manifolds of bounded geometry. In Sect. \ref{sec2} we recall the construction of two-point functions of generic Hadamard states from \cite{GW,bounded}. We then introduce the time-dependent pseudodifferential operator classes $\Psi_\td^{m,\delta}$, $\Psi_\std^{m,\delta}$ and state some of their properties, in particular we give a variant of Seeley's theorem on powers of pseudo\-differential operators elliptic in the standard $\Psi^m$ sense.

In Sect. \ref{sec2} we first recall the approximate diagonalization of the Cauchy evolution used in \cite{bounded} to construct generic Hadamard states. We then give a refinement in the setup of assumptions $(\Htd)$ and $(\Hstd)$ (the $\Psi^{m,\delta}_\std$ analogue of the former) by showing decay of various remainder terms.

Sect. \ref{inout} contains the construction of \textit{in}/\textit{out} states and the proof of their Hadamard property in the case of asymptotically static spacetimes (assumptions $(\ast)$ and $(\pos)$). The key ingredients are the reduction to the setup of assumption $(\Htd)$ and the estimates obtained in Sect. \ref{sec2}.

In Sect. \ref{sec:abstract} we set up a Fredholm problem for the Klein-Gordon operator, assuming hypothesis $(\Hstd)$. We also construct a parametrix with Feynman type wave front set and prove that the remainder terms are compact operators. An important role is played by the approximate diagonalization and the estimates from Sect. \ref{sec2}.

Finally, in Sect. \ref{sec:ams} we consider asymptotically Minkowski spacetimes $(\aM)$. We show that in this case, using the procedure from Sect. \ref{inout} one is reduced to assumption $(\Hstd)$. This allows us to adapt the results from Sect. \ref{sec:abstract} and to prove Thm. \ref{thm:main2}.

Various auxiliary proofs are collected in Appendix \ref{secapp1}. 

\section{Preliminaries}\init\label{sec1}

\subsection{Notation}\label{sec1.1}The space of differential operators (of order $m$) over a smooth manifold $M$ (here always without boundary) is denoted  $\Diff(M)$ ($\Diff^m(M)$). The space of smooth functions on $M$ with compact support is denoted $\coinf(M)$.

The operator of multiplication by a function $f$ will be denoted by $f$, while the operators of partial differentiation will be denoted by $\pe_{i}$, so that $[\pe_{i}, f]= \p_{i}f$.

\begin{notations}
\item If $a, b$ are selfadjoint operators on a Hilbert space $\cH$, we write $a\sim b$ if
\[
  a,b >0, \ \ \Dom a^{\12}= \Dom b^{\12}, \ \ c^{-1}b\leq a \leq cb,
\]
for some  constant $c>0$.

\item Similarly, if  $I\subset \rr$ is an open interval and $\{\cH_{t}\}_{t\in I}$ is a family of Hilbert spaces with $\cH_{t}= \cH$ as topological vector spaces, and $a(t), b(t)$ are two selfadjoint operators on   $\cH_{t}$, we write $a(t)\sim b(t)$ if for each $J\Subset I$ there exist constants $c_{1, J}, c_{2, J}>0$  such that
  \beq\label{equitd}
  a(t), b(t) \geq c_{1, J}>0, \ \  c_{2, J}b(t)\leq a(t) \leq c_{2, J}^{-1}b(t), \ t\in J.
\eeq
\end{notations}

\subsection{Klein-Gordon operator}\label{ssec:classical}

Let $(M,\altg)$ be a Lorentzian spacetime (we use the convention $(-,+,\dots,+)$ for the Lorentzian signature).  We consider the Klein-Gordon operator with a real-valued potential $\altV\in\cf(M)$ 
\[
P=-\Box_g + \altV \in\Diff^2(M), 
\]
Since $\altV$ is real-valued we have $P=P^*$ in the sense of formal adjoints with respect to the $L^2(M,g)$ scalar product, naturally defined using the volume form.  

For $K\subset M$ we denote $J^\pm(K)\subset M$ its causal future/past, see e.g. \cite{BF,W}. Let $C^\infty_\pm(M)$ be the space of smooth functions whose support is future or past compact, that is
\[
C^\infty_\pm(M) = \{ f\in \cf(M) : \  \supp f\subset J^\pm(K) \mbox{ \ for\ some\ compact\ } K\subset M\}.
\]
We assume that $(M,\altg)$ is globally hyperbolic, i.e. admits a foliation by Cauchy surfaces\footnote{Let us recall that a Cauchy surface is a smooth hypersurface that is intersected by every inextensible, non-spacelike (i.e. causal) curve exactly once.} (in the next sections we will impose more restrictive conditions on $(M,\altg)$, but these are irrelevant for the moment). It is well known that $P$ has then unique \emph{advanced}/\emph{retarded propagators}, i.e. operators $G_\pm:C^\infty_\pm(M)\to C^\infty_\pm(M)$ s.t.
\beq\label{eq:Pinverse}
 P G_\pm = \one \mbox{ \ on \ } C^\infty_\pm(M).
\eeq
The domain of definition of $G_\pm$ on which \eqref{eq:Pinverse} holds true can actually be increased, this will be shown in a more specific setup in later sections.

A standard duality argument using $P=P^*$, (\ref{eq:Pinverse}), and the fact that $C^\infty_+(M)\cap C^\infty_-(M)=\cof(M)$ on globally hyperbolic spacetimes,  gives $G_+^*=G_-$ as sesquilinear forms on $\cof(M)$. The \emph{causal propagator} (often also called \emph{Pauli-Jordan commutator function}) of $P$ is by definition $G\defeq G_+-G_-$, interpreted here as a map from  $\cof(M)$ to  $C^\infty_+(M)+C^\infty_-(M)$, the space of space-compact smooth functions.

\subsection{Symplectic space of solutions} In what follows we recall the relation between quasi-free states, two-point functions, and field quantization. The reader interested only in the analytical aspects can skip this discussion and move directly to equations \eqref{enlambda1}--\eqref{enlambda3}, which can be taken as the definition of two-point functions in the present context. 

By a \emph{phase space} we will mean a pair $(\cV,q)$ consisting of a complex vector space $\cV$ and a non degenerate hermitian form $q$ on $\cV$. In our case the phase space of interest (i.e. the phase space of the classical non-interacting scalar field theory) is
\beq\label{defo}
\cV\defeq \frac{\cof(M)}{P\cof(M)}, \quad \bar u \,q v\defeq \i^{-1}(u| G v),
\eeq
where $(\cdot|\cdot)$ is the $L^2(M,\altg)$ pairing, canonically defined using the volume form. The sesquilinear form $q$ is indeed well-defined on the quotient space $\cof(M)/P\cof(M)$ because $PG=GP=0$ on test functions. Using that $G_+^*=G_-$ one shows that $q$ is hermitian, and it is also not difficult to show that it is non-degenerate.

 Note that in contrast to most of the literature, we work with hermitian forms rather than with real symplectic ones, but the two approaches are equivalent.

\subsection{States and their two-point functions}\label{ss:qfree}

Let $\cV$ be a complex vector space, $\cV^{*}$ its anti-dual and $L_{\rm h}(\cV, \cV^{*})$ the space of hermitian sesquilinear forms on $\cV$. If $q\in L_{\rm h}(\cV, \cV^{*})$ then we can define the { polynomial  CCR $*$-algebra ${\rm CCR}^{\rm pol}(\cV,q)$ (see e.g. \cite[Sect. 8.3.1]{derger}) \footnote{See also \cite{GW,wrothesis} for remarks on the transition between real and complex vector space terminology.}. It is constructed as the span of the so-called {\em abstract complex fields} $\cV\ni v\mapsto \psi(v), \psi^{*}(v)$, which are taken to be anti-linear, resp. linear in $v$ and are subject to  the canonical commutation relations
\[
[\psi(v), \psi(w)]= [\psi^{*}(v), \psi^{*}(w)]=0,  \ \ [\psi(v), \psi^{*}(w)]=  \bar{v} q w \one, \ \ v, w\in \cV.
\]
Our main object of interests are the states\footnote{Let us recall that a state $\omega$ is a linear functional on ${\rm CCR}^{\rm pol}(\cV,q)$ such that $\omega(a^* a)\geq 0$ for all $a$ in ${\rm CCR}^{\rm pol}(\cV,q)$, and $\omega(\one)=1$.} on ${\rm CCR}^{\rm pol}(\cV,q)$.

The \emph{complex covariances}  $\Lambda^\pm\in L_{\rm h}(\cV,\cV^*)$ of a state $\omega$ on ${\rm CCR}^{\rm pol}(\cV,q)$ are defined in terms of the abstract field operators by
\beq\label{eq:lambda}
\bar{v}\Lambda^+ w = \omega\big(\psi(v)\psi^*(w)\big), \quad \bar{v}\Lambda^- w = \omega\big(\psi^*(w)\psi(v)\big), \quad v,w\in \cV
\eeq
Note that both $\Lambda^\pm$ are positive and by the canonical commutation relations one always has $\Lambda^+ - \Lambda^- = q$. We are  interested in the reverse construction, namely if one has a pair of hermitian forms $\Lambda^\pm$ such that  $\Lambda^+ - \Lambda^- = q$ and $\Lambda^\pm\geq 0$ then there is a unique \emph{quasi-free }state $\omega$ such that (\ref{eq:lambda}) holds. 
We will thus further restrict our attention to quasi-free states and more specifically to their complex covariances $\Lambda^\pm$. 

In QFT (at least for scalar fields) the phase space of interest is the one  defined in (\ref{defo}). In that specific case it is convenient to consider instead of complex covariances a pair of operators $\Lambda^\pm:\cof(M)\to\cf(M)$ such that 
\beq\label{eq:lambdab}
(v|\Lambda^+ w) = \omega\big(\psi(v)\psi^*(w)\big), \quad (v|\Lambda^- w) = \omega\big(\psi^*(w)\psi(v)\big), \quad v,w\in \cof(M).
\eeq
We call $\Lambda^\pm$ the \emph{two-point functions} of the state $\omega$ and identify them with the associated complex covariances whenever possible. Note that because  $(\cdot|\Lambda^\pm\cdot)$ has to induce a hermitian form on the quotient space $\cof(M)/P \cof(M)$, the two-point functions have to satisfy $P\Lambda^\pm=\Lambda^\pm P=0$ on $\cof(M)$. By the Schwartz kernel theorem we can further identify $\Lambda^\pm$ with a pair of distributions on $M\times M$, these are then bi-solutions of the Klein-Gordon equation. 

In QFT on curved spacetime one is especially interested in the subclass of quasi-free \emph{Hadamard states} \cite{KW,radzikowski}. These can be defined as in the introduction \eqref{eq:had1}, or equivalently just by requiring that the primed wave front set of the  Schwartz kernel of $\Lambda^\pm$ is contained in $\cN^\pm\times \cN^\pm$ (cf. \cite{SV,sanders}), $\cN^\pm\subset T^*M\setminus\zero$ being the two connected components of the characteristic set of $P$ (and $\zero\subset T^*M$ the zero section). To sum this up, specifying a Hadamard state amounts to constructing a pair of operators $\Lambda^\pm:\cof(M)\to\cf(M)$ satisfying the properties\footnote{Especially in the literature on QFT on curved spacetimes one uses frequently the following alternative convention: one assumes that the Schwartz kernel $\Lambda^-(x,y)$ equals $\Lambda^+(y,x)$ (this can be always ensured by taking an appropriate average if necessary), in which case one positivity condition $\Lambda^+\geq 0$ implies the other one. One rather speaks then of one two-point function (often denoted $\omega_2(x,y)$ or $W_2(x,y)$) instead of a pair.}:
\begin{flalign}
\label{enlambda1} &P\Lambda^\pm=\Lambda^\pm P=0, \ \ \Lambda^+-\Lambda^-=\i G,\\
\label{enlambda2} &\Lambda^\pm\geq 0,\\
\label{enlambda3} &\wf'(\Lambda^\pm)\subset \cN^\pm\times\cN^\pm.
\end{flalign}
Existence of generic two-point functions as above was proved in \cite{FNW}, and an alternative argument was given in \cite{GW}. Here we will be interested in showing \eqref{enlambda3} for specific two-point functions with prescribed asymptotic properties.

\subsection{Cauchy data of two-point functions}\label{ssec:cauchy} We will need a version of two-point functions acting on Cauchy data of $P$ instead of spacetime quantities such as $\Lambda^\pm$. To this end, let $\{ \Sigma_s\}_{s\in \rr}$ be a foliation of $M$ by Cauchy surfaces (since all $\Sigma_s$ are diffeomorphic we occasionally write $\Sigma$ instead). We define the map
\[
\varrho_s   u \defeq (u, \i^{-1} n^a\nabla_a u)\traa{\Sigma_s},
\]
acting on distributions $u$ such that the restriction ${}\traa{\Sigma_s}$ makes sense, where $n^a$ is the unit normal vector to $\Sigma_s$. It is well-known that $\varrho_s\circ G$ maps $\cof(M)$ to $\cof(\Sigma_{s})$ and that there exists an operator $G(s)$ acting on $\cof(\Sigma)\otimes\cc^2$ (not to be confused with $G$) that satisfies
\beq\label{eq:idGs}
G \eqdef (\varrho_s G)^* \circ  G(s) \circ \varrho_s G,
\eeq
where $(\varrho_s G)^*$ is the formal adjoint of $\varrho_s \circ G$ wrt. the $L^2$ inner product on $\Sigma_s\sqcup\Sigma_s$ respective to some density (that can depend on $s$, later on we will make that choice more specific).  We also set
\[
q(s)\defeq \i^{-1} G(s),
\]
so that $q(s)^*=q(s)$. 

The next result provides a Cauchy surface analogue of the two-point functions $\Lambda^\pm$, cf.  \cite{GW2} for the proof.
\begin{proposition}\label{minusu}
For any $s\in\rr$ the maps:
 \beq\label{eq:nimnim}\lambda_{}^{\pm}(s)\mapsto \Lambda^{\pm}\defeq  (\varrho_s G)^{*}\lambda^{\pm}(s)(\varrho_s G),
 \eeq
 and 
 \begin{equation}
\label{eq:nim}
\Lambda^{\pm}\mapsto\lambda^{\pm}(s)\defeq (\varrho^{*}_s G(s))^{*} \Lambda^{\pm} (\varrho^{*}_s G(s))
\end{equation}
are bijective and inverse from one another.
\end{proposition}
It is actually convenient to make one more definition and set:
\beq\label{eq:nimnimnim}
c^\pm(s)=\pm \i^{-1} G(s) \lambda^\pm(s) : \cof(\Sigma)\otimes B(\cc^2)\to \cf(\Sigma)\otimes B(\cc^2).
\eeq
We will simply call $c^\pm(s)$ the \emph{(time-$s$) covariances of the state $\omega$}. A pair of operators $c^\pm(s)$ are covariances of a state iff the operators $\Lambda^\pm$ defined by (\ref{eq:nimnim}) and (\ref{eq:nimnimnim}) satisfy (\ref{enlambda1})-(\ref{enlambda2}), which is equivalent to the conditions
\begin{flalign}
&c^+(s)+c^-(s)=\one,\label{eq:secondcondlam0}\\
&\lambda^\pm(s)\geq 0,\label{eq:secondcondlam}
\end{flalign}
where we identified the operators $\lambda^\pm(s)$ with hermitian forms using the same pairing as when we took the formal adjoint in \eqref{eq:idGs}. Note that \eqref{eq:secondcondlam0} can also be expressed as $\lambda^+(s)-\lambda^-(s)=q(s)$. 
 
Additionally, a state (recall that we consider only quasi-free states) is \emph{pure} iff its covariances $c^\pm(s)$ extend to projections on the completion of $\cof(\Sigma)\otimes\cc^2$ w.r.t. the inner product given by $\lambda^+ + \lambda^-$. In practice it is sufficient to construct $c^\pm(s)$ as projections acting on a space that is big enough to contain $\cof(\Sigma)\otimes\cc^2$, but small enough to be contained in the Hilbert space associated to $\lambda^+ + \lambda^-$.
 
\subsection{Propagators for the Cauchy evolution}\label{ssec:propaCauchy}

Recall that we have defined the operator $G(s)$ via the identity
\beq\label{eq:idGs2}
G \eqdef (\varrho_s G)^* \circ  G(s) \circ \varrho_s G.
\eeq
A direct consequence is that the operator $G^* \varrho_s G(s)$ assigns to Cauchy data on $\Sigma_s$ the corresponding solution. Similarly, for $t,s\in\rr$ the operator
\beq\label{eq:defce}
\cU(s,t)\defeq \varrho_s G^* \varrho_t^* G(t)
\eeq
produces Cauchy data of a solution on $\Sigma_s$ given Cauchy data on $\Sigma_t$. We will call $\{ \cU(s,t)\}_{s,t\in\rr}$ the \emph{Cauchy evolution} of $P$. A straightforward computation gives the group property
\beq
\cU(t,t)=\one, \quad \cU(s,t')\cU(t',t)=\cU(s,t), \ \ t'\in\rr;
\eeq
and the conservation of the symplectic form by the evolution
\beq\label{eq:conserveq}
\cU^*(s,t)q(s)\cU(s,t)=q(t).
\eeq
These identities allow to conclude that the covariances $c^\pm(t)$ (and two-point functions $\lambda^\pm(t)$) at different `times' of a quasi-free state are related by
\beq\label{eq:tiatc}
\bea
\lambda^\pm(t)&=\cU(s,t)^* \lambda^\pm(s) \cU(s,t),\\
c^\pm(t) &=\cU(t,s) c^\pm(s) \cU(s,t).
\eea
\eeq
Notice that this induces a splitting of the evolution in two parts:
\[
\cU(s,t)=\cU^+(s,t)+\cU^-(s,t), \mbox{ \ with \ } \cU^\pm(s,t)=\cU(s,t)c^\pm(t).
\]
If the state is pure then $c^\pm(t)$ are projections for all $t$ and the operators $\cU^\pm(s,t)$ obey the composition formula
\[
\cU^\pm(s,t')\cU^\pm(t',t)=\cU^\pm(s,t), \ \ \cU^\pm(s,t')\cU^\mp(t',t)=0, \ \ t'\in\rr.
\]
Let us stress that $\cU^\pm(t,t)$ is not the identity, but rather equals $c^\pm(t)$. Furthermore, if the state is Hadamard then $\cU(s,t)c^\pm(t)$ propagate singularities along $\cN^\pm$ (see the discussion in \cite{GW2}). In Sect. \ref{sec2} we will be interested in the reversed argument, namely we will construct covariances $c^\pm(t)$ of pure Hadamard states from a splitting of the evolution $\cU(t,s)$ into two parts that propagate singularities along respectively $\cN^+$, $\cN^-$.  

\section{Pseudodifferential calculus on manifolds of bounded geometry}\init\label{secbounded}
\subsection{Manifolds of bounded geometry}\label{ss:mbg} In the present section we introduce manifolds of bounded geometry and review the pseudodifferential calculus of Kordyukov and Shubin \cite{Ko,Sh2}, making also use of some results from \cite{bounded}.

Let us denote by $\altdelta$ the flat metric on $\rr^{d}$ and by $B_{d}(\ry, r)\subset \rr^{d}$ the open ball of center $\ry$ and radius $r$.


 

If $(\Sigma, \alth)$ is a $d-$dimensional Riemannian manifold and $X$ is a  $(p,q)$ tensor  on $\Sigma$, we can define the canonical norm of $X(\rx)$, $\rx\in \Sigma$, denoted by $\|X\|_{\rx}$, using appropriate tensor powers of $\alth(\rx)$ and $\alth^{-1}(\rx)$.  $X$ is {\em bounded} if $\sup_{\rx\in \Sigma}\| X\|_{\rx}<\infty$.

If $U\subset \Sigma$ is open, we denote by ${\rm BT}^{p}_{q}(U, \altdelta)$ the Fr\'echet space of $(p,q)$ tensors on $U$, bounded with all covariant derivatives in the above sense. Among several equivalent definitions of manifolds of bounded geometry (see \cite{Sh2,bounded}), the one below is particularly useful in applications.

\begin{definition}\label{thp0.1}
A Riemannian manifold $(\Sigma,\alth)$ is of bounded geometry iff for each $\rx\in \Sigma$, there exists an open neighborhood of $\rx$, denoted $U_{\rx}$, and a smooth diffeomorphism
\[
\psi_{\rx}: U_{\rx} \xrightarrow{\sim} B_{d}(0,1)\subset \rr^{d}
\]
with $\psi_{\rx}(\rx)=0$, and such that if $\alth_{\rx}\defeq (\psi_{\rx}^{-1})^{*}\alth$ then:

\noindent
{\rm (C1)} the family $\{\alth_{\rx}\}_{\rx\in \Sigma}$ is  bounded in ${\rm BT}^{0}_{2}(B_{d}(0,1), \altdelta)$,

\noindent
 {\rm (C2)} there exists $c>0$ such that :
\[
c^{-1}\altdelta\leq \alth_{\rx}\leq c \altdelta, \ \rx\in \Sigma.
\]

 A family $\{U_{\rx}\}_{\rx\in \Sigma}$ resp. $\{\psi_{\rx}\}_{\rx\in \Sigma}$ as above will be called a family of {\em good chart neighborhoods}, resp. {\em good chart diffeomorphisms}.
\end{definition}

A known result (see \cite[Lemma 1.2]{Sh2}) says that one can find a covering $\Sigma=\bigcup_{i\in \nn}U_{i}$ by good chart neighborhoods $U_{i}= U_{\rx_{i}}$ $(\rx_{i}\in \Sigma)$ which is {\em uniformly finite}, i.e. there exists $N\in \nn$ such that $\bigcap_{i\in I}U_{i}= \emptyset$ if $\sharp I> N$. Setting $\psi_{i}= \psi_{\rx_{i}}$, we will call the sequence $\{U_{i}, \psi_{i}\}_{i\in \nn}$ a {\em good chart covering} of $\Sigma$. 

Furthermore, by \cite[Lemma 1.3]{Sh2} one can associate to a good chart covering a partition of unity:
\[
1=\sum_{i\in \nn}\chi_{i}^{2},  \ \ \chi_{i}\in \coinf(U_{i})
\]
such that $\{(\psi_{i}^{-1})^{*}\chi_{i}\}_{i\in \nn}$ is a bounded sequence in $\cinf_{\rm b}(B_{d}(0,1))$. Such a partition of unity will be called a {\em good partition of unity}.
\subsection{Bounded tensors and bounded diffeomorphisms}
\begin{definition}\label{defp0.2}
 Let $(\Sigma,\alth)$ be of  bounded geometry. We denote by ${\rm BT}^{p}_{q}(\Sigma,\alth)$ the spaces of  smooth $(q,p)$ tensors $X$ on $\Sigma$ such that if $X_{\rx}= (\exp_{\rx}^{\alth}\circ e_{\rx})^{*}X$, where $e_{\rx}: (\rr^{d}, \delta)\to (T_{\rx}\Sigma, h(\rx))$ is an isometry,  then the family 
 $\{X_{\rx}\}_{x\in \Sigma}$ is  bounded in ${\rm BT}^{p}_{q}(B_{d}(0, \frac{r}{2}), \altdelta)$. We equip ${\rm BT}^{p}_{q}(\Sigma, \alth)$ with its natural Fr\'echet space topology.
 
  We denote by $\cinfb(\rr; \BT^{p}_{q}(\Sigma, \alth))$ the space of smooth maps $\rr\in t\mapsto X(t)$ such that $\p_{t}^{n}X(t)$ is  uniformly bounded in $\BT^{p}_{q}(\Sigma, \alth)$ for $n\in \nn$. 
  
  We denote by $S^{\delta}(\rr; \BT^{p}_{q}(\Sigma, \alth))$, $\delta\in \rr$ the space of smooth maps $\rr\in t\mapsto X(t)$
  such that $\langle t\rangle^{-\delta+ n}\p_{t}^{n}X(t)$ is uniformly bounded in $\BT^{p}_{q}(\Sigma, \alth)$ for $n\in \nn$. 
 \end{definition}
It is well known (see e.g. \cite[Subsect. 2.3]{bounded}) that  we can replace in Def. \ref{defp0.2} the geodesic maps $\exp_{\rx}^{\alth}\circ e_{\rx}$ by $\psi_{\rx}^{-1}$, where $\{\psi_{\rx}\}_{\rx\in \Sigma}$ is any family of good chart diffeomorphisms as in Thm. \ref{thp0.1}. 
\begin{definition}\label{defdeboun}
 Let $(\Sigma, \alth)$ be an $n-$dimensional  Riemannian manifold of bounded geometry and $\chi: \Sigma\to \Sigma$ a smooth diffeomorphism. One says that $\chi$ is a {\em bounded diffeomorphism} of $(\Sigma, \alth)$ if for some some family of good chart diffeomorphisms $\{U_{x}, \psi_{x}\}_{x\in \Sigma}$,  
the maps
\[
\chi_{x}= \psi_{\chi(x)}\circ \chi\circ \psi_{x}^{-1}, \ \chi^{-1}_{x}= \psi_{\chi^{-1}(x)}\circ \chi^{-1}\circ \psi_{x}: B_{n}(0, 1)\to B_{n}(0,1)
\]
are bounded in $\cinfb(B_{n}(0,1))$ uniformly with respect to $x\in M$.
\end{definition}
It is easy to see that if the above properties are satisfied for some family of good chart  diffeomorphisms  then they are satisfied for any such family, furthermore bounded diffeomorphisms are stable under composition.

\subsection{Symbol classes}\label{symbolo}
We recall some well-known definitions about symbol classes on manifolds of bounded geometry, following \cite{Sh2, Ko, alnv1}.
\subsubsection{Symbol classes on $\rr^{n}$}\label{secp1.2.1}
Let $U\subset \rr^{d}$ be  an open set, equipped with the flat metric $\altdelta$ on $\rr^{d}$.

 We denote by $S^{m}(T^{*}U)$, $m\in \rr$,  the space of $a\in \cinf(U\times \rr^{d})$ such that
\[
 \langle \spexi\rangle^{-m+|\beta|}\p_{x}^{\alpha}\p_{\xi}^{\beta}a(\rx, \spexi)\hbox{  is  bounded on }U\times \rr^{d}, \ \forall \alpha, \beta\in \nn^{n}, 
\]
equipped with its canonical semi-norms $\| \cdot\|_{m,\alpha, \beta}$. 

We set 
\[
S^{-\infty}(T^{*}U)\defeq  \bigcap_{m\in \rr}S^{m}(T^{*}U),\ \ S^{\infty}(T^{*}U)\defeq  \bigcup_{m\in \rr}S^{m}(T^{*}U),
\]
 with their canonical Fr\'echet space topologies. If $m\in \rr$ and $a_{m-i}\in S^{m-i}(T^{*}U)$ we write
$a\simeq \sum_{i\in \nn}a_{m-i}$ if for each $p\in \nn$ 
\begin{equation}
\label{ep1.1}
r_{p}(a)\defeq  a- \sum_{i=0}^{p}a_{m-i}\in S^{m-p-1}(T^{*}U).
\end{equation}
It is well-known (see e.g. \cite[Sect. 3.3]{shubin}) that if $a_{m-i}\in S^{m-i}(T^{*}U)$, there exists $a\in S^{m}(T^{*}U)$, unique modulo $S^{-\infty}(T^{*}U)$ such that $a\simeq \sum_{i\in \nn}a_{m-i}$.

We denote by $S^{m}_{\rm h}(T^{*}U)\subset S^{m}(T^{*}U)$ the  space of $a$ such that $a(\rx, \lambda k)= \lambda^{m}a(\rx, \spexi)$, for $\rx\in U$, $|\spexi|\geq C$, $C>0$ and by $S^{m}_{\rm ph}(T^{*}U)\subset S^{m}(T^{*}U)$ the space of $a$ such that $a\simeq \sum_{i\in \nn}a_{m-i}$ for a sequence $a_{m-i}\in S^{m-i}_{\rm h}(T^{*}U)$ ($a$ is then called a \emph{poly-homogeneous}\footnote{These are also called classical symbols in the literature.} symbol). 
Following \cite{alnv1} one equips  $S^{m}_{\rm ph}(T^{*}U)$ with  the topology defined by  the semi-norms of $a_{m-i}$ in $S^{m-i}(T^{*}U)$ and $r_{p}(a)$ in $S^{m-p-1}(T^{*}U)$, (see \eqref{ep1.1}). This topology is strictly stronger than the topology induced by $S^{m}(T^{*}U)$.

The space  $S^{m}_{\rm ph}(T^{*}U)/S^{m-1}_{\rm ph}(T^{*}U)$ is isomorphic to $S^{m}_{\rm h}(T^{*}U)$, and the image of $a$ under the quotient map is called the {\em principal symbol} of $a$ and denoted by $\sigma_{\rm pr}(a)$. 

If $U= B_{n}(0,1)$ (more generally, if $U$ is relatively compact with smooth boundary), there exists a  continuous extension map $E: S^{m}(T^{*}U)\to S^{m}(T^{*}\rr^{d})$ such that $E a\traa{T^{*}U}=a$. Moreover $E$ maps $S^{m}_{\rm ph}(T^{*}U)$  into $S^{m}_{\rm ph}(T^{*}\rr^{d})$  and is continuous for the topologies of $S^{m}_{\rm ph}(T^{*}U)$ and $S^{m}_{\rm ph}(T^{*}\rr^{d})$, which means that all the maps
\[
a\mapsto (Ea)_{m-i},  \ \ a\mapsto r_{p}(Ea),
\]
are continuous.
\subsubsection{Time-dependent symbol classes on $\rr^{d}$}\label{timdep}
 We will also need to consider various classes of {\em time-dependent} symbols $a(t, \rx, \spexi)\in \cinf(\rr\times T^{*}U)$. 
 First of all the space $\cinf(\rr; S^{m}(T^{*}U))$  is defined as the 
  space of $a\in  \cinf(\rr\times T^{*}U)$ such that
 \[
 \langle \spexi\rangle^{-m+|\beta|}\p_{t}^{\gamma}\p_{\rx}^{\alpha}\p_{\spexi}^{\beta}a(t,\rx, \spexi)\hbox{  is bounded on }I\times U\times \rr^{d}, \ \forall \alpha, \beta\in \nn^{n}, \ \gamma\in \nn,
\]
for any interval $I\Subset \rr$.
We denote by  $\cinf_{\rm b}(\rr; S^{m}(T^{*}U))$  the subspace of symbols which are uniformly bounded   in $S^{m}(T^{*}U)$  with all time derivatives.

Furthermore, anticipating the need for some additional decay in $t$ in Sect. \ref{secscat}, we denote by $S^{\delta}(\rr; S^{m}(T^{*}U))$ the space of $a\in  \cinf(\rr\times T^{*}U)$ such that
 \[
\langle t\rangle^{\delta-\gamma} \langle \spexi\rangle^{-m+|\beta|}\p_{t}^{\gamma}\p_{\rx}^{\alpha}\p_{\spexi}^{\beta}a(t,\rx, \spexi)\hbox{  is bounded on }\rr\times U\times \rr^{d}, \ \forall \alpha, \beta\in \nn^{n}, \ \gamma\in \nn.
\]
The  notation $a\sim \sum_i {a_{m-i}}$ and the poly-homogeneous spaces   
\[
\cinf_{({\rm b})}(\rr; S^{m}_{\rm ph}(T^{*}U)), \ \ S^{\delta}(\rr; S^{m}_{\rm ph}(T^{*}U)),
\]
are defined analogously, by requiring estimates on the time derivatives of the $a_{m-i}$ and $r_{p}$ in \eqref{ep1.1}.

\subsubsection{Symbol classes on $\Sigma$}\label{secp1.2.2}
Let $(\Sigma, \alth)$ be a Riemannian manifold of bounded geometry and $\{\psi_{\rx}\}_{\rx\in \Sigma}$ a family of good chart diffeomorphisms.
\begin{definition}\label{defp.1}
 We denote by $S^{m}(T^{*}\Sigma)$  for $m\in \rr$ the space of $a\in \cinf(T^{*}\Sigma)$ such that for each $\rx\in \Sigma$,  $a_{\rx}\defeq (\psi_{\rx}^{-1})^{*}a\in S^{m}(T^{*}B_{n}(0,1))$ and the family $\{a_{\rx}\}_{\rx\in \Sigma}$ is bounded
 in $S^{m}(T^{*}B_{n}(0,1))$.  We equip  $S^{m}(T^{*}\Sigma)$ with the semi-norms
 \[
\| a\|_{m, \alpha, \beta}= \sup_{\rx\in \Sigma}\| a_{\rx}\|_{m, \alpha, \beta}.
\]
Similarly we denote by $S^{m}_{\rm ph}(T^{*}\Sigma)$ the space of $a\in S^{m}(T^{*}\Sigma)$ such that for each $\rx\in \Sigma$, $a_{\rx}\in S^{m}_{\rm ph}(T^{*}B_{n}(0,1))$ and the family $\{a_{\rx}\}_{\rx\in \Sigma}$ is bounded
 in $S^{m}_{\rm ph}(T^{*}B_{n}(0,1))$. We equip $S^{m}_{\rm ph}(T^{*}\Sigma)$ with the semi-norms
 \[
\| a\|_{m, i, p, \alpha, \beta}= \sup_{\rx\in \Sigma}\| a_{\rx}\|_{m,i, p,\alpha, \beta}.
\]
where $\| \cdot \|_{m,i, p,\alpha, \beta}$ are the semi-norms defining the topology of $S^{m}_{\rm ph}(T^{*}B_{n}(0,1))$.

We also set $S^{\infty}_{({\rm ph})}(T^{*}\Sigma)= \bigcup_{m\in \rr}S^{m}_{({\rm ph})}(T^{*}\Sigma)$.
\end{definition}
The definition of $S^{m}(T^{*}\Sigma)$, $S^{m}_{\rm ph}(T^{*}\Sigma)$ and their  Fr\'echet space topologies are  independent on the choice of  the family $\{\psi_{\rx}\}_{\rx\in \Sigma}$ of good chart diffeomorphisms.

The notation $a\simeq \sum_{i\in \nn}a_{m-i}$ for $a_{m-i}\in S^{m-i}_{\rm ph}(T^{*}\Sigma)$ is defined as before.  If $a\in S^{m}_{\rm ph}(T^{*}\Sigma)$, we denote again by $a_{\rm pr}$ the image of $a$ in $S^{m}_{\rm ph}(T^{*}\Sigma)/S^{m-1}_{\rm ph}(T^{*}\Sigma)$.

The spaces $\cinf_{({\rm b})}(\rr; S^{m}_{({\rm ph})}(T^{*}\Sigma))$,  $S^{\delta}(\rr; S^{m}_{({\rm ph})}(T^{*}\Sigma))$ are defined as in \ref{timdep} and equipped with their  natural Fr\'echet space topologies.

\subsection{Sobolev spaces and smoothing operators}\label{sobolo}
Using the metric $\alth$ one defines the Sobolev spaces $H^{\sobo}(\Sigma)$ as follows.

\begin{definition} For $s\in \rr$ the {\em Sobolev space} $H^{\sobo}(\Sigma)$ is:
 \[
H^{\sobo}(\Sigma)\defeq \langle -\Delta_{\alth}\rangle^{-\sobo/2}L^{2}(\Sigma),
\]
with its natural Hilbert space topology, where $-\Delta_{\alth}$ is the Laplace-Beltrami operator on $(\Sigma,\alth)$, strictly speaking the closure of its restriction to $\coinf(\Sigma)$. 
\end{definition}

We further set 
\[
H^{\infty}(\Sigma)\defeq \textstyle\bigcap_{m\in \zz}H^{m}(\Sigma), \quad H^{-\infty}(\Sigma)\defeq \textstyle\bigcup_{m\in \zz}H^{m}(\Sigma),
\]
equipped with their Fr\'echet space topologies.  

We denote by $\cW^{-\infty}(\Sigma)$ the Fr\'echet space  $B(H^{-\infty}(\Sigma), H^{\infty}(\Sigma))$ with its Fr\'echet space topology, given by the semi-norms 
\[
\| a\|_{m}= \| a\|_{B(H^{-m}(\Sigma),  H^{m}(\Sigma))}, \ \ m\in \nn.
\]
This allows us to define $\cinf_{({\rm b})}(\rr; \cW^{-\infty}(\Sigma))$,  $S^{\delta}(\rr; \cW^{-\infty}(\Sigma))$, the latter consisting of operator-valued functions $a(t)$ such that
\[
\|\p_{t}^{\gamma}a(t)\|_{m}\in O(\langle t\rangle^{\delta- \gamma}), \ \ \forall \gamma, m\in \nn.
\]
\subsection{Pseudodifferential operators}\label{pdosec}
Starting from the well-known Weyl quantization on open subsets of $\rr^d$, one constructs a quantization map $\Op$ for symbols in $S^m(T^*\Sigma)$ using a  good chart covering of $\Sigma$ and good chart diffeomorphisms. More precisely let $\{U_{i}, \psi_{i}\}_{i\in \nn}$ be a good chart covering of $M$   and  
\[
\sum_{i\in \nn}\chi_{i}^{2}= \one
\]
a subordinate good partition of unity, see 
 Subsect. \ref{ss:mbg}. If
\[
(\psi_{i}^{-1})^{*}dg\eqdef  m_{i}dx,
\]
we set
\[
\begin{array}{rl}
T_{i}:&L^{2}(U_{i}, dg)\to L^{2}(B_{n}(0,1), dx),\\[2mm]
&u\mapsto m_{i}^{\12}(\psi_{i}^{-1})^{*}u,
\end{array}
\]
so that $T_{i}:L^{2}(U_{i}, dg)\to L^{2}(B_{n}(0,1), dx)$ is unitary.  We then fix an extension map 
\[
E: S^{m}_{\rm ph}(T^{*}B_{d}(0, 1))\to S^{m}_{\rm ph}(T^{*}\rr^{d}).
\]
\begin{definition}\label{def-de-op}
 Let $a= a(t)\in \cinf(\rr; S^{m}_{\rm ph}(T^{*}M))$. We set
 \[
\Op(a)\defeq  \sum_{i\in \nn}\chi_{i}T_{i}^{*}\circ \Op^{\rm w}(Ea_{i})\circ T_{i}\chi_{i},
\]
where $a_{i}\in S^{m}_{\rm ph}(T^{*}B_{d}(0,1))$ is the push-forward of $a\traa{T^{*}U_{i}}$ by $\psi_{i}$ and $\Op^{\rm w}$ is the Weyl quantization.
\end{definition}

 If $\Op'$ is another such quantization map for different choices of $U_{i}, \psi_{i}, \chi_{i}$ and $E$  then
\begin{flalign*}
&S^{m}_{\rm ph}(T^{*}\Sigma)\to \cW^{-\infty}(\Sigma)\\[2mm]
\Op - \Op': \ \ & \cinf_{({\rm b})}(\rr; S^{m}_{\rm ph}(T^{*}\Sigma))\to \cinf_{({\rm b})}(\rr; \cW^{-\infty}(\Sigma)),\\[2mm]
&S^{\delta}(\rr; S^{m}_{\rm ph}(T^{*}\Sigma))\to S^{\delta}(\rr; \cW^{-\infty}(\Sigma)),
\end{flalign*}
are bounded. Then  one defines the classes
 \[
\bea 
\Psi^{m}(\Sigma)&\defeq \Op (S^{m}_{\rm ph}(T^{*}\Sigma))+ \cW^{-\infty}(\Sigma),\\[2mm]
\cinf_{({\rm b})}(\rr; \Psi^{m}(\Sigma))&\defeq \Op(\cinf_{({\rm b})}(\rr; S^{m}_{\rm ph}(T^{*}\Sigma)))+ \cinf_{({\rm b})}(\rr; \cW^{-\infty}(\Sigma)),\\[2mm]
S^{\delta}(\rr; \Psi^{m}(\Sigma))&\defeq \Op (S^{\delta}(\rr; \Psi^{m}_{\rm ph}(T^{*}\Sigma))+ S^{\delta}(\rr; \cW^{-\infty}(\Sigma)).
\eea 
\]
Thanks to including the  ideal   $\cW^{-\infty}(\Sigma)$ of smoothing operators, the so-obtained pseudodifferential classes are stable under composition,  for example  $\Psi^{m_1}(\Sigma)\circ \Psi^{m_2}(\Sigma)\subset \Psi^{m_1+m_2}(\Sigma)$.

Note that $S^{\delta}(\rr;\Psi^{m}(\Sigma))= \langle t\rangle^{\delta}S^{0}(\rr; \Psi^{m}(\Sigma))$ and similarly with $\Psi^{m}(\Sigma)$ replaced by $\cW^{-\infty}(\Sigma)$  so in what follows one can assume without loss of generality that  $\delta=0$.

The spaces $\cW^{-\infty}(\Sigma)$, $\cinf_{({\rm b})}(\rr; \cW^{-\infty}(\Sigma))$ and $S^{\delta}(\rr; \cW^{-\infty}(\Sigma))$ have natural Fr\'echet space topologies. If necessary we equip the spaces $\Psi^{m}(\Sigma)$, $\cinf_{({\rm b})}(\rr; \Psi^{m}(\Sigma))$ and $S^{\delta}(\rr; \Psi^{m}(\Sigma))$ with the quotient topology obtained from the map:
\[
(c, R)\mapsto \Op(c)+R
\]
between the appropriate spaces.

If $a\in \Psi^{m}(\Sigma)$, the {\em principal symbol} $\sigma_{\rm pr}(a)\in S^{m}_{\rm h}(T^{*}\Sigma)$ is defined in analogy to the case $\Sigma=\rr^d$. The operator $a$ is {\em elliptic} if  there exists $C>0$ such that
\beq\label{eq:elliptic}
| \sigma_{\rm pr}(a)|\geq C |\spexi|^{m}, \ \ |\spexi|\geq 1,
\eeq
uniformly in the chart open sets. If $a\in \cinf(\rr; \Psi^{m}(\Sigma))$ we say that $a$ is {\em  elliptic} if $a(t)$ is elliptic for all $t\in\rr$ and the constant $C$ in (\ref{eq:elliptic}) is locally uniform in $t$. For $a\in \cinf_{\rm b}(\rr; \Psi^{m}(\Sigma))$ or $S^{0}(\rr; \Psi^{m}(\Sigma))$ there is also a corresponding notion of ellipticity, where we require $C$ to be uniform in $t$.

As shown in \cite{bounded}, the pseudodifferential classes $\Psi^m(\Sigma)$ fit into the general framework of Ammann, Lauter, Nistor and Vasy \cite{alnv1}, and consequently they have many convenient properties that generalize well-known facts for say, pseudodifferential operators on closed manifolds, such as the existence of complex powers for elliptic, bounded from below operators. 

We state below a particular case of {\em Seeley's theorem} for real powers, partly proved in \cite[Sect. 5]{bounded}, based on a general result from \cite{alnv1}.
\begin{theorem}[Seeley's theorem]\label{seeley}
Let $a\in \cinf(\rr; \Psi^{m}(\Sigma))$ be elliptic, selfadjoint with $a(t)\geq c(t)\one$, $c(t)>0$. Then $a^{\alpha}\in \cinf(\rr; \Psi^{m\alpha}(\Sigma))$ for any $\alpha\in \rr$ and $\sigma_{\rm pr}(a^{\alpha})(t)= \sigma_{\rm pr}(a(t))^{\alpha}$.

The same result holds replacing  $\cinf(\rr; \Psi^{m}(\Sigma))$ by $\cinf_{({\rm b})}(\rr; \Psi^{m}(\Sigma))$ or $S^{0}(\rr; \Psi^{m}(\Sigma))$ if one assumes $a(t)\geq c_{0}\one$ for $c_{0}>0$.
\end{theorem}
\proof The $\cinf_{({\rm b})}$ cases  are proved in \cite[Thm. 5.12]{bounded}, by checking that the general framework of \cite{alnv1} applies to these two situations.  The $S^{\delta}$ case can be proved similarly. The only point deserving special care is the {\em spectral invariance} of the ideal $S^{\delta}(\rr; \cW^{-\infty}(\Sigma))$, which we explain in some detail. Let $r_{-\infty}\in S^{0}(\rr; \cW^{-\infty}(\Sigma))
$, considered as a bounded operator  on $L^{2}(\rr_{t}\times \Sigma_{\rx})$. The spectral invariance property is the fact that if $\one - r_{-\infty}$ is invertible in $B(L^{2}(\rr_{t}\times \Sigma_{\rx}))$ then $(\one - r_{-\infty})^{-1}= \one- r_{1, -\infty}$ for $r_{1, -\infty}\in S^{0}(\rr; \cW^{-\infty}(\Sigma))$.
This can be however proved exactly as in \cite[Lemma 5.5]{bounded}. \qed

\subsection{Egorov's theorem} If $b(t)\in C^\infty(\rr;\Psi^m(\Sigma))$ (or more generally, if $b(t)$ is a square matrix consisting of elements of $C^\infty(\rr;\Psi^m(\Sigma))$ and $H^{-\infty}(\Sigma)$ is tensorized by powers of $\cc$ accordingly) we denote by 
\[
\cU_{b}(t,s): H^{-\infty}(\Sigma)\to H^{-\infty}(\Sigma)
\]
the evolution generated by $b(t)$, i.e. the Cauchy evolution of $\pe_t-\i b(t)$, or put in other words, the unique solution (if it exists) of the system
\beq\label{eq:defUb}
\begin{cases}
\frac{\p}{\p t}\cU_{b}(t,s)= \i  b(t)\cU_{b}(t,s),\\[2mm]
\frac{\p}{\p s}\cU_{b}(t,s)= -\i  \cU_{b}(t,s)b(s),\\[2mm]
\cU_{b}(t,s)=\one.
\end{cases}
\eeq
The existence of $\cU_{b}(t,s)$ can typically be established if $b(t)$ defines a differentiable family of self-adjoint operators on a Hilbert space, or a small perturbation of such family. Specifically, consider $b(t)\in \cinf(\rr; \Psi^{1}(\Sigma))$ such that $b(t)=b_1(t)+b_0(t)$ with $b_i(t)\in \cinf(\rr; \Psi^{i}(\Sigma))$ and:
\[
({\rm E})\quad b_1(t)\hbox{ is  elliptic and bounded from below on } H^\infty(\Sigma), \hbox{ locally uniformly in }t.
\]
Using \cite[Prop. 2.2]{alnv1} it follows that $b(t)$ is closed with domain $\Dom b(t)= H^{1}(\Sigma)$. Moreover  the map $\rr\ni t\mapsto b(t)\in B(H^{1}(\Sigma), L^{2}(\Sigma))$ is norm continuous. It follows that we can define $\cU_{b}(t,s)$, using for instance \cite[Thm. X.70]{RS}. In the present setup one can prove a result known generally as Egorov's theorem, we refer to \cite{bounded} for the details and proofs.

\begin{lemma}\label{ego.1}
Assume $({\rm E})$. Then:
\ben
\item 
 $\cU_{b}(t,s)\in B(H^{m}(\Sigma))$ for $m\in \rr$ or $m= \pm \infty$.  
 \item  if $r\in \cW^{-\infty}(\Sigma)$ then
$\cU_{b}(t,s)r, \ r\cU_{b}(s,t)\in \cinf(\rr^{2}_{t,s}, \cW^{-\infty}(\Sigma))$.

 \item if moreover $b(t)\in S^{0}(\rr; \Psi^{1}(\Sigma))$ and $b(t)- b^{*}(t)\in S^{-1- \delta}(\rr; \Psi^{0}(\Sigma))$ for $\delta>0$ then 
 $\cU_{b}(t,s)$ is uniformly bounded in $B(L^{2}(\Sigma))$.
 \een
\end{lemma}
\begin{theorem}[Egorov's theorem] 
Let $c\in \Psi^{m}(\Sigma)$ and $b(t)$ satisfying $({\rm E})$. Then
\[
c(t,s)\defeq  \cU_{b}(t,s)c \cU_{b}(s,t)\in \cinf(\rr^{2}_{t,s}, \Psi^{m}(\Sigma)).
\]
Moreover 
\[
\sigma_{\rm pr}(c)(t,s)= \sigma_{\rm pr}(c)\circ \Phi(s,t),
\]
where $\Phi(t,s): T^{*}\Sigma\to T^{*}\Sigma$ is the flow of the time-dependent Hamiltonian $\sigma_{\rm pr}(b)(t)$.
\end{theorem}
\subsection{Scattering pseudodifferential calculus on $\rr^{d}$}\label{scatcat}
In this subsection we consider a smaller class of pseudodifferential calculus for $\Sigma= \rr^{d}$ which is the natural class on asymptotically Minkowski spacetimes. For $m, \delta\in \rr$  we denote by $S^{m, \delta}_{\std}(\rr; T^{*}\rr^{d})$ the space of functions $a(t, \rx, \spexi)$ such that
\[
 \p_{t}^{\gamma}\p_{\rx}^{\alpha}\p_{\spexi}^{\beta}a(t, \rx, \spexi)\in O((\langle t\rangle+ \langle\rx\rangle)^{\delta- \gamma- |\alpha|}\langle \spexi\rangle^{m-|\beta|}),  \ \gamma\in \nn, \ \alpha, \beta\in \nn^{d}.
\]
The subscript $\std$ refers to the space-time decay properties of symbols in $(t, \rx)$.
The subspace of symbols which are poly-homogeneous in $\spexi$ will be denoted by $S^{m, \delta}_{\std, {\rm ph}}(\rr; T^{*}\rr^{d})$.

We denote by  $\cW_{\std}^{-\infty}(\rr;\rr^{d})$ the space of operator-valued functions $a(t)$ such that 
\[
\|(D_{\rx}^{2}+\rx^{2})^{m}\p_{t}^{\gamma}a(t)(D_{\rx}^{2}+\rx^{2})^{m}\|_{B(L^{2}(\rr^{d}))}\in O(\bra t\ket^{-n}), \ \forall\ m,n\in \nn.
\]
Finally we set:
\[
\Psi^{m, \delta}_{\std}(\rr; \rr^{d})\defeq\Op^{\rm w}(S^{m, \delta}_{\std, {\rm ph}}(\rr; T^{*}\rr^{d}))+ \cW_{\std}^{-\infty}(\rr;\rr^{d}),
\]
where $\Op^{\rm w}$ is the Weyl quantization. 

Omitting the variable $t$ in the above conditions,  we also obtain classes of (time-independent) symbols and pseudodifferential operators on $\rr^{d}$, which will be denoted by $S^{m, \delta}_{\rm sd}(T^{*}\rr^{d})$, $\Psi^{m, \delta}_{\rm sd}(\rr^{d})$ and $\cW^{-\infty}_{\sd}(\rr^{d})$, where the subscript $\sd$ refers to space decay properties of the symbols or operators.  

The classes $\Psi^{m, \delta}_{\sd}(\rr^{d})$ are the well-known `scattering  pseudodifferential operators', see e.g. \cite{cordes,parenti,moreshubin}.

We will only need the definitions of the principal symbol  and of ellipticity on  $\Psi^{m, 0}_{\rm sd}(\rr^{d})$ resp.  $\Psi_{\std}^{m, 0}(\rr; \rr^{d})$, which are taken here to be identical\footnote{Thus, we do \emph{not} consider here ellipticity in the sense of scattering pseudo\-differential calculus.} with  the ones in $\Psi^{2}(\rr^{2})$, resp. $\cinf_{\rm b}(\rr; \Psi^{2}(\rr^{d}))$.  Seeley's theorem is still valid for the $\Psi_{\std}^{m, 0}(\rr; \rr^{d})$ classes, proved similarly as before by a reduction to \cite{alnv1}, see the arguments in \cite[Subsect. 5.3]{bounded}.
\begin{theorem}\label{seeley-std}
 Let $a\in\Psi_{\std}^{m, 0}(\rr; \rr^{d})$ be elliptic, selfadjoint with $a(t)\geq c_{0}\one$ with $c_{0}>0$. Then $a^{\alpha}\in \cinf_{({\rm b})}(\rr; \Psi^{m\alpha}(\Sigma))$ for any $\alpha\in \rr$ and $\sigma_{\rm pr}(a^{\alpha})(t)= \sigma_{\rm pr}(a(t))^{\alpha}$.
\end{theorem}

\subsection{Some auxiliary results} \label{ss:auxr}
For the sake of unifying the notation with the classes $\Psi_{\std}^{m, \delta}(\rr; \rr^{d})$ introduced in Subsect. \ref{scatcat} we set for  $(\Sigma, \alth)$ of bounded geometry: 
\[
\bea
\Psi_{\td}^{m, \delta}(\rr; \Sigma) &\defeq S^{\delta}(\rr; \Psi^{m}(\Sigma)).
\eea
\]
for pseudodifferential operator classes with time decay $(\td)$ of the symbols.  When writing $\Psi^{m, \delta}_{\std}(\rr; \Sigma)$ it is assumed implicitly that $\Sigma= \rr^{d}$.
\subsubsection{Difference of fractional powers}
We now state an auxiliary result about fractional powers of elliptic operators that will be needed later on.
\begin{proposition}\label{l5.1}
 Let   $a_{i}\in\Psi_{(*)}^{2,0}(\rr; \Sigma)$ for $*= \td, \std$, $i=1,2$ elliptic with $a_{i}=a_{i}^{*}$ and $a_{i}(t)\geq c_{0}\one$ for some $c_{0}>0$. Assume that $a_{1}- a_{2}\in \Psi_{(*)}^{k, -\delta}(\rr; \Sigma)$ with $\delta>0$, $k= 0, 1, 2$. Then  for each $\alpha\in \rr$ one has:
 \[
 a_{1}^{\alpha}- a_{2}^{\alpha}\in \Psi_{(*)}^{2(\alpha-1)+k, -\delta}(\rr;\Sigma).  
 \]
 \end{proposition}

Prop. \ref{l5.1} is proved in Subsect. \ref{ssecap1}.
\subsubsection{Ressummation of symbols}
We now examine the ressummation of symbols.  In the $(\td)$ case one can think of this as a statement about the uniform symbol classes on $\rr^{d}$, after applying a chart diffeomorphism. 

We denote $\Psi^{-\infty,-\delta}_{(*)}(\rr; \Sigma)\defeq\bigcap_{m\in\rr}\Psi^{m,-\delta}_{(*)}(\rr; \Sigma)$ for $*= \td, \std$.

\begin{lemma}\label{l5.2}
Let $\delta\in \rr$ and let $(m_{j})$ be a real sequence decreasing to $-\infty$. Then
if  $a_{j}\in \Psi_{(*)}^{ m_{j},\delta}(\rr; \Sigma)$ for $*= \td, \std$ there exists $a\in \Psi_{(*)}^{ m_{0},\delta}(\rr; \Sigma)$, unique modulo $\Psi^{-\infty,-\delta}_{(*)}(\rr; \Sigma)$, such that
 \[
 a\sim \sum_{j=0}^{\infty}a_{j}, \hbox{ i.e. }
 a-\sum_{j=0}^{N}a_{j}\in\Psi_{(*)}^{ m_{N+1},\delta}(\rr; \Sigma), \ \forall N\in \nn.
  \]
 \end{lemma}
\proof By introducing the new variable $s=\int_{0}^{t}\langle \sigma\rangle^{-1}d\sigma$ (so that $\langle t\rangle \p_{t}= \p_{s}$) and putting the extra variable $s$ together with the $\bx$ variables we can reduce ourselves to the situation covered by the standard proof (see e.g. \cite[Prop. 3.5]{shubin}).\qeds

\section{Parametrix for the Cauchy evolution and Hadamard states}\init\label{sec2}

\subsection{Model Klein-Gordon equation}\label{sec2.1}

In the present section we outline the approximate diagonalization and the parametrix construction that are used in \cite{bounded} to construct covariances of generic Hadamard states. 
We fix a $d-$dimensional manifold $\Sigma$ equipped with a reference Riemannian metric $\altk$  such that $(\Sigma, \altk)$ is of bounded geometry. We equip $M= \rr\times \Sigma$, whose elements are denoted by $x= (t, \rx)$, with a  Lorentzian metric  $\altg$ and a real function $\altV$ such that:

\beq\label{as:FLRW}
\begin{array}{l}
\altg= - dt^{2}+ \alth_{ij}(t,\rx )d\rx^{i}d\rx^{j},\\[2mm]
\alth\in \cinf(\rr, \BT^{0}_{2}(\Sigma, \altk)), \ \alth^{-1}\in \cinf(\rr; \BT^{2}_{0}(\Sigma, \altk)),\\[2mm]
\altV\in \cinf(\rr; \BT^{0}_{0}(\Sigma, \altk)).
\end{array}
\eeq
Although the first assumption may look restrictive, we will give in Subsects. \ref{s10.1}, \ref{s11.1} a reduction procedure that will allow us to treat more general cases.

The Klein-Gordon operator $P=-\Box_\altg+\altV$ equals
\beq\label{eq:modelP}
\bea
P &= |\alth|^{-\12}\pe_{t}|\alth|^{\12}\pe_{t}- |\alth|^{-\12}\pe_{i}\alth^{ij}|\alth|^{\12}\pe_{j}+\altV \\
   &= \pe_{t}^{2}+ r(t,\rx )\pe_{t}+ a(t, \rx, \pe_{\rx}),
\eea
\eeq
where
\[
a(t, \rx, \pe_{\rx})= - |\alth|^{-\12}\pe_{i}\alth^{ij}|\alth|^{\12}\pe_{j}+ \altV(t,\rx ) 
\]
is formally self-adjoint with respect to the $t$-dependent $L^{2}(\Sigma, |\alth|^{\12}dx)$-inner product and
\[
r(t,\rx )= |\alth|^{-\12}\p_{t}(|\alth|^{\12})(t,\rx ).
\]
Note that the above function is closely related to the extrinsic curvature of $\Sigma$ in $M$. 

In the sequel we will often abbreviate $a(t, \rx, \pe_{\rx})$ by $a(t)$ or $a$.

\subsection{Construction of parametrix}\label{sec2.3}

Following \cite{bounded} we now explain how one obtains a parametrix for the Cauchy evolution (and a splitting of it) by means of an approximate time-dependent diagonalization.  We will then adapt it to the setup of scattering theory. 

The first step consists of observing that the Klein-Gordon equation $(\p_t^2+ r(t)\p_t + a(t))\phi(t)=0$ is equivalent to
\beq\label{eq:evA}
\i^{-1}\p_t \psi(t) = \AH(t) \psi(t), \quad \mbox{ where \ } \AH(t)=\mat{0}{\one}{a(t)}{\i r(t)}, 
\eeq
by setting
\beq\label{eq:idpsi}
\psi(t)=\begin{pmatrix}\phi(t) \\ \i^{-1}\p_t \phi(t)\end{pmatrix}.
\eeq
Let us  denote by $\cU(s,t)$ the evolution generated by $\AH(t)$, cf. (\ref{eq:defUb}). 
Recall that on Cauchy data on $\Sigma_{s}= \{s\}\times \Sigma$, we have a symplectic form induced from an operator $G(s)$,  defined by:
\[
G= (\varrho_{s}G)^{*}\circ G(s)\circ (\varrho_{s}G).
\]
Here the formal adjoint will be always taken wrt. the $L^{2}(\Sigma, |\alth|^{\12}dx)$-inner product. We have also introduced the hermitian operator $q(s)= \i G(s)$. It is well known that with these choices, $q(s)$ equals specifically
\beq\label{defdeq(s)}
q(s)= \mat{0}{\one}{\one}{0},
\eeq
in particular it does not depend on $s$. Furthermore,
\beq\label{eq-e2}
\cU^{*}(t,s)q(s)\cU(t,s)= q(t),
\eeq
(the Cauchy evolution is symplectic).
\subsubsection{Riccati equation}\label{sec2.2} The approximate diagonalization of $\cU(s,t)$ will be based on solving the Riccati equation
\begin{equation}
\label{enp.11c}
\i \p_{t}b- b^{2}+a + \i rb=0,
\end{equation}
modulo smoothing terms, where the unknown is $b(t)\in C^{\infty}(\rr;\Psi^1(\Sigma))$. By repeating the arguments in \cite{GW,GW2} this can be solved modulo terms in $\cinf(\rr; \cW^{-\infty}(\Sigma))$. Concretely, supposing for the moment that $a(t)\geq c(t)\one$  for $c(t)>0$, one sets $\epsilon= a^{\12}$, $b= \epsilon+ b_{0}$ and obtains the equations:
\[
\bea 
b_{0} &= \frac{\i }{2}(\epsilon^{-1}\p_{t}\epsilon+ \epsilon^{-1}r \epsilon)+ F(b_{0}),\\[2mm]
F(b_{0}) &= \12 \epsilon^{-1}(\i \p_{t}b_{0}+ [\epsilon, b_{0}]+ \i r b_{0}- b_{0}^{2}).
\eea 
\]
These can be solved  by substituting a poly-homogeneous expansion of the symbol of $b_0$, yielding an approximate solution of (\ref{enp.11c}) in the sense that
\beq\label{eq:ricattiwrem}
\i \p_{t}b- b^{2}+a + \i rb=r_{-\infty}\in \cinf(\rr; \cW^{-\infty}(\Sigma)).
\eeq
 Set 
\beq
b^{+}=b, \quad b^{-}= - b^{*}.
\eeq
Taking the adjoint of the (\ref{eq:ricattiwrem}) with respect to the $t$-dependent inner product $L^{2}(\Sigma, |\alth|^{\12}dx)$ and using that
 \[
( \p_{t}b)^{*}= \p_{t}(b^{*})+ r b^{*}- b^{*}r,
 \]
 we obtain
 \beq\label{eq-ric}
 \i \p_{t}b^{\pm}- b^{\pm 2}+ a + \i rb^{\pm}=r_{-\infty}^{\pm},
 \eeq
 with $r^{+}_{-\infty}= r_{-\infty}$, $r^{-}_{-\infty}= r_{-\infty}^{*}\in\cinf(\rr; \cW^{-\infty}(\Sigma))$.  
 
In general  we can, using the locally uniform ellipticity of $a(t)$, find a cutoff function $\chi\in \coinf(\rr)$ such that $a(t)+ \chi(a(t))\geq c(t)\one$ for $c(t)$ as above. Since $\chi(a(t))$ is a smoothing operator,  replacing $a(t)$ by $a(t)+ \chi(a(t))$ is a harmless modification.

 A redefinition of $b(t)$ involving a cutoff in low frequencies as in \cite{GW2,bounded} gives then control of the norm sufficient to obtain in addition
\beq\label{eq:addi}
(b^+(t)- b^{-}(t))^{-1}\geq C(t) \epsilon(t)^{-1}
\eeq
for some $C(t)>0$, while keeping the property that $b^\pm(t)=\pm\epsilon(t)+\cf(\rr;\Psi^0(\Sigma))$, and with  \eqref{eq-ric} still valid for some $r_{-\infty}^{\pm}\in \cinf(\rr; \cW^{-\infty}(\Sigma))$.

Observe now that the Riccati equation \eqref{eq-ric} implies the following approximate factorization of the Klein-Gordon operator:
\begin{equation}
\label{eq-e1}
(\pe_{t}+ \i b^{\pm}(t)+ r(t))\circ (\pe_{t}- \i b^{\pm}(t))= \pe_{t}^{2}+ r\pe_t + a   - r_{-\infty}^{\pm}.
\end{equation}
The benefits of having such a factorization were already recognized by Junker and Schrohe \cite{junker,JS} in the context of Hadamard states (although it was obtained only in the special case of FRLW spacetimes with compact Cauchy hypersurface). Here we use (\ref{eq-e1}) to diagonalize (\ref{eq:evA}) by setting
\[
\tilde\psi(t)\defeq \begin{pmatrix}\pe_t-\i b^{-}(t) \\ \pe_t-\i b^{+}(t)\end{pmatrix}\phi(t).
\]
A direct computation yields then $\tilde{\psi}(t)= S^{-1}(t)\psi(t)$ with
 \beq\label{e4.01}
 S^{-1}(t)= \i \mat{- b^{-}(t)}{\one}{-b^{+}(t)}{\one}, \ \  S(t)= \i^{-1}\mat{\one}{-\one}{b^{+}(t)}{-b^{-}(t)}(b^{+}(t)- b^{-}(t))^{-1},
 \eeq
where well-definiteness and invertibility of $S(t)$ rely on the fact that $b^{+}(t)- b^{-}(t)$ is invertible by \eqref{eq:addi}. We obtain from \eqref{eq-e1} that
  \[
  \bea
& \mat{\pe_{t}+ \i b^{-}+ r}{0}{0}{\pe_{t}+ \i b^{+}+r}\tilde{\psi}(t)
 =\begin{pmatrix}
 \pe_{t}^{2}+ a+r\pe_{t}- r_{-\infty}^{-}\\ \pe_{t}^{2}+ a+ r\pe_{t}- r_{-\infty}^{+}\end{pmatrix}\phi(t)\\[2mm]
 &=\mat{r_{-\infty}^{-}}{0}{r_{-\infty}^{+}}{0}S(t)\tilde{\psi}(t)
 = \i^{-1}\mat{r_{-\infty}^{-}}{-r_{-\infty}^{-}}{r_{-\infty}^{+}}{-r_{-\infty}^{+}}(b^{+}- b^{-})^{-1}\tilde{\psi}(t).
  \eea
 \]
Therefore, $\tilde\psi(t)$ solves a diagonal matrix equation modulo smooth terms. More precisely, we have $\tilde{\psi}(t)= \cU_{B}(t, s)\tilde{\psi}(s)$ for
\beq\label{e4.02}
B(t)=\tilde{B}(t)+ R_{-\infty}(t),
\eeq
\beq\label{e4.02p}
\tilde{B}(t)= \mat{-b^{-}+ \i r}{0}{0}{-b^{+}+\i r}, \quad R_{-\infty}(t)=- \mat{r_{-\infty}^{-}}{-r_{-\infty}^{-}}{r_{-\infty}^{+}}{-r_{-\infty}^{+}}(b^{+}- b^{-})^{-1},
\eeq
Ultimately, we can thus conclude that
\beq\label{e4.00}
\cU(t,s)= S(t)\cU_{B}(t,s)S(s)^{-1}.
\eeq
\subsection{Improved approximate diagonalization}\label{ss:iad}
It is convenient to modify $S(t)$ to obtain a simple formula for the symplectic form $S^{*}(t)q(t)S(t)$ preserved by the almost diagonalized evolution.  Namely, setting
\beq\label{eq:defT}
\bea
&T(t)\defeq S(t)(b^{+}- b^{-})^{\12}(t)= \i^{-1}\mat{\one}{-\one}{b^{+}}{-b^{-}}(b^{+}- b^{-})^{-\12}, \\
&T^{-1}(t)= \i (b^{+}- b^{-})^{-\12}\mat{-b^{-}}{\one}{-b^{+}}{\one},
\eea
\eeq
we find that for $q(t)$ defined in \eqref{defdeq(s)} one has:
\[
T^{*}(t) q(t)T(t)=\mat{\one}{0}{0}{-\one}\eqdef q^{\adg}.
\]
We define 
\beq\label{intolo}
\cU(t,s)\eqdef T(t)\cU^{\adg}(t,s)T(s)^{-1}, 
\eeq
and we obtain that $\cU^{\adg}(t,s)^{*}q^{\adg} \cU^{\adg}(t,s)=q^{\adg}$, and the generator of $\{\cU^{\adg}(t,s)\}_{t,s\in\rr}$ is:
\beq\label{eq:defH}
\bea
H^{\adg}(t)&=(b^{+}- b^{-})^{-\12}B(t)(b^{+}- b^{-})^{\12}-\i \p_{t}(b^{+}- b^{-})^{-\12}(b^{+}- b^{-})^{\12}\\
&=\mat{- b^{-}+ r_{b}^-}{0}{0}{-b^{+}+ r_{b}^+}- (b^{+}- b^{-})^{-\12}\mat{r_{-\infty}^{-}}{-r_{-\infty}^{-}}{r_{-\infty}^{+}}{-r_{-\infty}^{+}}(b^{+}- b^{-})^{-\12},
\eea
\eeq
where $r^\pm_{-\infty}\in C^\infty(\rr;\cW^{-\infty}(\Sigma))$ are the remainder terms from \eqref{eq-ric}, and
\beq\label{eq:rbpm}
r_{b}^\pm= \i r+ [(b^{+}- b^{-})^{-\12}, b^{\pm}]-\i \p_{t}(b^{+}- b^{-})^{-\12}(b^{+}- b^{-})^{\12}\in \Psi^{0}(\Sigma).
\eeq  
This way, denoting by $H^{\dg}$ the diagonal part of $H^{\adg}(t)$ we have, using that $H^{\adg}(t)^{*}q^{\adg}= q^{\adg} H^{\adg}(t)$:
\[
H^{\dg}(t)= H^{\dg*}(t), \ \ H^{\dg}(t)= \mat{\epsilon^{+}(t)}{0}{0}{\epsilon^{-}(t)},
\]
where
\[
\epsilon^{\pm}= - b^{\mp}+ r_{b}^{\mp} +\cinf(\rr; \cW^{-\infty}(\Sigma)), 
\]
and 
$H^{\adg}(t)=H^{\dg}(t)+ V^{\adg}_{-\infty}(t)$, where $V^{\adg}_{-\infty}(t)\in \cinf(\rr; \cW^{-\infty}(\Sigma)\otimes B(\cc^2))$. The evolution $\cU^{\dg}(t,s)$ generated by $H^{\dg}(t)$ is diagonal, in fact:
\beq\label{e4.103}
\cU^{\dg}(t,s)= \mat{\cU_{\epsilon^+}(t,s)}{0}{0}{\cU_{\epsilon^-}(t,s)}.
\eeq
Moreover:
\begin{equation}
\label{turlututu}
\bea
\cU(t,s)&=T(t)\cU^{\adg}(t,s)T(s)^{-1}\\
&= T(t)\cU^{\dg}(t,s)T(s)^{-1}+ \cinf(\rr^{2}; \cW^{-\infty}(\Sigma)).
\eea
\end{equation}
This is shown   by an `interaction picture' argument explained in detail in \cite{bounded}, we omit the proof here.

\begin{remark}\label{remrem} One easily sees  that $S(t)$ is an isomorphism from  $L^{2}(\Sigma)\oplus L^{2}(\Sigma)$ to $H^{1}(\Sigma)\oplus L^{2}(\Sigma)$ (the so-called {\em energy space}  of Cauchy data of \eqref{eq:evA}), while $T(t)$ is an isomorphism from $L^{2}(\Sigma)\oplus L^{2}(\Sigma)$  to 
 $H^{\12}(\Sigma)\oplus H^{-\12}(\Sigma)$ (this is the {\em charge space} that appears naturally in the quantization of the Klein-Gordon equation).
 \end{remark}

\subsection{Splitting of the parametrix and of the Cauchy evolution}\label{sec2.5}
Let us set
\beq\label{defdepiplusmoins}
\pi^{+}= \mat{\one}{0}{0}{0}, \ \ \pi^{-}=\mat{0}{0}{0}{\one}.
\eeq
Since $\cU^{\dg}(t,s)$ is diagonal  we have:
\[
\cU^{\dg}(t,s)= \cU^{\dg}(t,s) \pi^++ \cU^{\dg}(t,s)\pi^{-},
\] 
with $\cU^{\dg}(t,s)\pi^\pm$ propagating with wave front set contained in $\cN^\pm$ (this follows from $b^\pm$ being $\pm\epsilon$ modulo terms of lower order). 
This suggests that at least modulo smoothing terms, the splitting of $\cU(t,s)$ at time $s$ should be given by a pair of operators $c_{\rf}^\pm(s)$ defined as follows. We first fix a reference time $t_0\in\rr$.

\begin{definition}\label{defdec} We set:
 \[
c_{\rf}^{\pm}(t_0)\defeq  T(t_0)\pi^{\pm}T^{-1}(t_0)=  \mat{\mp(b^{+}- b^{-})^{-1}b^{\mp}}{\pm(b^{+}- b^{-})^{-1}}{\mp b^{+}(b^{+}- b^{-})^{-1}b^{-}}{\pm b^{\pm}(b^{+}- b^{-})^{-1}}(t_0).
\]
\end{definition}
Then $c_{\rf}^{\pm}(t_0)$ is a $2\times 2$ matrix of pseudo\-differential operators and 
\[
c_{\rf}^{\pm}(t_0)^{2}= c_{\rf}^{\pm}(t_0), \  \ c_{\rf}^{+}(t_0)+ c_{\rf}^{-}(t_0)= \one.
\]
We set:
\beq\label{e2.10}
\cU^{\pm}(t,s)\defeq  \cU(t,t_0 )c_{\rf}^{\pm}(t_0)\cU(t_0,s),
\eeq
so that 
\beq\label{e2.11}
\cU(t,s)= \cU^{+}(t,s)+ \cU^{-}(t,s).
\eeq
This splitting has the following properties (see \cite{bounded}):

\begin{proposition}\label{newprop.2}
 \beq\label{eq-e-2}
\bea 
i)& \ \cU^{\pm}(t, s)\cU^{\pm}(s,t')= \cU^{\pm}(t, t'),\\[2mm]
ii)&\ (\pe_{t}- \i \AH(t))\cU^{\pm}(t,s)= \cU^{\pm}(t,s)(\pe_{s}+ \i \AH(s))=0,\\[2mm]
iii)& \ \WF(\cU^{\pm}(t,s))'=\{(X, X')\in T^{*}\Sigma\times T^{*}\Sigma : \  X= \Phi^{\pm}(t,s)(X')\},
\eea 
\eeq
where $\Phi^{\pm}(t,s): T^{*}\Sigma\to T^{*}\Sigma$ is the symplectic flow generated by the time-dependent Hamiltonian $\pm(\alth^{ij}(t,\rx )\spexi_{i}\spexi_{j})^{\12}$.
\end{proposition}

If we set for $t\in\rr$:
\beq\label{eq-e3}
\cU^{\pm}(t,t)\eqdef  c^{\pm}_{\rf}(t)=\cU(t, t_0)c_{\rf}^{\pm}(t_0)\cU(t_0,t),
\eeq
then 
\[
c^{\pm}_{\rf}(t)^{2}= c^{\pm}_{\rf}(t), \quad c^{+}_{\rf}(t)+ c^{-}_{\rf}(t)= \one, \quad c^{\pm}_{\rf}(t)= \cU(t,s)c^{\pm}_{\rf}(s)\cU(s,t).
\]
As a consequence, one gets that $c^\pm_\rf(t)$ are  the time-$t$ covariances of a Hadamard state \cite{bounded}. In general, we say that a state is a \emph{regular Hadamard state} if its time-$t$ covariances differ from $c_{\rf}^\pm(t)$ by terms in $\cW^{-\infty}(\Sigma)\otimes B(\cc^2)$, and one can show that it suffices to check that property for one value of $t$ \cite{bounded}. In summary:

\begin{theorem}[\cite{bounded}]The pair of operators $c^{\pm}_\rf(t)$ defined in \eqref{eq-e3} are the covariances of a pure, regular Hadamard state.
\end{theorem}

We stress that in general $c^\pm_\rf(t)$ are not `canonical' nor `distinguished', because they depend on the choice of the reference time $t_0$ and on the precise choice of the operators $b^\pm(t)$ (to which one can always add suitable regularizing terms). On the other hand, in Sect. \ref{secscat} we will construct covariances $c^\pm_{\inn}(t)$ and $c^\pm_{\out}(t)$ of the distinguished \emph{in} and \emph{out} states, and the operators $c^\pm_{\rf}(t)$ will play an important role in the proof of their Hadamard property: a suitable sufficient condition for that is in fact that
\beq
c^\pm_{\sca}(t)-c_{\rf}^\pm(t)\in \cW^{-\infty}(\Sigma)\otimes B(\cc^2)
\eeq
for some (and hence all) $t\in\rr$.

\subsection{Further estimates in scattering settings}\init\label{secscat}
 In what follows we give a refinement of the constructions in Sect. \ref{sec2} for the model Klein-Gordon equation in a scattering situation, corresponding to a situation when the metric $\altg$, resp. the potential $\altV$ converge to ultra-static metrics $\altg_{\rm out/in}= -dt^{2}+ \alth_{\rm out/in, ij}(\rx)d\rx^{i}d\rx^{j}$, resp. time-independent potentials $\altV_{\outin}$ as $t\to \pm\infty$.
We start by fixing two classes of assumptions on the model Klein-Gordon equation \eqref{eq:modelP}.
\medskip

We  will often abbreviate  the classes $\Psi_{(*)}^{m, \delta}$ (introduced in Subsect. \ref{pdosec}-\ref{ss:auxr}) for $(*)=\td,\std$   by $\Psi_{(*)}^{m, \delta}$. We make the following assumptions in the two respective cases.\medskip

{\em Case $(\td)$}:
\[
(\Htd) \ \ \beal
a(t, \rx, D_{\rx})= a_{\outin}(\rx, D_{\rx})+ \Psi_{\td}^{2, -\delta}(\rr; \Sigma)  \hbox{ on }\rr_{\pm}\times \Sigma,\ \delta>0,\\[2mm]
r(t)\in \Psi_{\td}^{0, -1-\delta}(\rr; \Sigma),\\[2mm]
a_{\outin}(\rx, D_{\rx})\in \Psi^{2}(\Sigma) \hbox{\ elliptic}, \ a_{\outin}(\rx, D_{\rx})= a_{\outin}(\rx, D_{\rx})^{*}\geq C_{\infty}>0.
\eeal  
\]

{\em Case $(\std)$}: $\Sigma= \rr^{d}$ and
\[
(\Hstd) \  \ \beal
a(t, \rx, D_{\rx})= a_{\outin}(\rx, D_{\rx})+ \Psi_{\std}^{2, - \delta}(\rr_{\pm}; \rr^{d})  \hbox{ on }\rr_{\pm}\times \Sigma,\ \delta>0,\\[2mm]
r(t)\in \Psi_{\std}^{0, -1-\delta}(\rr; \rr^{d}),\\[2mm]
a_{\outin}(\rx, D_{\rx})\in \Psi_{\scc}^{2,0}(\rr^{d})\hbox{\ elliptic}, \ a_{\outin}(\rx, D_{\rx})= a_{\outin}(\rx, D_{\rx})^{*}\geq C_{\infty}>0,\\[2mm]
\eeal  
\]
 From the definitions of $\Psi_{\std}^{m,\delta}$ and $\Psi_{\td}^{m,\delta}$ one easily sees that $(\Hstd)$ is a special case of $(\Htd)$. 

\medskip
   
Below, we give estimates on the solution of the Riccati equation, taking now into account the decay in time that follows from either $(\Htd)$ or $(\Hstd)$. To simplify notation we write simply $a(t)=b(t)+ \Psi_{(*)}^{m, \delta}(\rr_{\pm}; \Sigma)$ when $a(t)=b(t)+ \Psi_{(*)}^{m, \delta}(\rr; \Sigma)$ in $\rr_{\pm}\times \Sigma$. We also abbreviate $\Psi_{(*)}^{m, \delta}(\rr_{\pm}; \Sigma)$ by $\Psi_{(*)}^{m, \delta}$ when it is clear from the context whether the future or past case is meant.

From hypotheses $(*)$  there exists $c(t)\in \coinf(\rr)$ such that $a(t)+ c(t)\one\sim a_{\outin}$, uniformly in $t\in\rr_{\pm}$. By functional calculus we can find $\chi\in \coinf(\rr)$ such that $a(t)+ \chi(\frac{a(t)}{c(t)})\sim a_{\outin}$, uniformly in $t\in \rr_{\pm}$.  The error term  $ \chi(\frac{a(t)}{c(t)})$ belongs to $\coinf(\rr; \cW^{-\infty}(\Sigma))$, resp. $\coinf(\rr; \cW^{-\infty}_{\sd}(\Sigma))$. 

We can hence replace $a(t)$ by $a(t)+ \chi(\frac{a(t)}{c(t)})$ in the Riccati  equation \eqref{enp.11c} and assume that
\[
a(t)\sim  a_{\outin}\hbox{ uniformly in }t\in\rr_{\pm}.
\]
If $\epsilon_{\outin}\defeq  a_\outin^{\12}$, then from Prop.  \ref{l5.1} we deduce that if $(*)$ holds then
\begin{equation}
\label{e5.1}
\epsilon(t)\defeq  a(t)^{\12}=  \epsilon_\outin+ \Psi^{1, -\delta}_{(*)}, \ *=\td,\std.
\end{equation}

\begin{proposition}
 \label{p5.1}
 {\em Case $(\td)$}: There exists $b(t)= \epsilon(t)+ \Psi_{\td}^{0, -1-\delta}(\rr; \Sigma)= \epsilon_{\outin}+ \Psi^{1,-\delta}_{\td}(\rr_{\pm}; \Sigma)$ that solves 
\[
 \i\p_{t}b- b^{2}+ a + \i rb\in \Psi_{\td}^{-\infty, -1-\delta}(\rr; \Sigma).
 \]
 {\em Case $(\std)$}:  There exists $b(t)= \epsilon(t)+ \Psi_{\std}^{0,-1-\delta}(\rr; \Sigma)= \epsilon_{\outin}+ \Psi^{1,-\delta}_{\std}(\rr_{\pm}; \Sigma)$ that solves 
\[
 \i\p_{t}b- b^{2}+ a + \i rb\in \Psi_{\std}^{-\infty,-1- \delta}(\rr; \Sigma).
 \]
 \end{proposition}
 The proof is given in Appendix \ref{to1}.

\begin{proposition}\label{propoesti}
 Assume $(*)$ for $*= \td, \std$ and let $r_{b}^{\pm}$ be defined in \eqref{eq:rbpm} and $r_{-\infty}^{\pm}$ in \eqref{eq-ric}. Then
 \[
 r_{b}^{\pm}\in \Psi^{0, -1- \delta}_{(*)}(\rr; \Sigma), \ \ r_{-\infty}^{\pm}\in \Psi^{-\infty, -1- \delta}_{(*)}(\rr; \Sigma).
 \]
 \end{proposition}
 The proof is given in Appendix \ref{to2}.

\section{The $\outin$ states on asymptotically static spacetimes}\init\label{inout}
\subsection{Assumptions}\label{ss:asast} In what follows we introduce a class of asymptotically static spacetimes on which we will construct the \emph{out}/\emph{in} states and prove their Hadamard property. One of the key ingredients is the reduction to a model Klein-Gordon operator that satisfies the assumptions $(\Htd)$ considered in Subsect. \ref{secscat}.

We will use the framework of manifolds and diffeomorphisms of bounded geometry introduced in Defs. \ref{defp0.2}, \ref{defdeboun}.

We fix a  $d-$dimensional manifold $\Sigma$ equipped with a reference Riemannian metric $\altk$ such that $(\Sigma, \altk)$ is of bounded geometry, and consider $M= \rr_{t}\times\Sigma_{\ry}$, setting $y= (t, \ry)$, $n= 1+d$.  We equip $M$ with a Lorentzian metric $\altg$ of the form
\begin{equation}
\label{e10.1}
\altg= - \altc^{2}(y)dt^{2}+ (d\ry^{i}+ \altb^{i}(y)dt)\alth_{ij}(y)(d\ry^{j}+ \altb^{j}(y)dt),
\end{equation} 
where we assume:
\[
(\bg)\ \ \beal
\alth_{ij}\in \cinfb(\rr; \BT^{0}_{2}(\Sigma, \altk)), \  \alth_{ij}^{-1}\in \cinfb(\rr; \BT^{2}_{0}(\Sigma, \altk)),\\[2mm]
\altb\in \cinfb(\rr; \BT^{1}_{0}(\Sigma, \altk)), \\[2mm]
\altc, \ \altc^{-1}\in \cinfb(\rr; \BT^{0}_{0}(\Sigma, \altk)).
\eeal
\]
We recall that  $\tilde{t}\in \cinf(M)$ is called a  \emph{time function} if  $\nabla \tilde{t}$ is a timelike vector field. It is called a \emph{Cauchy time function} if its level sets are Cauchy hypersurfaces.
By \cite[Thm. 2.1]{CC} we know that $(M, \altg)$ is globally hyperbolic and  $t$ is  a Cauchy time function.

We will consider the Klein-Gordon operator on $(M, \altg)$:
\begin{equation}
\label{e10.2}
P= - \Box_{\altg}+ \altV,
\end{equation}
with $\altV\in \cinfb(\rr; \BT^{0}_{0}(\Sigma, \altk))$ a smooth real-valued function.
We consider two static metrics 
\[
\altg_{\outin}= -\altc^{2}_{\outin}(\ry)dt^{2}+ \alth_{\outin}(\ry)d\ry^{2}
\] 
and time-independent potentials $\altV_{\outin}$ and assume the following conditions:
\[
(\ast)\ \ \beal
\alth(y)- \alth_{\outin}(\ry)\in S^{-\mu}(\rr_{\pm}; \BT^{0}_{2}(\Sigma, \altk)), \\[2mm]
\altb(y)\in S^{-\mu'}(\rr; \BT^{1}_{0}(\Sigma, \altk)), \\[2mm]
\altc(y)- \altc_{\outin}(\ry)\in S^{-\mu}(\rr_{\pm}; \BT^{0}_{0}(\Sigma, \altk)),\\[2mm]
\altV(y)- \altV_{\outin}(\ry)\in S^{-\mu}(\rr_{\pm}; \BT^{0}_{0}(\Sigma, \altk)),
\eeal
\]
\[
(\pos)\ \ \frac{n-2}{4(n-1)}(\altR_{\altc_{\outin}^{-2}\alth_{\outin}}- \altc_{\outin}^{2}\altR_{\altg_\outin})+ \altc_{\outin}^{2}\altV_{\outin}\geq \altm^{2}.
\]
for some $\mu>0$, $\mu'>1$ and $\altm>0$.  Above, $\altR_{\altg}$, resp. $\altR_{\alth}$ denotes the scalar curvature of $\altg$, resp. $\alth$.

Condition $(\ast)$ means that $\altg$, resp. $\altV$ are asymptotic to the static metrics $\altg_{\outin}$, resp. to the time-independent potentials  $\altV_{\outin}$ as $t\to \pm \infty$. Condition $(\pos)$ means that the asymptotic Klein-Gordon operators $\pe_{t}^{2}+ a_{\outin}(\rx, \pe_{\rx})$ introduced in Lemma \ref{l10.2} below   are {\em massive}. 

 It follows from $(\bg)$ that $\alth_{\outin}\in \BT^{0}_{2}(\Sigma, \altk)$, $\alth_{\outin}^{-1}\in \BT^{2}_{0}(\Sigma, \altk)$ and $\altV_{\outin}, \altV_{\outin}^{\,-1}\in \BT^{0}_{0}(\Sigma, \altk)$.
\subsection{Reduction to the model case}\label{s10.1}
In this subsection we perform the reduction of the Klein-Gordon operator $P$ to the model case considered in Sect. \ref{secscat}. 
We start with the well-known orthogonal decomposition of $\altg$ associated with the time function $t$. Namely, we set
\[
v\defeq  \dfrac{\altg^{-1}dt}{dt\cdot \altg^{-1}dt}=\pe_{t}+ \altb^{i}\pe_{\ry^{i}},
\]
which using  $(\bg)$ is a complete vector field. Furthermore, we denote by $\phi_{t}$ its flow, so that 
\[
\phi_{t}(\rx)= (t,0,  \ry(t,0, \rx)),\ t\in \rr, \ \rx\in \Sigma,
\]
where $\rx(t, s, \cdot)$ is the flow of the time-dependent vector field $b$ on $\Sigma$.
We also set
\beq\label{e10.0}
\chi: \rr\times \Sigma\ni(t, \rx)\mapsto (t, \ry(t, 0, \rx))\in \rr\times \Sigma.
\eeq

\begin{lemma}\label{l10.1}Assume $(\bg)$, $(\ast)$. Then
\[
\altgh:= \chi^{*}\altg= -  \altch^{2}(t, \rx)dt^{2}+ \hat  \alth(t, \rx)dy^{2}, \ \ \chi^{*}\altV=  \altVh, 
\]
where:
\[
\begin{array}{l}
 \altch,  \altch^{-1},  \altVh\in \cinfb(\rr; \BT^{0}_{0}(\Sigma, \altk)),\\[2mm]
 \hat \alth\in \cinfb(\rr; \BT^{0}_{2}(\Sigma, \altk)), \  \hat \alth^{-1}\in \cinfb(\rr; \BT^{2}_{0}(\Sigma, \altk)).
\end{array}
\]
Moreover there exist bounded diffeomorphisms $\ry_{\outin}$ of $(\Sigma, \altk)$ such that if:
\[
\begin{array}{l}
 \hat \alth_{\outin}\defeq \ry_{\outin}^{*}\alth_{\outin},\\[2mm]
   \altch_{\outin}\defeq  \ry_{\outin}^{*}\altc_{\outin}, \  \altVh_{\outin}\defeq  \ry_{\outin}^{*}\altV_{\outin},
\end{array}
\]
then we have:
\[
\begin{array}{l}
 \hat \alth_{\outin}\in \BT^{0}_{2}(\Sigma, \altk), \  \hat \alth_{\outin}^{-1}\BT^{2}_{0}(\Sigma, \altk),\\[2mm]
 \altch_{\outin},  \altch_{\outin}^{-1},  \altVh_{\outin}\in \BT^{0}_{0}(\Sigma, \altk),
\end{array}
\]
and furthermore,
\[
\begin{array}{rl}
 \hat \alth- \hat  \alth_{\outin}\in S^{-\min(1-\mu', \mu)}(\rr_{\pm}, \BT^{0}_{2}(\Sigma, \altk)), \\[2mm]
 \altch- \altch_{\outin}\in S^{-\min(1-\mu', \mu)}(\rr_{\pm}, \BT^{0}_{0}(\Sigma, \altk)),\\[2mm]
 \altVh-  \altVh_{\outin}\in S^{-\mu}(\rr_{\pm}, \BT^{0}_{0}(\Sigma, \altk)).
\end{array}
\]
\end{lemma}
Lemma \ref{l10.1} is proved in Appendix \ref{apoti}.

Writing $P$ as $-\Box_{\altg}+  \frac{n-2}{4(n-1)}\altR_{\altg}+ \altW\,$ for $\altW= \altV-  \frac{n-2}{4(n-1)}\altR_{\altg}$, and using the conformal invariance of $-\Box_{\altg}+  \frac{n-2}{4(n-1)}\altR_{\altg}$ and the estimates in Lemma \ref{l10.1},
we obtain the following result, which completes the reduction to the model case.
\begin{lemma}\label{l10.2}
Assume $(\bg)$, $(\ast)$, $(\pos)$ and consider the Klein-Gordon operator $P$  in \eqref{e10.2}. Let  $\hat{\alth}, \altch,\altVh$ be as in Lemma \ref{l10.1} and set:
\[
\hat P\defeq  \chi^{*}P, \ \ \tilde{P}\defeq  \altch^{1-n/2}\hat P \altch^{1+ n/2}, \ \altgt= \altch^{-2}\altgh, \ \tilde{\alth}=  \altch^{-2}\hat{\alth}.
\]
Then 
\[
\tilde{P}= \pe_{t}^{2}+r(t, \rx)\pe_{t}+ a(t, \rx, \pe_{\rx}),
\]
for
\[
\begin{array}{l}
a(t, \rx, \pe_{\rx})=- \Delta_{\tilde{\alth}_{t}}+\altVt, \ \ r=  |\tilde{\alth}_{t}|^{-\12} \p_{t}|\tilde{\alth}_{t}|^{\12},\\[2mm]
\altVt= \frac{n-2}{4(n-1)}(R_{\altgt}- \altch^{2}R_{\altgh})+ \altch^{2}\altVh.
\end{array}
\]
Moreover $a, r$ satisfy $(\Htd)$ with $\delta= \min (\mu, \mu'-1)$ and
\[
a_{\outin}(\rx, \pe_{\rx})= - \Delta_{\tilde{\alth}_{\outin}}+\altVt_{\outin}(\rx),
\]
where
\[
\altVt_{\outin}= \left(\frac{n-2}{4(n-1)}(\altR_{\altc^{-2}_{\outin}h_{\outin}}- \altc^{2}_{\outin }\altR_{\altg_{\outin}})+ \altc_{\outin}^{2}\altV_{\outin}\right)\circ \ry_{\outin}.
\]\end{lemma}
Note that condition $(\pos)$ simply means that $\altVt_{\outin}\geq \altm^{2}>0$.
\subsection{Cauchy evolutions}\label{cauchycauchy}
In this subsection we relate the Cauchy evolutions of $P$ and  of the model Klein-Gordon operator $\tilde{P}$.

The trace operator for $P$ associated to the time function $t$ is given by:
\beq\label{defdetrace}
\varrho_{t}\phi= \col{u(t, \cdot)}{\i^{-1}n\cdot\nabla \phi(t, \cdot)},
\eeq
where $n$ is the future directed unit normal to $\Sigma_{t}$. The corresponding trace operator for $\hat{P}= \chi^{*}P$ is:
\[
\hat{\varrho}_{t}\phi= \col{\phi(t, \cdot)}{ \i^{-1}\altch^{-1}\p_{t}\phi(t, \cdot)},
\]
so that denoting $\chi^{*}\phi= \phi\circ \chi$, we have:
\[
\hat{\varrho}_{t}\chi^{*}\phi= \chi_{t}^{*}\varrho_{t}\phi \hbox{ for }\chi_{t}^{*}\col{u_{0}}{u_{1}}= \col{u_{0}\circ \chi_{t}}{u_{1}\circ \chi_{t}},
\]
and $\chi_{t}(\rx)= \ry(t, 0, \rx)$, see \eqref{e10.0}. Finally the trace operator for $\tilde{P}$ as in Lemma \ref{l10.2} is
\[
\tilde{\varrho}_{t}\phi= \col{\phi(t, \cdot)}{\i^{-1}\phi(t, \cdot)}
\]
so that if $\tilde{\phi}= \altch^{n/2-1}\phi$  is the conformal transformation in Lemma \ref{l10.2} we have:
\[
\tilde{\varrho}_{t}\tilde{\phi}= R(t)\hat{\varrho}_{t}\phi,\hbox{ for }R(t)= \altch^{n/2-1}\mat{1}{0}{-\i(n/2-1)\p_{t}\ln (\altch)}{1}.
\]
Let us denote by $\cU(t,s)$ the Cauchy evolution for $P$ associated to $\varrho_{t}$ and by $\cU^{\adg}(t,s)$ the almost diagonal Cauchy evolution introduced in 
Subsect. \ref{ss:iad} for the model Klein-Gordon operator $\tilde P$.  The following lemma  follows from the above computations and \eqref{intolo}.
\begin{lemma}\label{defdeZ}
Let $Z(t)\defeq (\chi_{t}^{*})^{-1} R(t)T(t)$, where $T(t)$ is defined in  \eqref{eq:defT}. Then
\beq\label{e10.5}
\cU(t,s)= Z(t)\cU^{\adg}(t,s)Z^{-1}(s).
\eeq
\end{lemma}

We have a similar reduction for  the asymptotic Klein-Gordon operators:
\[
P_{\outin}= - \Box_{\altg_{\outin}}+\frac{n-2}{4(n-1)}\altR_{g_{\outin}}+ \altV_{\outin},
\]
for $\altg_{\outin}= -\altc^{2}_{\outin}(\ry)dt^{2}+ \alth_{\outin}(\ry)d\ry^{2}$, where $\alth_{\outin}, \altc_{\outin}, \altV_{\outin}$ were introduced in $(\ast)$. The associated trace operator is
\[
\varrho_{t,\outin}\phi= \col{\phi(t, \cdot)}{\i^{-1}\altc_{\outin}^{-1}\p_{t}\phi(t, \cdot)}.
\]
We also set
\[
\chi_{\outin}^{*}\col{u_{0}}{u_{1}}= \col{u_{0}\circ \ry_{\outin}}{u_{1}\circ \ry_{\outin}},\ \ R_{\outin}=\altch_{\outin}^{(d-1)/2}\one,
\]
and for $\epsilon_{\outin}= a_{\outin}^{\12}$:
\[
T_{\outin}= (\i \sqrt{2})^{-1}\mat{\epsilon_{\outin}^{-\12}}{-\epsilon_{\outin}^{-\12}}{\epsilon_{\outin}^{\12}}{\epsilon_{\outin}^{\12}}, \ \ Z_{\outin}= (\chi_{\outin}^{*})^{-1}R_{\outin}T_{\outin},
\]
so that the Cauchy evolution of $P_{\outin}$ is given by
\begin{equation}
\label{e10.6}
\cU_{\outin}(t,s)= Z_{\outin}\circ \cU_{\outin}^{\adg}(t,s)\circ Z_{\outin}^{-1}, 
\end{equation}
where $\cU_{\outin}^{\adg}$ stands for the evolution generated by
\begin{equation}
\label{e10.7}
 H^{\adg}_{\outin}= \mat{\epsilon_{\outin}}{0}{0}{\epsilon_{\outin}}.
\end{equation}
The following fact will be needed in the sequel.
\begin{lemma}\label{l10.3}We have:
\[
Z^{-1}(t)Z_{\outin}-\one, \ Z_{\outin}^{-1}Z(t)-\one \to 0\hbox{ in }B(L^{2}(\Sigma)\otimes \cc^{2})\hbox{ as }t\to \pm \infty.
\]
\end{lemma}
\proof
From Prop. \ref{p5.1} we obtain that $T^{-1}_{\outin}T(t)-\one$ tends to $0$ in norm as $t\to \pm \infty$.
By Lemma \ref{l10.1}, $R(t)$ tends to $R_{\outin}$ in norm. Finally, from the proof of Lemma \ref{l10.1}, see in particular \eqref{e10.4}, we obtain that 
$(\chi_{\outin}^{*})^{-1}\chi_{t}^{*}$ tends  to $\one$ in norm. This implies the lemma. \qed
\subsection{Construction of Hadamard states by scattering theory}
In this subsection we construct the \emph{out}/\emph{in} states $\omega_{\outin}$ for the Klein-Gordon operator $P$ and show that they are Hadamard states. We assume hypotheses $(\bg)$, $(\ast)$, $(\pos)$.

By the positivity condition $(\pos)$, the asymptotic Klein-Gordon operators $P_{\outin}$ admit {\em vacuum states} (that is, {\em ground states} for the dynamics $\cU_\outin$) $\omega^{\rm vac}_\outin$.  In terms of $t=0$ Cauchy data  their covariances are the projections:
\[
c^{\pm, {\rm vac}}_{\outin}= Z_{\outin}\pi^{\pm}Z_{\outin}^{-1}, 
\]
where  $\pi^{\pm}$ are defined in \eqref{defdepiplusmoins}.  Clearly we have
\[
\cU_{\outin}(t,s)c^{\pm, {\rm vac}}_{\outin}\cU_{\outin}(s,t)= c^{\pm, {\rm vac}}_{\outin},
\]
i.e. $\omega^{\rm vac}_{\outin}$ are invariant under the asymptotic dynamics.
 For $t\in\rr$ we now consider the projections:
 \beq\label{eq:defcp}
\bea
c^{\pm,t}_{\outin}(0)&\defeq  \cU(0, t)c_{\outin}^{\pm, {\rm vac}}\cU(t,0)\\
&= \cU(0, t)\cU_{\outin}(t, 0)c_{\outin}^{\pm, {\rm vac}}\cU_{\outin}(0, t)\cU(t,0).
\eea
\eeq
By taking the $t\to\pm\infty$ limit of $c^{\pm,t}_{\outin}(0)$ we  obtain the time-$0$ covariances  $c^{\pm}_{\outin}(0)$  of  a state $\omega_\outin$ (for the Klein-Gordon operator $P$) that `equal $\omega^{\rm vac}_{\outin}$ asymptotically' at $t=\pm\infty$. The main new result that we prove is that  $\omega_\outin$ are Hadamard states.

  Before stating the theorem let us recall that the Sobolev spaces $H^{\sobo}(\Sigma)$ are naturally defined using the reference Riemannian metric $\altk$ on $\Sigma$. The {\em charge space} $H^{\12}(\Sigma)\oplus H^{-\12}(\Sigma)$ is the natural space of Cauchy data in connection with quantized Klein-Gordon fields. 

\begin{theorem}\label{thm.scat1}Assume hypotheses $(\bg)$, $(\ast)$, $(\pos)$. Then 
 \beq
\lim_{t\to \pm\infty}c^{\pm,t}_{\outin}(0)\eqdef c^{\pm}_{\outin}(0)=  c_{\rf}^{\pm}(0)+ \cW^{-\infty}(\Sigma), \hbox{ in }B(H^{\12}(\Sigma)\oplus H^{-\12}(\Sigma)),
\eeq
where $c^{\pm}_{\rm ref}(0)= Z(0)\pi^{\pm}Z^{-1}(0)$. The operators $c^{\pm}_{\outin}(0)$ are pairs of projections defining a pure state $\omega_{\outin}$ for the Klein-Gordon operator $P$. Moreover $\omega_{\outin}$ is a Hadamard state.
\end{theorem}
\proof  From \eqref{e10.5}, \eqref{e10.6}  we obtain:
\begin{equation}
\label{e5.3}\bea 
\cU_{\outin}(0, t)\cU(t, 0)&= Z_{\outin}(0)\cU^{\adg}_{\outin}(0, t)Z^{-1}_{\outin}Z(t)\cU^{\adg}(t,0)Z^{-1}(0),\\[2mm]
\cU(0, t)\cU_{\outin}(t, 0)&= Z(0)\cU^{\adg}(0, t)Z^{-1}(t)Z_{\outin}\cU^{\adg}_{\outin}(t,0)Z_{\outin}^{-1}.
\eea 
\end{equation}
 It follows that:
\begin{equation}
\label{e5.3c}
\bea
c^{\pm,t}_{\outin}(0)&=Z(0)\cU^{\adg}(0,t)Z^{-1}(t)Z_{\outin}\cU^{\adg}_{\outin}(t, 0)\\
&\phantom{=}\times\pi^{\pm}\cU^{\adg}_{\outin}(0, t)Z^{-1}_{\outin}Z(t)\cU^{\adg}(t, 0)Z^{-1}(0).
\eea
\end{equation}
Since $Z(0):L^{2}(\Sigma)\otimes \cc^{2}\to H^{\12}(\Sigma)\oplus H^{-\12}(\Sigma)$ is boundedly invertible it suffices to  show the existence of the limit
\[
d^{\pm}_{\outin}= \lim_{t\to \pm\infty}\cU^{\adg}(0,t)Z^{-1}(t)Z_{\outin}\cU^{\adg}_{\outin}(t,0)\pi^{\pm}\cU^{\adg}_{\outin}(0, t)Z^{-1}_{\outin}Z(t)\cU^{\adg}(t, 0)
\]
in $B(L^{2}(\Sigma)\otimes \cc^{2})$.

By Prop.  \ref{l.scat1} (1) below we know that $\cU^{\adg}(t,s)$, $\cU^{\adg}_{\outin}(t,s)$ are uniformly bounded in $B(L^{2}(\Sigma)\otimes \cc^{2})$. Hence using Lemma \ref{l10.3} we can replace $Z^{-1}(t)Z_{\outin}$ and $Z^{-1}_{\outin}Z(t)$ by $\one$ in the rhs of \eqref{e5.3c}, modulo an error of size $o(t^{0})$ in $B(L^{2}(\Sigma)\otimes \cc^{2})$, i.e.  we are reduced to prove the existence of the limit
\[
\bea 
d^{\pm}_{\outin}&\defeq \lim_{t\to \pm\infty}\cU^{\adg}(0,t)\cU^{\adg}_{\outin}(t, 0)\pi^+\cU^{\adg}_{\outin}(0, t)\cU^{\adg}(t, 0)\\[2mm]
&=\lim_{t\to \pm\infty}W_{\outin}(t)\pi^+ W_{\outin}^{-1}(t),
\eea 
\]
where $W_{\outin}(t)= \cU^{\adg}(0, t)\cU^{\adg}_{\outin}(t, 0)$.  By Prop. \ref{l.scat1} the limit exists in $B(L^{2}(\Sigma)\otimes \cc^{2})$ and equals $\pi^++ \cW^{-\infty}(\Sigma)$.  The limit operators $d^{\pm}_{\outin}$ are  projections as  norm limits of projections. It follows that
\beq\label{ebou}
c^{\pm}_{\outin}(0)= Z(0)d^{\pm}_{\outin}Z(0)^{-1}+ \cW^{-\infty}(\Sigma)= c^{+}_{\rf}(0) + \cW^{-\infty}(\Sigma)
\eeq
is a projection.  The conditions \eqref{eq:secondcondlam0}, \eqref{eq:secondcondlam} are satisfied by $c^{\pm}_{\outin}(0)$ since they are satisfied by $c^{\pm,t}_{\outin}(0)$ for each finite $t$.  Therefore $c^{\pm}_{\outin}$ are the covariances of two pure states $\omega_{\outin}$ for $P$.  Finally as in \cite{bounded} we obtain from \eqref{ebou} that $\omega_{\outin}$ are Hadamard states. \qeds

In the proof of Thm. \ref{thm.scat1}, the crucial ingredient is the following proposition.

\begin{proposition}\label{l.scat1}
 Let $H^{\adg}(t),H^{\adg}_{\outin}$ be  as in \eqref{eq:defH}, \eqref{e10.7}. Then:
 \ben
 \item $\cU^{\adg}_{\outin}(t,s)$ and $\cU^{\adg}(t,s)$ are uniformly bounded in $B(H^{m}(\Sigma)\otimes \cc^{2})$, for all $m\in \rr.$
 \item Let $W_{\outin}(t)= \cU^{\adg}(0, t)\cU^{\adg}_{\outin}(t, 0)$. Then  
 \[
\lim_{t\to +\infty} W_{\outin}(t)\pi^+ W_{\outin}(t)^{-1}= \pi^+ + \cW^{-\infty}(\Sigma)\otimes L(\cc^{2}), \hbox{ in }B(L^{2}(\Sigma)\otimes \cc^{2}).
 \]
  \een
\end{proposition}
\proof  {\it Proof of (1)}: we can assume without loss of generality that $s=0$.  The statement for $\cU^{\adg}_{\outin}(t,0)$ is obvious since $H^{\adg}_{\outin}= \mat{\epsilon_{\outin}}{0}{0}{-\epsilon_{\outin}}$. Let us prove it for $\cU^{\adg}(t,0)$. We have: 
\beq\label{e5.5b}
\bea 
H^{\adg}(t)&=  \mat{-b^{-}(t)+ \i r_{b}^- (t)}{0}{0}{-b^{+}(t)+\i r_{b}^+(t)}+ \Psi_{\td}^{-\infty, -1-\delta}(\rr; \Sigma)\otimes B(\cc^{2})\\[2mm]
&=\mat{\epsilon(t)}{0}{0}{-\epsilon(t)}+  \Psi_{\td}^{0, -1-\delta}(\rr; \Sigma)\otimes B(\cc^{2}),
\eea 
\eeq
 by  Props. \ref{p5.1}, \ref{propoesti}. Since $\epsilon(t)$ is selfadjoint, this implies that $\cU^{\adg}(t,0)$  is uniformly bounded in $B(L^{2}(\Sigma))$, which proves (1) for $m=0$.

We now note that $\| u\|_{H^{m}(\Sigma)}\sim \| \epsilon^{m}(t)u\|_{L^{2}(\Sigma)}$, uniformly for $t\in \rr$, since  $\epsilon(t)$ is elliptic uniformly for $t\in \rr$.  Therefore to prove (1) it suffices, using the uniform boundedness of $\cU^{\adg}(t,0)$ in $B(L^{2}(\Sigma))$, to show that
\beq\label{e5.10}
\cU^{\adg}(0,t)\left(\epsilon(t)^{m}\otimes \one_{\cc^{2}}\right)\cU^{\adg}(t,0) \left(\epsilon(0)^{-m}\otimes \one_{\cc^{2}}\right)\hbox { is uniformly bounded in }B(L^{2}(\Sigma)).
\eeq
 We have by \eqref{e5.5b}: 
\[
\bea 
&\p_{t}\cU^{\adg}(0,t)\left(\epsilon(t)^{m}\otimes \one_{\cc^{2}}\right)\cU^{\adg}(t,0) \left(\epsilon(0)^{-m}\otimes \one_{\cc^{2}}\right)\\[2mm]
&= \cU^{\adg}(0,t)\left(\p_{t}\epsilon^{m}(t)\otimes\one_{\cc^{2}}+ \i [H^{\adg}(t), \epsilon^{m}(t)\otimes \one_{\cc^{2}}]\right)\cU^{\adg}(t,0)\left(\epsilon(0)^{-m}\otimes \one_{\cc^{2}}\right)\\[2mm]
&= \cU^{\adg}(0,t)\left(\p_{t}\epsilon^{m}(t)\otimes \one_{\cc^{2}}+ \i [H^{\adg}(t), \epsilon^{m}(t)\otimes \one_{\cc^{2}}]\right)\left(\epsilon(t)^{-m}\otimes\one\right)\cU^{\adg}(t,0)\\[2mm]
&\phantom{=}\times \cU^{\adg}(0,t)\left(\epsilon(t)^{m}\otimes \one_{\cc^{2}}\right)\cU^{\adg}(t,0)\left(\epsilon(0)^{-m}\otimes \one_{\cc^{2}}\right)\\[2mm]
&\eqdef  M(t) \cU^{\adg}(0,t)\left(\epsilon(t)^{m}\otimes \one_{\cc^{2}}\right)\cU^{\adg}(t,0)\left(\epsilon(0)^{-m}\otimes \one_{\cc^{2}}\right).
\eea 
\]
By  $(\Htd)$ and Prop.  \ref{l5.1} we  see that  $\p_{t}\epsilon^{m}(t)\in \Psi_{\td}^{m, -1- \delta}$, and by \eqref{e5.5b}   that $ [H^{\adg}(t), \epsilon^{m}(t)\otimes \one_{\cc^{2}}]\in \Psi_{\td}^{m, -1-\delta}$. Therefore 
$\|M(t)\|_{B(L^{2}(\Sigma)\otimes \cc^{2})}\in O(\langle t\rangle^{-1- \delta})$. Hence, setting
\[
f(t)\defeq\|\cU^{\adg}(0,t)\left(\epsilon(t)^{m}\otimes \one_{\cc^{2}}\right)\cU^{\adg}(t,0) \left(\epsilon(0)^{-m}\otimes \one_{\cc^{2}}\right)\|_{B(L^{2}(\Sigma))},
\]
we have $f(0)=1$, $|\p_{t}f(t)|\in O(\langle t\rangle^{-1- \delta})f(t)$. If $f(t)\neq +\infty$  for each $t$, an application of 
 Gronwall's inequality  would immediately imply \eqref{e5.10}.  If $m\leq 0$ the  use of Gronwall's inequality is justified by applying the above time dependent operator to a vector $u\in H^{m}(\Sigma)$. If $m>0$ we replace the unbounded operator $A= \epsilon(t)\otimes \one_{\cc^{2}}$ by the bounded operator
 $A_{\delta}= A(1+ \i \delta A)$, for $\delta>0$. For the corresponding function $f_{\delta}(t)$ we obtain that
  $f_{\delta}(0)\leq 1$, $|\p_{t}f_{\delta}(t)|\in O(\langle t\rangle^{-1- \delta})f_{\delta}(t)$ uniformly for $0<\delta\leq 1$.
  Then \eqref{e5.10} follows using that $\|A^{m}u\|= \sup_{0<\delta\leq 1}\| A_{\delta}^{m}u\|$.

{\it Proof of  (2)}: note first that $[\pi^+, A]=0$ for any diagonal operator $A$. Therefore:
 \[
W_{\outin}(t)\pi^+ W_{\outin}(t)^{-1}= \cU(0,t)\pi^+ \cU(t,0),
\]
and by \eqref{e5.5b}
\beq\label{eq:rin}
\begin{array}{l}
\p_{t}(W_{\outin}(t)\pi^+ W_{\outin}(t)^{-1})= - \i \cU(0,t)[H^{\adg}(t), \pi^+]\cU(t,0)\\[2mm]
= \cU(0,t)[R_{-\infty}(t), \pi^+]\cU(t,0), \ R_{-\infty}\in \Psi_{\td}^{-\infty, -1- \delta}(\rr; \Sigma)\otimes B(\cc^{2}).
\end{array}
\eeq
By (1), this implies that $\p_{t}(W_{\outin}(t)\pi^+ W_{\outin}(t)^{-1})\in \Psi_{\td}^{-\infty, -1- \delta}(\rr; \Sigma)\otimes B(\cc^{2})$, hence:
\[
\lim_{t\to +\infty}W_{\outin}(t)\pi^+ W_{\outin}(t)= \pi^++ \int_{0}^{+\infty}\p_{t}(W_{\outin}(t)\pi^+ W_{\outin}(t)^{-1}) dt \hbox{ in }B(L^{2}(\Sigma)\otimes \cc^{2}).
\]
The integral term belongs to $\cW^{-\infty}(\Sigma)$. \qeds

\section{Feynman inverses from scattering data for model Klein-Gordon equations}\init\label{sec:abstract}

\subsection{Setup} In this section we consider again the model Klein-Gordon operator studied in Subsect. \ref{sec2.1}:
\beq\label{eq:defPstd}
P= \pe_{t}^{2}+ r(t, \rx)\pe_{t}+ a(t, \rx, \pe_{\rx}),
\eeq
and  denote by $P_{\outin}$  the asymptotic Klein-Gordon operators
\[
P_{\outin}\defeq \pe_{t}^{2}+ a_{\outin}(\rx, \pe_{\rx}).
\]
We will assume conditions $(\Hstd)$ for $\delta>1$, which corresponds to a {\em short-range} situation. Recall that in particular $\Sigma=\rr^d$.  

By a parametrix for $P$ we will mean an operator $G_I$ such that $PG_I-\one$ and $G_I P-\one$ have smooth Schwartz kernel. Duistermaat and H\"ormander proved the existence of a Feynman parametrix $G_\F$, or parametrix with \emph{Feynman type wave front set}, i.e.
\[
\WF'(G_\F)= (\diag_{T^*M})\cup\textstyle\bigcup_{t\leq 0}(\Phi_t(\diag_{T^*M})\cap \pi^{-1}\cN).
\]
This means that up to singularities on the full diagonal $\diag_{T^*M}$ of $T^*M\times T^*M$, $\WF'(G_\F)$ is contained in the backward flowout of $\diag_{T^*M}$ by the bicharacteristic flow (here acting on the left component of $T^*M\times T^*M$, accordingly $\pi$ is the projection to that component).
 
Our primary goal will be to prove that for suitably chosen Hilbert spaces of distributions $\cX_I^m, \cY^m$, the operator $P:\cX_I^m\to\cY^m$ is Fredholm, i.e. its kernel and cokernel are of finite dimension. This guarantees the existence of pseudo-inverses, i.e. operators $G_I:\cY^m \to \cX_I^m$ such that $P G_I-\one$ and $G_I P - \one$ are compact. 

We will be interested in constructing a pseudo-inverse that is at the same time a Feynman parametrix. This will be based on the reduction to the almost diagonalized dynamics $\cU^{\adg}(t,s)$ obtained in  Subsect. \ref{ss:iad}. 

 \subsection{Notation}
As a rule, all objects related to the almost diagonalized situation will be decorated with a superscript $\adg$. We recall that $L^{2}(\rr^{1+d})$ is equipped with the  scalar product
\[
(u|v)\defeq\int \bar{u}v|h_{t}|^{\12}dtd\rx.
\]
\subsubsection{Operators}
Let us recall that  the operators $\AH(t)$, $H^{\adg}(t)$, $T(t)$ are defined respectively in \eqref{eq:evA}, \eqref{eq:defH}, \eqref{eq:defT}.   

\begin{notations}
\item We set for $u\in \cinf(\rr; \cD'(\rr^{d}))$, $u^{\adg}\in \cinf(\rr; \cD'(\rr)\oplus \cD'(\rr))$:
\[
\begin{array}{l}
\varo_{t}u= (u(t), \i^{-1}\p_{t}u(t)), \ \ \varo^{\adg}_{t}u^{\adg}= u^{\adg}(t),\\[2mm]
(Tu^{\adg})(t)\defeq T(t)u^{\adg}(t), \ \ (\varo u)(t)\defeq  \varo_{t}u(t).
\end{array}
\]
 Setting also $\pi_{i}(u_{0}, u_{1})= u_{i}$ we have:
\begin{equation}
\label{e100.-1}
P= - \pi_{1} (D_{t}- \AH(t))\varrho,
\end{equation}
where as usual $D_{t}= \i^{-1}\pe_{t}$.

\item We set:
\[
P^{\adg}\defeq  D_{t}- H^{\adg}(t),
\]
and an easy computation shows that:
\beq\label{e100.0}
TP^{\adg}T^{-1}= D_{t}- \AH(t), \hbox{\ hence }P= -\pi_{1}TP^{\adg}T^{-1}\varo. 
\eeq

\item We  denote by $\AH_{\outin},H^{\adg}_{\outin},T_{\outin}$, the analogues of $\AH(t),H^{\adg}(t), T(t)$ with $a(t), r(t)$ replaced by $a_{\outin}, 0$.

\item The Cauchy evolutions generated by  $\AH(t)$, $\AH_{\outin}$, $H^{\adg}(t)$, $H^{\adg}_{\outin}$ are denoted by $\cU(t,s)$,  $\cU_{\outin}(t,s)$, $\cU^{\adg}(t,s)$, $\cU_{\outin}^{\adg}(t,s)$. We recall that:
\beq\label{e100.10}
\cU(t,s)= T(t)\cU^{\adg}(t,s)T^{-1}(s),  \ \ \cU_{\outin}(t,s)= T_{\outin}\cU^{\adg}_{\outin}(t,s)T^{-1}_{\outin}.
\eeq

\item The symplectic forms for $P$, $P^{\adg}$ are denoted by
\[
q\defeq \mat{0}{\one}{\one}{0}, \ \ q^{\adg}\defeq\mat{\one}{0}{0}{-\one},
\]
and we have:
\beq\label{e100.00} 
T^{*}(t)q T(t)= q^{\adg},
\eeq
i.e. $T$ is symplectic. 
\end{notations}

The adjoint of an operator $A$ for $q, q^{\adg}$ will be denoted by $A^{\dag}$. We recall that $\cU_{(\outin)}^{(\adg)}(t,s)$, $\cU_{(\outin)}^{(\adg)}(t,s)$ are symplectic for $q^{(\adg)}$.

\begin{notations}
\item If $a$ acting on $\cD'(\rr^{d})$ is a 'scalar' operator, the operator $a\otimes \one_{\cc^{2}}$  will be abbreviated by $a$ for simplicity.  
\end{notations}

\subsubsection{Properties of $H^{\adg}(t)$}
Let us  summarize the properties of $H^{\adg}(t)$ in the $(\Hstd)$ case, that follow from identity \eqref{eq:defH} and Props. \ref{p5.1}, \ref{propoesti}, namely:
\begin{equation}
\label{e11.3}
\begin{array}{l}
H^{\adg}(t)= \mat{-b^{-}(t)+ r_{b}^{-}(t)}{0}{0}{-b^{+}(t)+ r_{b}^{+}(t)}+R_{-\infty}(t), \hbox{ where}\\[3mm]
R_{-\infty}(t)\in \Psi^{-\infty, -1- \delta}_{\std}(\rr; \rr^{d})\otimes B(\cc^{2}),
\end{array}
\end{equation} 
\begin{equation}
\label{e11.4}
H^{\adg}(t)= \mat{\epsilon(t)}{0}{0}{-\epsilon(t)}+ \Psi^{0, -1- \delta}_{\std}(\rr; \rr^{d})\otimes B(\cc^{2}),
\end{equation} 
\begin{equation}
\label{e11.50}
\begin{array}{l}
b^{\pm}(t)\mp \epsilon(t), r_{b}^{\pm}(t)\in  \Psi^{0, 1- \delta}_{\std}(\rr; \rr^{d}),\\[2mm]
\epsilon(t)- \epsilon_{\outin}\in  \Psi^{1, - \delta}_{\std}(\rr; \rr^{d}).
\end{array}
\end{equation}

\subsubsection{Function spaces}
 We will abbreviate by $H^{\sobo}$ the Sobolev spaces $H^{\sobo}(\rr^{d})$. Furthermore we set:
 \[
\cE^{\sobo}\defeq H^{\sobo+1}\oplus H^{\sobo}, \ \ \cH^{\sobo}\defeq H^{\sobo}\oplus H^{\sobo}, \ \sobo\in \rr.
\]
As usual we  define $\cE^{ \infty}\defeq \bigcap_{\sobo\in \rr}\cE^{\sobo}$, $\cE^{-\infty}\defeq \bigcup_{\sobo\in \rr}\cE^{\sobo}$ and similarly for $\cH^{\infty}$, $\cH^{-\infty}$, equipped with their canonical topologies.

We will frequently use  the fact that  $T(t): \cE^{\sobo}\to \cH^{\sobo+\12}$ is boundedly invertible with $\|T(t)\|, \|T^{-1}(t)\|$ uniformly bounded in $t$.  Using  Prop. \ref{l.scat1}, we also obtain that:
\begin{equation}
\label{e11.51}
\sup_{t,s\in\rr}\|\cU^{\adg}(t,s)\|_{B(\cH^{\sobo})}<\infty, \ \ \sup_{t,s\in \rr}\| \cU(t,s)\|_{B(\cE^{\sobo})}<\infty.
\end{equation}

\begin{notations}
\item If $\cE$ is a Banach space, $k\in \nn$, we denote by $C^{k}(\rr; \cE)$ the Banach space of $\cE-$valued functions with norm
\[
\| u\|_{C^{k}(\rr; \cE)}= \sum_{0\leq l\leq k}\sup_{t\in \rr}\| \p_{t}^{l}u(t)\|_{\cE}.
\]
\end{notations}

\subsection{M{\o}ller (wave) operators}\label{orlov} We will consider $t=0$ as our fixed reference time. It is a standard fact, derived using \eqref{e11.4}, \eqref{e11.50}  and the so-called Cook argument (see e.g. \cite{dergersc}), that the M{\o}ller operators
\beq\label{azer}
W^{\adg}_\sca \defeq \lim_{t\to\pm\infty} \cU^{\adg}(0,t)\cU^{\adg}_{\sca}(t,0 )\in B(\cH^{\sobo})
\eeq
exist and are invertible with inverses given by
\beq\label{azor}
(W^{\adg}_\sca)^{-1} = (W_\sca^{\adg})^\dagger=\lim_{t\to\pm\infty} \cU^{\adg}_{\outin}(0,t)\cU^{\adg}(t,0)\in B(\cH^{\sobo}).
\eeq
Using then \eqref{e100.10} and that fact that $T^{-1}(t)T_{\outin}- \one$ tends to $0$ in  $B(\cH^{\sobo})$ when $t\to \pm\infty$, we obtain the existence of
\begin{equation}
\label{e100.11}
W_{\sca}\defeq \lim_{t\to \pm\infty}\cU(0,t)\cU_{\sca}(t,0)\in B(\cE^{\sobo}),
\end{equation}
with inverses
\beq\label{e100.12}
(W_\sca)^{-1} = (W_\sca)^\dagger=\lim_{t\to\pm\infty} \cU_{\outin}(0,t)\cU(t,0)\in B(\cE^{\sobo}),
\eeq
and satisfying the identities
\begin{equation}
\label{e100.12b}
W_{\outin}= T(0)W^{\adg}_{\outin}T^{-1}_{\outin}.
\end{equation}
\begin{remark}\label{remstupid}
 Strictly speaking  $W^{(\adg)}_{\outin}$ acting on $\cE^{\sobo}$ or $\cH^{\sobo}$ should be denoted by, e.g., $W_{\outin}^{(\adg), \sobo}$ to indicate its dependence on $\sobo$. However since $W_{\outin}^{(\adg), \sobo}$ is the closure of $W_{\outin}^{(\adg), \sobo'}$ for any $\sobo'>\sobo$, we will  often dispense with the exponent $\sobo$ in the sequel. The same remark applies to $(W_{\outin}^{(\adg)})^{-1}$.
\end{remark}
\subsection{Compactness properties of $W^{\adg}_{\outin}$}
 The additional space decay properties  implied by conditions $(\Hstd)$ have the following important consequence:

\begin{proposition}\label{prop:stdcase} Assume condition $(\Hstd)$ and let $\alpha<\delta/2$. Then 
\[
W^{\adg}_\outin \pi^+(W^{\adg}_\outin)^{-1}- \pi^+ \in \langle\rx\rangle^{-\alpha}\cW^{-\infty}(\rr^{d})\langle\rx\rangle^{-\alpha}\otimes B(\cc^{2}).
\]
It follows that $[W^{\adg}_{\outin}, \pi^{+}]$ is a compact operator on   $\cH^{\sobo}$ for $\sobo\in \rr$.
\end{proposition}

To prove Prop. \ref{prop:stdcase}, we will need the following lemma; its proof is given in Appendix \ref{apota}.
\begin{lemma}\label{turlututi}
 Assume conditions $(\Hstd)$ for $\delta>0$. Then for all $m, k\in \rr_{+}$:
 \[
\sup_{t\geq 0}\|\langle D_{\rx}\rangle^{m}\langle \rx\rangle^{k}\cU^{\adg}(0,t)(\langle \rx\rangle +\langle t\rangle)^{-k}\langle D_{\rx}\rangle^{-m}\|_{B(\cH^{0})}<\infty.
\]
\end{lemma}

{\bf\noindent Proof of Prop. \ref{prop:stdcase}.}  Recall that we have set  $W^{\adg}_{\outin}(t)= \cU^{\adg}(0,t)\cU^{\adg}_{\outin}(t, 0)$. We have for $m\in \nn$, $\alpha>0$:
\beq\label{taratata}
\begin{array}{l}
\langle D_{\rx}\rangle^{m}\langle\rx\rangle^{\alpha}\p_{t}(W^{\adg}_{\outin}(t)\pi^+ (W^{\adg}_{\outin}(t)^{-1})\langle\rx\rangle^{\alpha}\langle D_{\rx}\rangle^{m}\\[2mm]
=\langle D_{\rx}\rangle^{m}\langle\rx\rangle^{\alpha}\cU^{\adg}(0,t)[R_{-\infty}(t), \pi^+]\cU^{\adg}(t,0)\langle\rx\rangle^{\alpha}\langle D_{\rx}\rangle^{m}\\[2mm]
=\langle D_{\rx}\rangle^{m}\langle\rx\rangle^{\alpha}\cU^{\adg}(0,t)(\langle\rx\rangle+ \langle t\rangle)^{-\alpha}\langle D_{\rx}\rangle^{-m}\\[2mm]
\phantom{=}\times \langle D_{\rx}\rangle^{m}(\langle\rx\rangle+ \langle t\rangle)^{\alpha}[R_{-\infty}(t), \pi^+](\langle\rx\rangle+ \langle t\rangle)^{\alpha}\langle D_{\rx}\rangle^{m}\\[2mm]
\phantom{=}\times \langle D_{\rx}\rangle^{-m}(\langle\rx\rangle+ \langle t\rangle)^{-\alpha}\cU^{\adg}(t,0)\langle\rx\rangle^{\alpha}\langle D_{\rx}\rangle^{m}\\[2mm]
\eqdef  R_{m, \alpha}(t)\times M_{m, \alpha}(t)\times R_{m, \alpha}(t)^{\dag}.
\end{array}
\eeq

Since $R_{-\infty}(t)\in \Psi^{-\infty, -1- \delta}_\std(\rr; \Sigma)\otimes B(\cc^{2})$ we know that 
\[
\|M_{m, \alpha}(t)\|_{B(\cH^{\sobo})}\in O(\langle t\rangle^{-1- \delta+ 2\alpha}).
\] 
From Lemma \ref{turlututi} we know that  $\|R_{m, \alpha}(t)\|_{B(\cH^{\sobo})}\in O(1)$, which implies the same bound for 
$R_{m,\alpha}(t)^{\dag}$.
Thus from \eqref{taratata} we obtain that
\[
\langle D_{\rx}\rangle^{m}\langle\rx\rangle^{\alpha}\p_{t}(W^{\adg}_{\outin}(t)\pi^+ (W^{\adg}_{\outin}(t)^{-1})\langle\rx\rangle^{\alpha}\langle D_{\rx}\rangle^{m}\in O(\langle t\rangle ^{-1- \delta + 2 \alpha}).
\]
This is integrable for $\alpha< \delta/2$.
By integrating from $t=0$ to $t=\pm\infty$,   since $m$ is arbitrary this implies that:
\[
\lim_{t\to\pm\infty}W^{\adg}_{\outin}(t) \pi^+ W^{\adg}_{\outin}(t)^{-1}- \pi^+ \in \langle\rx\rangle^{-\alpha}\cW^{-\infty}(\Sigma)\langle\rx\rangle^{-\alpha}. 
\]
 Since $W^{\adg}_{\outin}= \lim_{t\to\pm\infty}W^{\adg}_{\outin}(t)$ this proves the proposition. \qeds

\subsection{Inhomogeneous Cauchy problem}\label{inhomomo}
Fixing  $\gamma$ with $\12<\gamma<\12 + \delta$, we set:
\[
\cY^{\sobo}\defeq \langle t\rangle^{-\gamma}L^{2}(\rr; H^{\sobo}), \ \ \cY^{\adg, \sobo}\defeq   \langle t\rangle^{-\gamma}L^{2}(\rr; \cH^{\sobo}).
\]
The exponent $\gamma$ is chosen so that  $\cY^{\sobo}\subset L^{1}(\rr; \cE^{\sobo})$, $\cY^{\adg, \sobo}\subset L^{1}(\rr; \cH^{\sobo})$. The benefit of working with $\cY^{(\adg),\sobo}$ is that these are Hilbert spaces; this will be needed in Subsect. \ref{ss:fredf}.

\begin{definition}\label{defowit}
 We denote by $\cX^{\sobo}$ the space of $u\in C^{0}(\rr; H^{\sobo+1})\cap C^{1}(\rr; H^{\sobo})$ such that  $Pu\in \cY^{\sobo}$, and similarly by $\cX^{\adg, \sobo}$ the space of $u^{\adg}\in C^{0}(\rr; \cH^{\sobo})$ such that $P^{\adg}u^{\adg}\in \cY^{\adg, \sobo}$.
 We equip $\cX^{(\adg), \sobo}$ with the Hilbert norms:
 \beq\label{defdenorme}\begin{array}{l}
 \|u^{\adg}\|^{2}_{\sobo}\defeq \| \varo^{\adg}_{0}u^{\adg}\|^{2}_{\cH^{\sobo}}+ \| P^{\adg}u^{\adg}\|^{2}_{\cY^{\adg, \sobo}},\\[2mm]
  \|u\|^{2}_{\sobo}\defeq \| \varo_{0}u\|^{2}_{\cE^{\sobo}}+ \| Pu\|^{2}_{\cY^{\sobo}}.
\end{array}
 \eeq
\end{definition}
The existence and uniqueness of the Cauchy problem for $P$ and $P^{\adg}$ implies  that $\cX^{(\adg), \sobo}$ are Hilbert spaces, as  stated  implicitly in the following lemma.
\begin{lemma}\label{lem:inho} The map
\beq
\bea
\varrho_{0}\oplus P: \cX^{\sobo} &\to  \cE^{\sobo}\oplus\cY^{\sobo}\\
u&\mapsto (\varrho_{0} u, Pu)
\eea
\eeq
is boundedly invertible  with inverse given by:
\beq\label{e12.1}
(\varrho_{0}\oplus P)^{-1}(v, f)= \pi_{0}\cU(t, 0)v-\i\pi_{0} \int_{0}^{t}\cU(t,s)\pi_{1}^{*}f(s)ds.
\eeq
Similarly, the map
\beq
\bea
\varrho^{\adg}_{0}\oplus P^{\adg}: \cX^{\adg, \sobo} &\to  \cH^{\sobo}\oplus\cY^{\adg, \sobo}\\
u^{\adg} &\mapsto (\varrho^{\adg}_{0} u^{\adg}, P^{\adg}u^{\adg})
\eea
\eeq
is boundedly invertible  with inverse given by:
\beq\label{e12.1b}
(\varrho^{\adg}_{0}\oplus P^{\adg})^{-1}(v^{\adg}, f^{\adg})= \cU^{\adg}(t, 0)v^{\adg}+ \i \int_{0}^{t}\cU^{\adg}(t,s)f^{\adg}(s)ds.
\eeq
\end{lemma}
It follows that \begin{equation}
\label{e12.1bb}
\begin{array}{l}
\cX^{\sobo}\hookrightarrow C^{k}(\rr; H^{\sobo+1-k}), \\[2mm]
\cX^{\adg, \sobo}\hookrightarrow C^{k}(\rr; \cH^{\sobo-k}),
\end{array}
\end{equation}
 continuously for $\sobo\in \rr, \ k\in \nn$.

The following facts are easy computations that make use of \eqref{e100.0}:
\beq\label{e100.1}
\begin{array}{l}
T^{-1}\varrho\in B(\cX^{\sobo}, \cX^{\adg, \sobo+\12}), \ \ -T^{-1}\pi_{1}^{*}\in B(\cY^{\sobo}, \cY^{\adg, \sobo+\12}),\\[2mm]
 \pi_{0}T\in B(\cX^{\adg, \sobo+ \12}, \cX^{\sobo}), \ \ -\pi_{1}T\in B(\cY^{\adg, \sobo+\12}, \cY^{\sobo}).
\end{array}
\eeq
In the sequel we will also need the auxiliary identities\beq\label{e100.2}
\begin{array}{l}
\Ran T^{-1}\varrho= \Ker (\varrho \pi_{0}- \one)T, \\[2mm]
(T^{-1}\varrho)^{-1}= \pi_{0}T\hbox{ on }\Ran T^{-1}\varrho,\\[2mm]
 \Ran T^{-1}\pi_{1}^{*}= \Ker \pi_{0}T,\\[2mm]
 (T^{-1}\pi_{1}^{*})^{-1}=  \pi_{1}T\hbox{ on }\Ran T^{-1}\pi_{1}^{*}.
\end{array}
\eeq

\subsection{Retarded and advanced propagators}\label{retard}
The retarded/advanced propagators for $P^{\adg}$ are defined as follows:
\beq\label{retadv}
(G^{\adg}_{+}f^{\adg})(t)\defeq\i\int_{-\infty}^{t}\cU^{\adg}(t,s )f^{\adg}(s)ds, \ \ (G_{-}^{\adg}f)(t)\defeq -\i \int_{t}^{+\infty}\cU^{\adg}(t, s)f^{\adg}(s)ds,
\eeq
for $f^{\adg}\in L^1(\rr;\cH^{\sobo})$. Using \eqref{e11.51}  one obtains:
\[
\begin{array}{l}
G^{\adg}_{\pm}\in B(L^{1}(\rr; \cH^{\sobo}), C^{0}(\rr; \cH^{\sobo})),\\[2mm]
(G^{\adg}_{\pm})^{\dag}= G^{\adg}_{\mp}\hbox{ on }L^{1}(\rr; \cH^{\sobo}), \ \ P^{\adg}G^{\adg}_{\pm}= \one\hbox{ on }L^{1}(\rr; \cH^{\sobo}).
\end{array}
\]
The analogous propagators for $P$ are:
\begin{equation}
\label{e100.13}
(G_{+}f)(t)=-\i\pi_{0}\int^{t}_{-\infty}\cU(t,s)\pi_{1}^{*}f ds, \ \ (G_{-}f)(t)= \i \pi_{0}\int_{t}^{+\infty}\cU(t, s)\pi_{1}^{*}f(s)ds,
\end{equation}
for $f\in L^{1}(\rr; H^{\sobo})$. One has:
\[
\begin{array}{l}
G_{\pm}\in B(L^{1}(\rr; H^{\sobo}), C^{0}(\rr; H^{\sobo+1})\cap C^{1}(\rr; H^{\sobo})),\\[2mm]
G_{\pm}^{*}= G_{\mp}\hbox{ on }L^{1}(\rr; H^{\sobo}),\ \ PG_{\pm}= \one\hbox{ on }L^{1}(\rr; H^{\sobo}).
\end{array}
\]
Using \eqref{e100.10} we have the relation:
\beq\label{e100.13c}
G_{\pm}= - \pi_{0} T G^{\adg}_{\pm}T^{-1}\pi_{1}^{*}.
\eeq
\subsection{Fredholm problems from scattering data}

We now want to define the maps that assign  to an element of $\cX^{(\adg), \sobo}$ its scattering data in the standard sense, as well as its Feynman and anti-Feynman data. By \emph{Feynman data} we mean positive-frequency data of a solution at $+\infty$ and negative-frequency data at $-\infty$, and by \emph{anti-Feynman} the reverse. 

\begin{proposition}\label{idiotic}
 The limits 
 \[
 \slim_{t\to \pm \infty}\cU_{\outin}^{\adg}(0, t)\varrho^{\adg}_{t}, \ \hbox{resp. }  \slim_{t\to \pm \infty}\cU_{\outin}(0, t)\varrho_{t},
 \]
exist in $B(\cX^{\adg, \sobo}, \cH^{\sobo})$,  resp. in $B(\cX^{\sobo}, \cE^{\sobo})$, 
and equal $(W_{\outin}^{\adg})^{-1}$ on  $\Ker P^{\adg}|_{\cX^{\adg, \sobo}}$, resp. $(W_{\outin})^{-1}$ on $\Ker P|_{\cX^{\sobo}}$.
\end{proposition}
\proof Let $u^{\adg}\in \cX^{\adg, \sobo}$. By Lemma \ref{lem:inho} we have
\[
\cU_{\out}^{\adg}(0,t)\varrho^{\adg}_{t}u^{\adg}= \cU_{\out}^{\adg}(0,t)\cU^{\adg}(t, 0)v^{\adg}+ \i\int_{0}^{t}\cU^{\adg}_{\out}(0, t)\cU^{\adg}(t, 0)\cU^{\adg}(0,s)f^{\adg}(s)ds
\]
which by dominated convergence tends to $(W^{\adg}_{\out})^{-1}(v^{\adg}- \varrho^{\adg}_{0}G^{\adg}_{-}f^{\adg})$ as $t\to +\infty$.  Similarly we obtain that $\cU^{\adg}_{\rm in}(0, t)\varrho^{\adg}_{t}u^{\adg}$ converges to $(W_{\rm in}^{\adg})^{-1}(v^{\adg}- \varrho^{\adg}_{0}G^{\adg}_{+}f^{\adg})$ as $t\to -\infty$. 
The proof in the scalar case is similar. 
\qeds
\medskip

We can now introduce  four scattering data maps $\varrho^{(\adg)}_I:\cX^{(\adg), \sobo}\to\cH^{\sobo}$. Note the presence of the operators $T_{\outin}^{-1}$ below; this simplifies some considerations later on.
\begin{definition}
We set:
\[
\begin{array}{l}
\varrho_{\outin }^{\adg}\defeq \slim_{t\to\pm\infty}\cU^{\adg}_{\outin}(0,t)\varrho^{\adg}_{t},\\
\varrho_{\outin }\defeq \slim_{t\to\pm\infty}T_{\outin}^{-1}\cU^{}_{\outin}(0,t)\varrho^{}_{t},\\
\varo^{(\adg)}_{\F}\defeq \pi^{+}\varo^{(\adg)}_{\out}+ \pi^{-}\varo^{(\adg)}_{\inn},\\
\varo^{(\adg)}_{\rm \overline{F}}\defeq \pi^{-}\varo^{(\adg)}_{\out}+ \pi^{+}\varo^{(\adg)}_{\inn}.
\end{array}
\]

\end{definition}

\begin{lemma}\label{l12.1}
 For $I\in\{\inn,\out,\F, \aF\}$ we have:
 \begin{equation}
\label{e100.14}
\varo_{I}= \varo^{\adg}_{I}T^{-1}\varo,
\end{equation}
\beq\label{e100.15}
\bea
\varrho^{\adg}_I &= W_{I}^{\adg\dagger} \circ \varrho^{\adg}_{0} \hbox{ on  } \Ker P^{\adg}|_{\cX^{\adg, \sobo}},\\[2mm]
\varo_{I}&= W_{I}^{\adg\dagger} T^{-1}(0)\varo_{0}\hbox{ on }\Ker P|_{\cX^{\sobo}},
\eea
\eeq
for $I\in\{\inn,\out,\F, \aF\}$, where
\beq\label{eq:defR}
W^{\adg\dagger}_{\F}\defeq  \pi^+W_{\out}^{\adg\dagger}+ \pi^-W_{\inn}^{\adg\dagger}, \quad
W^{\adg\dagger}_{\rm \overline{F}}\defeq  \pi^-W_{\out}^{\adg\dagger}+ \pi^+W_{\inn}^{\adg\dagger}.
\eeq
\end{lemma}
\proof To prove \eqref{e100.14} we write:
\[
\bea
\varo_{\outin}&= T_{\outin}^{-1}\cU_{\outin}(0, t)\varo_{t}+ o(1)
= \cU^{\adg}_{\outin}(0,t)T^{-1}_{\outin}\varo_{t}+ o(1)\\
&=\cU^{\adg}_{\outin}(0,t)T^{-1}(t)\varo_{t}+ o(1)
=\cU^{\adg}_{\outin}(0,t)\varo^{\adg}_{t}T^{-1}\varo+ o(1).
\eea
\]
This implies \eqref{e100.14} for $I= \outin$ and then for $I= \F/{\rm {\overline F}}$.  The first statement of \eqref{e100.15} follows then from the fact that $\varo^{\adg}_{\outin}= W^{\adg\dagger}_{\outin}$ on $\Ker P^{\adg}$,  the second from \eqref{e100.14} and the fact that $T^{-1}\varo: \Ker P\to \Ker P^{\adg}$. \qeds 
 \begin{lemma}\label{l12.2}
 Let $I\in\{ \F, {\rm \overline{F}} \}$. Then 
 $W^{\adg}_{I}  W_{I}^{\adg\dagger}-\one$ and $W_{I}^{\adg\dagger}W^{\adg}_{I} -\one$ are compact  on $\cH^{\sobo}$ and hence  $W^{\adg}_{I}$, $W^{\adg\dag}_{I}$ are Fredholm.  Moreover: 
 \[
 \Ker W^{\adg(\dagger)}_{I}|_{\cH^{\sobo}}= \Ker W^{\adg(\dagger)}_{I}|_{\cE^{\adg,\infty}}, \ \ \coKer W^{\adg(\dagger)}_{I}|_{\cH^{\sobo}}= \coKer W^{\adg(\dagger)}_{I}|_{\cE^{\adg, -\infty}},
  \]
and hence $\ind( W^{\adg(\dag)}_{I})|_{\cH^{\sobo}}$ is independent on $\sobo$.
\end{lemma}
 
\proof We consider  only the $\F$ case. We have
\[
\begin{array}{l}
W^{\adg}_{\F} W^{\adg\dagger}_{\F}=\one+K_{1}, \ \ K_{1} = W^{\adg}_\out \pi ^+ (W^{\adg}_\out)^{-1}-\pi^{+}+W^{\adg}_\inn \pi ^- (W^{\adg}_\inn)^{-1}-\pi^{-},\\[2mm]
W^{\adg\dagger}_{\F} W^{\adg}_{\F}= \one + K_{2}, \ \ K_{2}=\pi^+ (W^{\adg}_{\out})^{-1} W^{\adg}_{\inn} \pi^- + \pi^- (W^{\adg}_{\inn})^{-1}W^{\adg}_\out \pi^+.
\end{array}
\]
By Prop. \ref{prop:stdcase}   we see that $K_{1}$, $K_{2}$ are compact on $\cH^{\sobo}$ and moreover map $\cH^{\sobo}$ to $\cH^{\infty}$. Therefore  $\Ker W_{\F}^{\adg\dag}|_{\cH^{\sobo}}\subset \Ker (\one+ K_{1})|_{\cH^{\sobo}}\subset \cH^{\infty}$ hence $\Ker W_{\F}^{\adg\dag}|_{\cH^{\sobo}}= \Ker W_{\F}^{\adg\dag}|_{\cH^{ \infty}}$. Similarly, identifying $(\cH^{\sobo})^{*}$ with $\cH^{-\sobo}$ and $\coKer A$ with $\Ker A^{*}$ we have $\coKer W_{\F}^{\adg\dag}|_{\cH^{\sobo}}\subset \Ker (\one+ K_{2}^{*})|_{\cH^{\sobo}}\subset \cH^{\infty}$, hence $\coKer W_{\F}^{\adg\dag}|_{\cH^{\sobo}}= \coKer W_{\F}^{\adg\dag}|_{\cH^{-\infty}}$. \qeds



We will need the following lemma, see \cite[Prop. A.1]{BB} for its proof. The next few results are simple applications of it, following the  strategy  in \cite{BS} in the case of the Dirac equation on a compact cylinder.

\begin{lemma}\label{lem:fredholm} Let $\cK$ be a Hilbert space and $\cE$, $\cF$ Banach spaces. Let  $K:\cK\to\cE$, $Q:\cK\to\cF$ be bounded and assume that $Q$ is surjective. Then $K:\Ker Q \to \cE$ is Fredholm (of index $l$) iff $K\oplus Q:\cK\to\cE\oplus \cF$ is Fredholm (of index $l$).   
\end{lemma}

\begin{lemma}\label{lem:aux0} For $I\in\{\inn,\out,\F, \aF\}$, the operator
\[
\varrho^{(\adg)}_{I}: \ \{ u^{(\adg)}\in \cX^{(\adg), \sobo}: \ P^{(\adg)}u^{(\adg)}=0\}\to \cH^{\sobo}
\]
is Fredholm of index equal $\ind W_I^{\adg\dag}$ and  is invertible for $I\in\{\inn,\out\}$. 
\end{lemma}
\proof We apply  \eqref{e100.15} and the fact that $\varo_{0}^{\adg}: \Ker P^{\adg}|_{\cX^{\adg, \sobo}}\to \cH^{\sobo}$ and $T^{-1}(0)\varo_{0}: \Ker P|_{\cX^{\sobo}}\to \cH^{\sobo+\12}$ are bijections, by Lemma \ref{lem:inho}. \qeds

\begin{lemma}\label{lem:aux} 
The maps 
\[
\begin{array}{l}
\varrho^{\adg}_{I}\oplus P^{\adg} : \cX^{\adg, \sobo}\to \cH^{\sobo}\oplus \cY^{\adg, \sobo}, \\[2mm]
\varrho_{I}\oplus P : \cX^{\sobo}\to \cH^{\sobo+\12}\oplus \cY^{\sobo},
\end{array}
\] are Fredholm of index $\ind W_{I}^{\adg\dag}$.
\end{lemma}
\proof We use Lemma \ref{lem:fredholm} with $\cK=\cX^{\adg, \sobo}$ resp.  $\cX^{\sobo}$, $\cE=\cH^{\sobo}$ resp. $\cH^{\sobo+\12}$, $\cF=\cY^{\adg, \sobo}$ resp. $\cY^{\sobo}$, $K=\varrho^{\adg}_{I}$ resp. $\varo_{I}$, $Q=P^{\adg}$ resp. $P$. The assumptions of Lemma \ref{lem:fredholm} are satisfied in view of Lemma \ref{lem:aux0} and Lemma \ref{lem:inho} which gives surjectivity of $P^{\adg}$ resp. $P$.  \qeds

Let us introduce the following notation: if $I=\inn/\out$ then $I^{\rm c}\defeq\out/\inn$ and if $I=\F/\aF$ then $I^{\rm c}\defeq \aF/\F$.

\begin{theorem}\label{prop:fredholm1}Let $\cX^{(\adg), \sobo}_{I}\defeq\{ u\in \cX^{(\adg), \sobo}: \varrho^{(\adg)}_{I^{\rm c}}u=0\}$, equipped with the topology of $\cX^{(\adg), \sobo}$.  Then 
\[
\begin{array}{l}
P^{\adg}: \cX^{\adg, \sobo}_{I}\to\cY^{\adg, \sobo},\\[2mm]
P: \cX^{\sobo}_{I}\to \cY^{\sobo}
\end{array}
\] are  Fredholm of index $\ind W_{I^{\rm c}}^{\adg\dag}$.
\end{theorem}
\proof It suffices to  check the assumptions of Lemma \ref{lem:fredholm}  for $\cK=\cX^{\adg, \sobo}$ resp. $\cX^{\sobo}$,  $\cE=\cY^{\adg, \sobo}$ resp. $\cY^{\sobo}$, $\cF=\cH^{\sobo}$ resp. $\cH^{\sobo+\12}$,  $K=P^{\adg}$ resp. $P$, and $Q=\varrho^{\adg}_{I^{\rm c}}$ resp. $\varo_{I^{\rm c}}$. 
 The Fredholm property of $K\oplus Q$ follows from Lemma \ref{lem:aux}, so it remains to check that $\varrho^{\adg}_{I^{\rm c}}:\cX^{\adg, \sobo}\to\cH^{\sobo}$  and $\varo_{I^{\rm c}}: \cX^{\sobo}\to \cH^{\sobo+\12}$ are  surjective.  This is obvious if $I= \outin$ using \eqref{e100.15} and Lemma \ref{lem:inho}.  Let us now consider the case $I= \F$.

Let $\eta_\outin\in\cf(\rr)$ with  $\eta_\inn(t)+\eta_\out(t)=1$ and $\eta_{\outin}(t)=1$ for large $\pm t$. Then 
\[\varrho^{(\adg)}_{\outin}\circ\eta_{\outin}=\varrho^{(\adg)}_{\outin}, \ \varrho^{(\adg)}_{\inout}\circ\eta_{\outin}=0.
\] Furthermore $\eta_\outin \Ker P^{(\adg)}|_{\cX^{(\adg), \sobo}}\subset \cX^{(\adg), \sobo}$. It follows that
\[
\bea
\varrho^{\adg}_{\F} \cX^{\adg, \sobo} &\supset \varrho^{\adg}_{\F}(\eta_\inn\Ker P^{\adg}|_{\cX^{\adg, \sobo}}+\eta_\out \Ker P^{\adg}|_{\cX^{\adg, \sobo}} )\\
&= (\pi^+ \varrho^{\adg}_\out + \pi^-\varrho^{\adg}_\inn )(\eta_\inn\Ker P^{\adg}|_{\cX^{\adg, \sobo}}+\eta_\out \Ker P^{\adg}|_{\cX^{\adg, \sobo}} )\\
&=\pi^+ \varrho^{\adg}_\out  \Ker P^{\adg}|_{\cX^{\adg, \sobo}} + \pi^-\varrho^{\adg}_\inn   \Ker P^{\adg}|_{\cX^{\adg, \sobo}} = \pi^+\cH^{\sobo} + \pi^- \cH^{\sobo}=\cH^{\sobo}.
\eea
\] 
This proves $\varrho^{\adg}_{\F}:\cX^{\adg, \sobo}\to\cH^{\sobo}$ is surjective.  The same argument shows that $\varo_{\F}: \cX^{\sobo}\to \cH^{\sobo+\12}$ is surjective. In the analogous way we obtain surjectivity of $\varrho^{(\adg)}_\aF$. \qed
\subsection{Retarded/advanced propagators}
We now show that as anticipated, the retarded/ad\-vanced propagators $G^{\adg}_{\pm}$ are the inverses of $P^{\adg}: \cX^{\adg, \sobo}_{\outin}\to \cY^{\adg, \sobo}$, and a similar statement holds true in the scalar case.

\begin{proposition}\label{proporetard}
 $P^{(\adg)}: \cX^{(\adg), \sobo}_{\outin}\to \cY^{(\adg), \sobo}$ are boundedly invertible with inverse equal to $G^{(\adg)}_{\pm}$.
\end{proposition}
\proof We only treat the case of  $G^{(\adg)}_{+}$. We have seen in Subsect. \ref{retard} that 
\[
G^{\adg}_{+}\in B(L^{1}(\rr; \cH^{\sobo}),C^{0}(\rr; \cH^{\sobo}))
\]
and  $P^{\adg} G^{\adg}_{+}=\one$ on $L^{1}(\rr; \cH^{\sobo})$,  hence $G^{\adg}_{+}\in B(\cY^{\adg, \sobo}, \cX^{\adg, \sobo})$ and $P^{\adg}G^{\adg}_{+}=\one$ on $\cY^{\adg, \sobo}$. Since $\lim_{t\to-\infty}\varrho^{\adg}_{t} G^{\adg}_{+}f^{\adg}=0$,  we have $G^{\adg}_{+}\cY^{\adg, \sobo}\subset \cX^{\adg, \sobo}_{\out}$. It remains to show that $G^{\adg}_{+}P^{\adg}=\one$ on $\cX^{\adg, \sobo}_{\out}$. If $u^{\adg}\in \cX^{\adg, \sobo}_{\out}$  we have:
\[
\bea
(G^{\adg}_{+}P^{\adg}u^{\adg})(t)&= \int_{-\infty}^{t}\cU^{\adg}(t, s)(\pe_{s}- \i H^{\adg}(s))u^{\adg}ds\\
&=\lim_{\varT\to -\infty}\int_{-\infty}^{t}\cU^{\adg}(t, s)(\pe_{s}- \i H^{\adg}(s))u^{\adg}ds\\
&=\lim_{\varT\to -\infty}\left[\cU^{\adg}(t,s)u(s)\right]^{t}_{\varT}-\lim_{\varT\to -\infty}\int_{\varT}^{t}(-\pe_{s}+ \i H^{\adg}(s))\cU^{\adg}(t,s)u^{\adg}(s)ds\\
&=u^{\adg}(t),
\eea
\]
since $\lim_{\varT\to -\infty}u^{\adg}(\varT)=0$ in view of $u^{\adg}\in \cX^{\adg, \sobo}_{\out}$. 
In the scalar case we obtain from \eqref{e100.0} that $(D_{t}- \AH(t))TG^{\adg}_{+}T^{-1}=\one$ hence $(\varo \pi_{0}-\one)TG^{\adg}_{+}T^{-1}\pi_{1}^{*}=0$ which implies that $PG_{+}=\one$ on $\cY^{\sobo}$.  Conversely, by \eqref{e100.1}, \eqref{e100.14} we know that $T^{-1}\varo: \cX^{\sobo}_{\out}\to \cX^{\adg, \sobo+\12}_{\out}$. Since $G^{\adg}_{+}P^{\adg}= \one$ on $\cX^{\adg, \sobo+ \12}$ this yields
\[
TG^{\adg}_{+}T^{-1}(D_{t}- \AH(t))\varo=TG^{\adg}_{+}P^{\adg}T^{-1}\varo= \varo\hbox{ on }\cX^{\sobo}_{\out},  
\]
hence $G_{+}P= \one$ on $\cX^{\sobo}_{\out}$ using  $(D_{t}-\AH(t))\varo= \pi_{1}^{*}\pi_{1}(D_{t}- \AH(t))$. This completes the proof. \qeds
\subsection{ The Fredholm inverses for $P^{(\adg)}$ on $\cX^{(\adg), \sobo}_{\F}$}
From Thm.  \ref{prop:fredholm1} we know that $P^{(\adg)}: \cX^{(\adg), \sobo}_{\F}\to \cY^{(\adg), \sobo}$ are Fredholm. We will now construct explicit approximate inverses $G^{(\adg)}_{\F}$ of $P^{(\adg)}:\cX^{(\adg),\sobo}_{\F}\to\cY^{\sobo}$, which requires some special care because of the requirement $\varrho_\aF^{(\adg)} \circ    G^{(\adg)}_{\F}=0$ that follows from the definition of $\cX^{(\adg),\sobo}_{\F}$ (in fact, for instance the time-ordered Feynman propagators associated to the \emph{in} or \emph{out} state\footnote{These are analysed for instance by Isozaki in \cite{isozaki}.} fail to satisfy this condition in general). We will then show that $G_{\F}$ has Feynman-type wavefront set.

\subsubsection{Auxiliary diagonal Hamiltonian}\label{auxili} We denote by $H^{\dg}(t)$   the  diagonal part of $H^{\adg}(t)$, see Subsect. \ref{ss:iad}. We recall that:
\begin{equation}
\label{e11.5b}
\begin{array}{l}
V^{\adg}_{-\infty}(t)= H^{\dg}(t)- H^{\adg}(t)\in \Psi^{-\infty, -1- \delta}_{\std}(\rr; \rr^{d})\otimes \cc^{2},\\[2mm]
H^{\dg}(t)= \mat{\epsilon^{+}(t)}{0}{0}{\epsilon^{-}(t)}, \hbox{ where}\\[2mm]
\epsilon^{\pm}(t)= \epsilon^{\pm}(t)^{*}, \ \epsilon^{\pm}(t)\mp \epsilon(t)\in  \Psi^{0, -1- \delta}_{\std}(\rr; \rr^{d}).
\end{array}
\end{equation}

Let $\cU^{\dg}(t,s)$ be the evolution generated by the Hamiltonian $H^{\dg}(t)$ defined in \eqref{e11.3}.
Using \eqref{e11.5b}  we see that $\cU^{\dg}(t,s)$ is well defined  and moreover $\sup_{t, s\in \rr}\|\cU^{\dg}(t,s)\|_{B(\cH^{\sobo})}<\infty$ using the same argument as for $\cU(t,s)$.  Since $H^{\dg}(t)= H^{\dg\dag}(t)$ we also have
\beq\label{e11.5c}
\cU^{\dg}(t,s)^{\dag}= \cU^{\dg}(s,t).
\eeq
We set correspondingly
\[
P^{\dg}\defeq D_{t}- H^{\dg}(t)= P^{\adg} - V^{\adg}_{-\infty}(t).
\]
Note that since $\| V^{\adg}_{-\infty}(t)\|_{B(\cH^{\sobo})}= O(\langle t\rangle^{-1- \delta})$ and  we have assumed that $\gamma<\12 + \delta$, we see that for $u\in C^{0}(\rr; \cH^{\sobo})$ we have $ P^{\adg}u\in \cY^{\adg, \sobo}$ if and only if $P^{\dg}u\in \cY^{\adg, \sobo}$ and the two norms in \eqref{defdenorme} on $\cX^{\adg, \sobo}$ defined with $P^{\adg}$ and $P^{\dg}$ are equivalent.

Finally we define the operator $\cU^{\dg}: \cH^{\sobo}\to \cX^{\sobo}$ by: 
  \beq\label{defudiag}
  \cU^{\dg}v^{\adg}(t)\defeq  \cU^{\dg}(t,0)v^{\adg}, \ v^{\adg}\in \cH^{\sobo}.
  \eeq 
By the remark above, $\cU^{\dg}\in B(\cH^{\sobo}, \cX^{\adg, \sobo})$.

\subsubsection{Fredholm inverse for $P^{\adg}$ on $\cX^{\adg, \sobo}_{\F}$}
\begin{definition}We set for $f^{\adg}\in \cY^{\adg, \sobo}$:
 \[
G^{\adg}_{\F} f^{\adg}(t)\defeq  \i\int_{-\infty}^{t}\cU^{\dg}(t, 0)\pi^{+}\cU^{\dg}(0,s)f^{\adg}(s)ds-\i\int_{t}^{+\infty}\cU^{\dg}(t, 0)\pi^{-}\cU^{\dg}(0,s)f^{\adg}(s)ds.
\]
\end{definition}
Using the `time-kernel notation' $A(t,s)\defeq  \varrho_{t}\circ A\circ \varrho_{s}^{*}$ we can write:
\[
\bea
G^{\adg}_{\F}(t, s)&= \i \theta(t-s)\cU^{\dg}(t, 0)\pi^{+}\cU^{\dg}(0,s)-\i \theta(s-t)\cU^{\dg}(t, 0)\pi^{-}\cU^{\dg}(0,s)\\
&= \i \cU^{\dg}(t, 0)\pi^{+}\cU^{\dg}(0,s)-\i \theta(s-t)\cU^{\dg}(t,s),
\eea
\]
where $\theta$ is the Heaviside step function. Let us also observe that since $[\cU^{\dg}(t,s), \pi^{+}]=0$, we have
\beq\label{asz}
G^{\adg}_{\F}= G^{\dg}_{+}\pi^{+}+ G^{\dg}_{-}\pi^{-},
\eeq
where $G^{\dg}_{\pm}$ are the retarded/advanced propagators for $H^{\dg}(t)$, defined in analogy to $G^{\adg}_{\pm}$.

\begin{theorem}\label{theotheo}
Let $m\in\rr$. We have:
\[
\begin{array}{rl}
i)&G^{\adg}_{\F}\in B(\cY^{\adg, \sobo}, \cX^{\adg, \sobo}_{\F}), \  P^{\adg}G^{\adg}_{\F}= \one_{\cY^{\adg, \sobo}}+ K_{\cY^{\adg, \sobo}}, \hbox{where }K_{\cY^{\adg, \sobo}}\hbox{ is compact on }\cY^{\adg, \sobo},\\[2mm]
ii)&G^{\adg}_{\F} P^{\adg}= \one_{\cX^{\adg, \sobo}_{\F}}+ K_{\cX^{\adg, \sobo}_{\F}}, \hbox{where }K_{\cX^{\adg, \sobo}_{\F}}\hbox{ is compact on }\cX^{\adg, \sobo}_{\F},\\[2mm]
iii)&\i^{-1}q^{\adg}(G^{\adg}_{\F}- (G^{\adg}_{\F})^{\dag})\geq 0\hbox{ on }\cY^{\adg, \sobo}, \hbox{ for }\sobo\geq 0.
\end{array}
\] 
\end{theorem}
To prove Thm. \ref{theotheo} we will need the following lemma.
\begin{lemma}\label{compact1}
$V^{\adg}_{-\infty}: \cX^{\adg, \sobo}\to \cY^{\adg, \sobo}$ is compact.
\end{lemma}
\proof 
From \eqref{e12.1bb} we first obtain that the injection $\cX^{\adg, \sobo}\hookrightarrow C^{k}(\rr; \cH^{\sobo-k})$ is bounded for any $k\in \nn$, $\sobo\in \rr$. We pick $\varepsilon>0$ such that $\gamma<\12 + \delta -\varepsilon$ and write $V^{\adg}_{-\infty}(t)$ as $\langle t\rangle^{-1- \delta+\varepsilon}\langle \rx\rangle^{-\varepsilon}Y^{\adg}(t)$, where 
$Y^{\adg}(t)\in C^{\infty}(\rr; \cW^{-\infty}(\rr^{d})\otimes B(\cc^2))$. It follows that $Y^{\adg}: C^{k}(\rr; \cH^{\sobo})\to C^{k}(\rr; \cH^{\sobo'})$ is bounded for any $\sobo,\sobo'$, hence 
\beq\label{compacto}
V^{\adg}_{-\infty}: \cX^{\sobo}\to \langle t\rangle^{-1- \delta + \varepsilon}C^{k}(\rr; \cH^{\sobo'})\hbox{ is compact for any }k\in \nn, \ \sobo, \sobo'\in \rr. 
\eeq
We use \eqref{compacto} for   $k=0,s'=s$,   and the fact that  the injection  $\langle t\rangle^{-1- \delta + \varepsilon}C^{0}(\rr; \cH^{\sobo})\hookrightarrow \langle t\rangle^{-\gamma}L^{2}(\rr; \cH^{\sobo})= \cY^{\adg, \sobo}$ is bounded since $\gamma<\12+ \delta- \varepsilon$. It follows that $V^{\adg}_{-\infty}: \cX^{\adg, \sobo}\to \cY^{\adg, \sobo}$ is compact.  \qeds

{\noindent\bf Proof of Thm. \ref{theotheo}}
{\it Proof of $i)$}: note first that  $G^{\adg}_{\F}= \gdia_{+}\pi^{+}+ \gdia_{-}\pi^{-}\in B(\cY^{\adg, \sobo}, \cX^{\adg, \sobo})$ since $\gdia_{\pm}\in B(\cY^{\adg, \sobo}, \cX^{\adg, \sobo})$.
We then have:
\[
\bea
P^{\adg} G^{\adg}_{\F}&= P^{\dg}G^{\adg}_{\F}+ V^{\adg}_{-\infty} G^{\adg}_{\F}\\[2mm]
&= P^{\dg}G^{\dg}_{+}\pi^{+}+ P^{\dg}G^{\dg}_{-}\pi^{-}+ V^{\adg}_{-\infty} G^{\adg}_{\F}\\[2mm]
&=\one_{\cY^{\adg, \sobo}}+ V^{\adg}_{-\infty} G^{\adg}_{\F},
\eea
\]
by Prop. \ref{proporetard} applied to $P^{\dg}$.  By Lemma \ref{compact1}, $V^{\adg}_{-\infty}G^{\adg}_{\F}$ is compact on $\cY^{\adg, \sobo}$.

 It remains to check that 
$G^{\adg}_{\F}: \cY^{\adg, \sobo}\to\cX^{\adg, \sobo}_{\F}$, i.e.  $\pi^{+}\varrho^{\adg}_{\rm in}G^{\adg}_{\F}= \pi^{-}\varrho^{\adg}_{\out}G^{\adg}_{\F}=0$. We have:
 \[
 \pi^{+}\varrho^{\adg}_{\rm in}G^{\adg}_{\F}= \varrho^{\adg}_{\rm in}\pi^{+}(\gdia_{+}\pi^{+}+ \gdia_{-}\pi^{-})= \varrho^{\adg}_{\rm in}\gdia_{+}\pi^{+}=0,
 \]
 since $[\gdia_{\pm}, \pi^{\pm}]=0$ and $\varrho^{\adg}_{\rm in }\gdia_{+}=0$. Similarly we obtain that $ \pi^{-}\varrho^{\adg}_{\out}G^{\adg}_{\F}=0$,  which completes the proof of {\it i)}.
 
 {\it Proof of $ii)$}:   we have  by \eqref{asz}:
 \[
G^{\adg}_{\F} P^{\adg}= G^{\adg}_{\F} P^{\dg}+ G^{\adg}_{\F} V^{\adg}_{-\infty}=\gdia_{+}P^{\dg}\pi^{+}+ \gdia_{-}P^{\dg}\pi^{-}+ G^{\adg}_{\F} V^{\adg}_{-\infty}.
\]
 If $u\in \cX^{\adg, \sobo}_{\F}$ we have $\pi^{+}\varrho^{\adg}_{\rm in }u= \varrho^{\adg}_{\rm in }\pi^{+}u=0$ and  $\pi^{-}\varrho^{\adg}_{\rm out }u= \varrho^{\adg}_{\rm out }\pi^{-}u=0$. By Prop. \ref{proporetard} applied to $P^{\dg}$ we have $\gdia_{+}P^{\dg}\pi^{+}u= \pi^{+}u$, $\gdia_{-}P^{\dg}\pi^{-}u= \pi^{-}u$, hence
 \[
G^{\adg}_{\F} P^{\adg}= \one_{\cX^{\adg, \sobo}_{\F}}+ G^{\adg}_{\F}V^{\adg}_{-\infty}.
\]
Again, by Lemma \ref{compact1} $G^{\adg}_{\F}V^{\adg}_{-\infty}$ is compact on $\cX^{\adg, \sobo}$. 
 
{\it Proof of iii)}: using the time-kernel notation we have by \eqref{e11.5c}
\[
\bea
&(G^{\adg}_{\F})^{\dag}(t,s)= G^{\adg}_{\F}(s,t)^{\dag}\\[2mm]
&=  \i \theta(t-s)\cU^{\dg}(t,0)\pi^{-}\cU^{\dg}(0,s)- \i \theta(s-t)\cU^{\dg}(t,0)\pi^{+}\cU^{\dg}(0,s),
\eea
\]
hence
\[
\i^{-1}(G^{\adg}_{\F}- (G^{\adg}_{\F})^{\dag})(t,s)= \cU^{\dg}(t,0)(\pi^{+}- \pi^{-})\cU^{\dg}(0,s)\eqdef \cU^{\dg}(t)q^{\adg}(\cU^{\dg})^{\dag}(s),
\]
where $\cU^{\dg}$ is defined in \eqref{defudiag}. It follows that
\[
\i^{-1}(f^{\adg}| q^{\adg}(G^{\adg}_{\F}- (G^{\adg}_{\F})^{\dag})f^{\adg})_{\cH^{0}}= ((\cU^{\dg})^{\dag}f^{\adg}| (q^{\adg})^{2}(\cU^{\dg})^{\dag}f^{\adg})_{\cH^{0}}\geq 0,
\]
hence $\i^{-1}q^{\adg}(G^{\adg}_{\F}- (G^{\adg}_{\F})^{\dag})\geq 0$ on $\cH^{0}$  hence on $\cH^{\sobo}$ for $\sobo\geq 0$. \qeds

\subsubsection{Fredholm inverse for $P$ on $\cX^{\sobo}_{\F}$}
\begin{theorem}\label{teuheuteuheu} Assume $(\std)$. Let \beq\label{e100.20}
G_{\F}\defeq- \pi_{0} T G^{\adg}_{\F}T^{-1}\pi_{1}^{*}.
\eeq
We have:
\[
\begin{array}{rl}
i)&G_{\F}\in B(\cY^{\sobo}, \cX^{\sobo}_{\F}), \  PG_{\F}= \one_{\cY^{\sobo}}+ K_{\cY^{\sobo}}, \hbox{where }K_{\cY^{\sobo}}\hbox{ is compact on }\cY^{\sobo},\\[2mm]
ii)&G_{\F}P= \one_{\cX^{\sobo}_{\F}}+ K_{\cX^{\sobo}_{\F}}, \hbox{where }K_{\cX^{\sobo}_{\F}}\hbox{ is compact on }\cX^{\sobo}_{\F},\\[2mm]
iii)&\i^{-1}(G_{\F}- G_{\F}^{*})\geq 0\hbox{ on }\cY^{\sobo}, \hbox{ for }\sobo\geq 0,\\[2mm]
iv)&P G_{\F}-\one, \ G_{\F}P-\one \hbox{ are smoothing operators},\\[2mm]
v)& \WF(G_{\F})'= (\diag_{T^*M})\cup\textstyle\bigcup_{t\leq 0}(\Phi_t(\diag_{T^*M})\cap \pi^{-1}\cN).
\end{array}
\] 
In particular $G_{\F}$ is a Feynman parametrix of $P$ in the sense of \cite{DH}.
\end{theorem}
\proof 
{\it Proof of $i)$:} from \eqref{e100.1} and Thm. \ref{theotheo} we see that $G_{\F}\in B(\cY^{\sobo}, \cX^{\sobo})$. Let us show that $G_{\F}$ maps $\cY^{\sobo}$ into $\cX^{\sobo}_{\F}$.   For $V^{\adg}_{-\infty}$ the operator introduced in \eqref{e11.5b} we have:
\[
P^{\adg} G^{\adg}_{\F}= \one +  V^{\adg}_{-\infty}G^{\adg}_{\F} \ \Rightarrow \ TP^{\adg}T^{-1}TG^{\adg}_{\F}T^{-1}= \one + TV^{\adg}_{-\infty} G^{\adg}_{\F}T^{-1}.
\]
Using \eqref{e100.0} this implies that:
\beq\label{e100.4}
(D_{t}- \AH(t))T G^{\adg}_{\F}T^{-1}\pi_{1}^{*}= \pi_{1}^{*}+ TV^{\adg}_{-\infty}G^{\adg}_{\F}T^{-1}\pi_{1}^{*}\eqdef \pi_{1}^{*}+ R_1,
\eeq
where $R_1\in B(\cY^{\sobo}, \langle t\rangle^{-1-\delta}C^{0}(\rr; \cE^{\sobo}))$, using that $V^{\adg}_{-\infty}\in \Psi^{-\infty, -1- \delta}_{\std}(\rr; \rr^{d})\otimes B(\cc^2)$.
This implies that
\[
\varrho \pi_{0} T G^{\adg}_{\F}T^{-1}\pi_{1}^{*}= T G^{\adg}_{\F}T^{-1}\pi_{1}^{*}+ R_2,
\]
where $R_2\in B(\cY^{\sobo}, \langle t\rangle^{-1- \delta}C^{0}(\rr; \cE^{\sobo}))$. We now have:
\[
\bea
\cU_{\outin}(0,t)\varrho_{t}G_{\F}f&=-\cU_{\outin}(0,t)\varrho_{t}\pi_{0} T G^{\adg}_{\F}T^{-1}\pi_{1}^{*}f\\[2mm]
&=-\cU_{\outin}(0,t)T(t)\varrho^{\adg}_{t} G^{\adg}_{\F}T^{-1}\pi_{1}^{*}f+ o(1)\\[2mm]
&=-T_{\outin}\cU^{\adg}_{\outin}(0,t)T_{\outin}^{-1}T(t)\varrho^{\adg}_{t}G^{\adg}_{\F}T^{-1}\pi_{1}^{*}f+ o(1)\\[2mm]
&=-T_{\outin}\cU^{\adg}_{\outin}(0,t)\varrho^{\adg}_{t}G^{\adg}_{\F}T^{-1}\pi_{1}^{*}f+ o(1)\\[2mm]
&=- T_{\outin}\varrho^{\adg}_{\outin} G^{\adg}_{\F}T^{-1}\pi_{1}^{*}f+ o(1),
\eea
\]
hence
\begin{equation}
\label{e100.3}
\varrho_{\outin}G_{\F}= - \varrho^{\adg}_{\outin} G^{\adg}_{\F}T^{-1}\pi_{1}^{*}.
\end{equation}
By Thm. \ref{theotheo} we have $\varrho^{\adg}_{\aF}G^{\adg}_{\F}=0$, i.e. $\pi^{-}\varrho^{\adg}_{\out}G^{\adg}_{\F}= \pi^{+}\varrho^{\adg}_{\inn}G^{\adg}_{\F}=0$, which by \eqref{e100.3} gives $\varrho_{\aF}G_{\F}=0$.
It follows that $G_{\F}$ maps $\cY^{\sobo}$ to $\cX^{\sobo}_{\F}$ as claimed.

From \eqref{e100.4}, we obtain by an easy computation:
\beq\label{e100.5}
P G_{\F}= \one - \pi_{1}R_1- D_{t}\pi_{0}R_1+ \i r\pi_{0}R_1.
\eeq
Using \eqref{compacto} we obtain that $R_1: \cY^{\sobo}\to \langle t\rangle^{-1- \delta + \varepsilon}C^{k}(\rr;  \cE^{\sobo'})$ is compact for any $\sobo, \sobo', k$, hence $PG_{\F}- \one$ is compact on $\cY^{\sobo}$. 

{\it Proof of $ii)$:} by Thm. \ref{theotheo} and \eqref{e100.0} we know that:
\[
G^{\adg}_{\F}T^{-1}(D_{t}- \AH(t))T= G^{\adg}_{\F}P^{\adg}= \one + G^{\adg}_{\F}V^{\adg}_{-\infty}\hbox{ on }\cX^{\adg, \sobo+ \12}_{\F},
\]
hence
\[
TG^{\adg}_{\F}T^{-1}(D_{t}- \AH(t))T= T+ TG^{\adg}_{\F}V^{\adg}_{-\infty}\hbox{ on }\cX^{\adg, \sobo+ \12}_{\F}.
\]
By \eqref{e100.1}, \eqref{e100.14} we know that $T^{-1}\varrho: \cX^{\sobo}_{\F}\to \cX^{\adg, \sobo+ \12}_{\F}$. It follows that
\[
TG^{\adg}_{\F}T^{-1}(D_{t}- \AH(t))\varrho= \varrho + T G^{\adg}_{\F}V^{\adg}_{-\infty}T^{-1}\varrho, \hbox{ on }\cX^{\sobo}_{\F}.
\]
Since $(D_{t}-\AH(t))\varo= \pi_{1}^{*}\pi_{1}(D_{t}- \AH(t))$, we obtain that
\beq\label{e100.5b}
\bea
G_{\F}P&= \pi_{0}TG^{\adg}_{\F}T^{-1}\pi_{1}^{*}\pi_{1}(D_{t}- \AH(t))\varo\\[2mm]
&=\pi_{0}\varo+   \pi_{0}T G^{\adg}_{\F}V^{\adg}_{-\infty}T^{-1}\varo\\[2mm]
&=\one +  \pi_{0}T G^{\adg}_{\F}V^{\adg}_{-\infty}T^{-1}\varo  \hbox{ on }\cX^{\sobo}_{\F}.
\eea
\eeq
Using \eqref{e100.1} and Lemma \ref{compact1} we obtain that $G_{\F}P-\one$ is compact on $\cX^{\sobo}_{\F}$, which proves {\it ii)}.

{\it Proof of $iii)$:} we note that for any operator $A^{\adg}$ one has  
\[
\pi_{0}A^{\adg}\pi_{1}^{*}= \pi_{1}q A^{\adg}\pi_{1}^{*}, \ \hbox{ hence } (\pi_{0}A^{\adg}\pi_{1}^{*})^{*}= \pi_{1}q A^{\adg\dag}\pi_{1}^{*}.
\]
This gives
\[
\i(G_{\F}- G_{\F}^{*})= \i^{-1}\pi_{1}q T(G^{\adg}_{\F}- G^{\adg\dag}_{\F})T^{-1}\pi_{1}^{*}=\i^{-1}(\pi_{1}T)q^{\adg}(G^{\adg}_{\F}- G^{\adg\dag}_{\F})(\pi_{1}T)^{*}\geq 0,
\]
by Thm. \ref{theotheo} {\it iii)}.

{\it Proof of $iv)$:} we first  see using $\cU^{\dg}(t,s)(D_{s}- H^{\dg}(s))=0$ and integration by parts that  $G^{\adg}_{\F}$ maps compactly supported elements of $H^{-p}(\rr; \cH^{-k})$ into $H^{p}(\rr; \cH^{-k-2p})$ for $k, p\in \nn$, hence $V^{\adg}_{-\infty}G^{\adg}_{\F}$ maps compactly supported elements of $H^{-p}(\rr; \cH^{k})$ into $H^{p}(\rr, \cH^{k})$. The same argument shows that $G^{\adg}_{\F}$ maps also compactly supported elements $H^{-p}(\rr; \cH^{k})$ into $H^{p}(\rr, \cH^{k})$. By \eqref{e100.4}, \eqref{e100.5}, \eqref{e100.5b} this implies {\it iv)}.

{\it Proof of $v)$:}  let $\omega_{\rm ref}$ be the Hadamard state given by  the projections $c^{\pm}_{\rm ref}(0)$, see Subsect. \ref{sec2.5}. If $\Lambda^{\pm}_{\rm ref}$ are its two-point functions (see Subsect. \ref{ss:qfree}), then as shown in \cite{bounded}, $G_{\F,\rm ref}= \i\Lambda^{+}_\rf+ G_{+}$ is a Feynman inverse for $P$, i.e. $P G_{\F,\rm ref}= G_{\F,\rm ref}P= \one$ on $\cof(\rr^{1+d})$ and 
\[
\WF'(G_{\F, \rf})=  (\diag_{T^*M})\cup\textstyle\bigcup_{t\leq 0}(\Phi_t(\diag_{T^*M})\cap \pi^{-1}\cN).
\]
 Moreover we know (see e.g. \cite[Thm. 7.10, Prop. 7.11]{bounded}) that $G_{\F, \rf}$ is given by the analog of \eqref{e100.20} with $\cU^{\dg}$ replaced by $\cU^{\adg}$ in the definition of $G^{\adg}_{\F}$. From \eqref{turlututu} we obtain that $\cU^{\dg}(t,s)- \cU^{\adg}(t,s)\in \cinf(\rr^{2}; \cW^{-\infty}(\rr^{d}))$, which implies that $G_{\F}- G_{\rm F, \rf}$ is smoothing and completes the proof of of {\it v)}. \qed

\section{Asymptotically Minkowski spacetimes}\init\label{sec:ams}

\subsection{Assumptions}\label{ss:nt}
In this final section we consider asymptotically Minkowski spacetimes and prove analogues of the results from Sect. \ref{sec:abstract} by using the reduction procedure from Sect. \ref{inout}.

We work on $M= \rr^{1+d}$, whose elements are denoted by $y=(t, \ry)$.

For $\delta\in \rr$ we denote by $S_{\std}^{\delta}(\rr^{1+d})$ the class of smooth functions such that 
\[
\p^{\alpha}_{y}f\in O(\langle y\rangle^{\delta- |\alpha|}), \ \alpha\in \nn^{1+d}.
\]
The analogous spaces on $\rr^{d}$ will be denoted by $S^{\delta}_{\rm sd}(\rr^{d})$.

We denote by $\alteta_{\mu\nu}$ the Minkowski metric on $\rr^{1+d}$, fix a Lorentzian metric $\altg$ on $\rr^{1+d}$ and  consider the Klein-Gordon operator
\begin{equation}
\label{e11.1}
P= -\Box_{\altg}+ \altV\,(y),
\end{equation}
where  $\altV$ is again a smooth real function.  We assume that $(M, \altg)$ is asymptotically Minkowski and $\altV$ is asymptotically constant in the following sense:
\[
(\aM)\ \begin{array}{rl}
i)&\altg_{\mu\nu}(y)- \alteta_{\mu\nu} \in S^{-\delta}_{\std}(\rr^{1+d}), \ \delta>1,\\[2mm]
ii)&\altV(y)- \altm^{2}\in S^{-\delta}_{\std}(\rr^{1+d}), \ \altm>0,\ \delta>1, \\[2mm]
iii)&(\rr^{1+d}, \altg) \hbox{ is globally hyperbolic},\\[2mm]
iv)&(\rr^{1+d}, \altg) \hbox{ has a  time function }\tilde{t}\hbox{ with  }\tilde{t}- t\in S^{1-\epsilon}_{\std}(\rr^{1+d})\hbox{ for }\epsilon>0.
\end{array}
\]
\begin{remark}
 We conjecture that $(\aM) \ iv)$  follows from $(\aM)\ i), \ iii)$.
\end{remark}
\subsection{Global hyperbolicity and non-trapping condition}
The null geodesics of $\altg$ coincide modulo reparametrization with  the projections on the base of  null bicharacteristics of 
$m(x, \xi)= |\xi|^{-1}\xi\cdot \altg^{-1}(x)\xi$.  We recall that 
$\Phi_{s}$  is  the flow of the Hamiltonian vector field $H_{p}$, $p(y, \xi)=\xi\cdot \altg^{-1}y\xi$, which acts naturally on $S^{*}\rr^{1+d}= T^{*}\rr^{1+d}\cap \{|\xi|=1\}$.  Null bicharacteristics stay in one of the two connected components $\cN^{\pm}$ of $\cN$. We set
\[
\Gamma_{\rm in /out}^{\pm}=\{X\in S^{\pm}: \ \phi_{s}(X)\not\to \infty \hbox{ as }s\to \pm \infty \}. 
\]
The familiar \emph{non-trapping condition} is:
\[
 (\nt)\hbox{ there are no trapped null geodesics of }\altg, \hbox{ i.e. } \Gamma^{\pm}= \Gamma^{\pm}_{\rm in}\cap \Gamma^{\pm}_{\rm out}= \emptyset.
 \]
 By a well-known argument, this actually implies that $\Gamma^{\pm}_{\rm in/out}= \emptyset$, see Lemma \ref{lem:traop} below, hence any null geodesic escapes to infinity  both when the affine parameter $s$ tends to $+\infty$ {\em and} to $-\infty$. 

\begin{lemma}\label{lem:traop}If $(\nt)$ holds then $\Gamma^{\pm}_{\rm in}= \Gamma^{\pm}_{\rm out}=\emptyset$.
\end{lemma}
\proof We  drop the $\pm$ superscript. We claim that $\Gamma_{\rm in}\neq \emptyset$  or $\Gamma_{\rm out}\neq \emptyset$ implies $\Gamma\neq \emptyset$. In fact Let $X_{0}\in \Gamma_{\rm in}$, $K_{1}$ a compact set such that $\{\Phi_{s}(X_{0}) : \ s\geq 0\}\subset K_{1}$. Let $s_{j}\to +\infty$ a sequence such that $X_{j}= \Phi_{s_{j}}(X_{0})\to X_{\infty}\in K_{1}$. Clearly $\Phi_{s}(X_{j})= \Phi_{s+s_{j}}(X_{0})\to \Phi_{s}(X_{\infty})$ for any $s\in \rr$. For $j$ large enough we have $\Phi_{s+s_{j}}(X_{0})\in K_{1}$ hence $\Phi_{s}(X_{\infty})\in K_{1}$ for any $s\in \rr$, which means that $X_{\infty}\in \Gamma$.
\qeds
\begin{proposition}\label{propnico}
 Assume $(\aM)\ i)$. Then 
 \ben
 \item $(\rr^{1+d}, \altg)$ is globally hyperbolic iff $(\nt)$ holds,
 \item if $(\aM)$, $iii)$ and $iv)$ hold then there exists a Cauchy time function $\tilde{t}$ such that $\tilde{t}- t\in \coinf(\rr^{1+d})$.
 \een
\end{proposition}
In the sequel we will work with the Cauchy time function $\tilde{t}$ obtain in  Prop. \ref{propnico} (2).\medskip

\proof First let us prove (1). By $(\aM)\  i)$ we have
\[
\{p, t\}=  \p_{\tau}(|\xi|^{-1}(\tau^{2}- k^{2})) + O(\langle x\rangle^{-\delta}|\xi|^{-1})\geq \tau|\xi|^{-1}+ O(\langle x\rangle^{-\delta}|\xi|^{-1}).
\]
It follows that there exist $c_{0}>0$ and compact sets $K^{\pm}\subset \cN^{\pm}$ such that 
\begin{equation}
\label{enice.1}
\pm\{m, t\}\geq c_{0} \hbox{ on }\cN^{\pm}\setminus K^{\pm}
\end{equation}
This implies that if $X\in \cN^{+}$ and $\phi_{s}(X)\to \infty$ when $s\to \pm \infty$ then $t\circ \phi_{s}(X)\to \pm\infty$ when $s\to \pm \infty$. Of course a similar statement  is true for $X\in \cN^{-}$ with  the reversed sign.

Let us set $\Sigma_{s}= t^{-1}(s)$.
Using   \eqref{enice.1} we obtain that there exists $T_{0}>0$ such that any null geodesic intersects $\Sigma_{\pm T}$ transversally for $T\geq T_{0}$ and hence enters $I^{\pm}(\Sigma_{\pm T})$.  Moreover  $\Sigma_{\pm T}$ is achronal for $T$ large enough, since $\p_{t}$ is a future directed time-like vector field in $\{\pm t\geq \pm T\}$ for $T$ large enough.  We can apply then the Geroch-S\'anchez  theorem  (see for instance \cite[Thm. 8.3.7]{W} for its basic version), which implies that $\Sigma_{\pm T}$ are Cauchy hypersurfaces for $T$ large enough, which completes the proof of (1) $\Leftarrow$.

 Assume now  that $(\rr^{1+d}, \altg)$ is globally hyperbolic  and $(\nt)$ is violated. Let  $\gamma=\{x(s) : \ s\in \rr\}$ be  an (affine parametrized) null geodesic  which is past and future trapped, ie $\gamma\subset K$ for some compact  set $K$.  Since $(\rr^{1+d}, \altg)$ is  strongly causal, for each $x\in K$ there exists an open neighborhood $V(x)$ of $x$ such that $\gamma$ enters $V(x)$ only once, ie $\{s\in \rr\ : \ x(s)\in V(x)\}=: I(x)$ is a bounded  open interval.  By compactness  of $K$ we have $x(s)\not\in K$ for $s\not\in \cup_{1}^{n}I(x_{i})$, which is a contradiction. This completes the proof of (1) $\Rightarrow$.
 
 Now let us prove  (2).
 Let $\tilde{t}$ be the  time function in $(\aM)\ iii)$. First of all, we note that it follows from $(\aM)$ that $-C^{-1}\leq d\tilde{t}\cdot \altg^{-1}d\tilde{t}\leq - C$ for some $C>0$. We fix a cutoff function $\chi\in \coinf(\rr^{1+d})$ with $0\leq \chi\leq 1$, $\chi= 1$ near $0$ and set  $\chi_{R}(y)= \chi(R^{-1}y)$, $\hat{t}_{R}=\chi_{R} \tilde{t}+ (1- \chi_{R})t$.  We have:
\[
d\hat{t}_{R}= \chi_{R} d\tilde{t}+ (1- \chi_{R})dt+ (\tilde{t}- t)d\chi_{R}.
\]
The covector $\alpha_{R}= \chi_{R} d\tilde{t}+ (1- \chi_{R})dt$ is a convex combination of future directed timelike covectors, which using $(\aM) \ i)$ implies that there exists $C>0$ such that $-C^{-1}\leq \alpha_{R} \cdot \altg^{-1}\alpha_{R}\leq -C$, uniformly for $R\geq 1$. The error term $(\tilde{t}- t)d\chi_{R}$ is of norm $O(R^{-\epsilon})$, which shows that $\hat{t}_{R}$ is a  time function for $R$ large enough.  Let us fix such an $R$ and denote $\hat{t}_{R}$ by 
$\hat{t}$.  Clearly $\hat{t}-t\in \coinf(\rr^{1+d})$. It remains to check that $\hat{t}$ is a Cauchy time function. First using  that $\hat{t}=t+ \coinf(\rr^{1+d})$  we obtain that
\beq\label{debiloff}
\lim_{T\to +\infty}\sup_{\Sigma_{-T}}\hat{t}= -\infty, \ \ \lim_{T\to +\infty}\inf_{\Sigma_{T}}\hat{t}= +\infty.
\eeq
Let now $\gamma$ be an inextendible future directed continuous causal curve and $s\in \rr$. Since $\hat{t}$ is a time function, $\gamma$ intersects $\hat{t}^{-1}(s)$ at most once.  By global hyperbolicity, $\gamma$ intersects the Cauchy hypersurfaces $\Sigma_{\pm T}$  for  $T$ large enough. By \eqref{debiloff} this implies choosing $T$ very large that $\gamma$ intersects $\hat{t}^{-1}(s^{\pm})$ for some $s_{-}<s<s_{+}$ hence  also $\hat{t}^{-1}(s)$. Therefore $\hat{t}^{-1}(s)$ is a Cauchy hypersurface for each $s$ and $\hat{t}$ is a Cauchy time function. \qeds

\subsection{Reduction to the model case}\label{s11.1}
We now  repeat the  constructions in Subsect. \ref{s10.1}, taking into account the additional space-time decay of $\altg$ and $\altV$.

After possibly redefining $\tilde{t}$ by adding a constant, we can assume that $\Sigma\defeq\tilde{t}^{-1}(\{0\})= \{0\}\times \rr^{d}$, so that $\Sigma$ is a Cauchy hypersurface both for $\altg$ and  $\eta$.

We set $v= \dfrac{\altg^{-1}d\tilde{t}}{d\tilde{t}\cdot \altg^{-1} d\tilde{t}}$, so that $v= \p_{t}$ outside a compact set. If $\phi_{t}$ is the flow of $v$, we set  as before:
\[
\chi: \rr\times \Sigma\in (t, \rx)\mapsto \phi_{t}(0, \rx)\in \rr^{1+d},
\]
so that $\tilde{t}(\chi(t, \rx))=t$. Due to the additional space decay properties, the diffeomorphism $\chi$ has  better properties than the ones stated in Lemma \ref{l10.1}.

\begin{lemma}\label{l11.1}Assume $(\aM)$. Then
\[
\chi^{*}\altg= -  \altch^{2}(t, \rx)dt^{2}+ \hat  \alth(t, \rx)d\rx^{2}, \ \ \chi^{*}\altV=  \altVh, 
\]
where:
\[
\hat \alth, \hat \alth^{-1}, \altch, \altch^{-1}, \altVh\in S^{0}_{\std}(\rr^{1+d}).
\]
Moreover there exist  diffeomorphisms $\ry_{\outin}$ of $\Sigma$ with 
\[
\ry_{\outin}(\rx)- \rx\in S^{1-\delta}_{{\rm sd}}(\rr^{d})
\]
such that if
\[
\hat{\alth}_{\outin}\defeq  \ry_{\outin}^{*}\altdelta,
\]
where $\altdelta$ is the flat Riemannian metric on $\rr^{d}$, we have:
\[
 \hat \alth- \hat \alth_{\outin}, \altch-1, \altVh- \altm^{2} \in S^{-\delta}_{\std}(\rr_{\pm}\times \rr^{d}).
\]
\end{lemma}
\proof  
We have $v= \p_{t}+ S^{-\delta}_{\std}$, which also implies that
\begin{equation}
\label{e11.1a}
\langle \phi_{s}(0, \rx)\rangle \geq C (\langle s\rangle+ \langle \rx\rangle), \ C>0.
\end{equation}
Setting $w\defeq  \pi_{\ry}v$, we have $\pi_{\ry}\chi(t, \rx)= \pi_{\ry}\rx+ \int_{0}^{t}w(\phi_{s}(\rx))ds$. Using that $w\in S^{-\delta}_{\std}(\rr^{1+d})$, we obtain that
\[
\ry_{\outin}(\rx)\defeq  \lim_{t\to \pm\infty}\pi_{\ry}\chi(t, \rx)
\]
exist and:
\begin{equation}
\label{e11.2}
\pi_{\ry}\chi(t, \rx)-\ry_{\outin}(\rx)\in S^{1-\delta}_{\std}(\rr_{\pm}\times \rr^{d}), \ \ry_{\outin}(\rx)- \rx\in S^{1-\delta}_{{\rm sd}}(\rr^{d}).
\end{equation}
By $(\aM)$, $iii)$, we also have $\chi(t, \rx)= (t, \pi_{\ry}\chi(t, \rx))$ for $|t|+ |\rx|\geq C$, hence
\[
D\chi(t, \rx)= \mat{1}{0}{0}{D\ry_{\outin}(\rx)}+ S^{-\delta}_{\std}(\rr_{\pm}\times \rr^{d}).
\]
This estimate and \eqref{e11.1a} imply the assertion. \qeds

As in Subsect. \ref{s10.1}, we set $\hat{P}= \chi^{*}P$, $\tilde{P}= \hat{c}^{1-n/2}\hat{P}\hat{c}^{1+n/2}$. In a similar vein we obtain that:
\beq\label{e11.333}
\tilde{P}= \pe_{t}^{2}+ r(t, \rx)\pe_{t}+ a(t, \rx, \pe_{\rx}), 
\eeq
where now by Lemma \ref{l11.1}, $r$, $a$ satisfy $(\Hstd)$  for $\delta>1$ with 
\beq\label{e11.3b}
a_{\outin}(\rx, \pe_{\rx})\defeq -\Delta_{\hat{\alth}_{\outin}}+ \altm^{2}= \chi_{\outin}^{*}(-\Delta_{\rx}+ \altm^{2}).
\eeq
There are several inconveniences related to the possibility that $a_\out\neq a_\inn$. It turns out, however, that they can be circumvented by considered the dynamics associated to the free Laplace operator $-\Delta_\rx+\altm^2$ instead of $\cU_\outin(t,s)$.
 
\subsection{Wave operators}\label{wavu}
We use the same notation as  in Sect. \ref{sec:abstract}. Let  $\Sigma_{s}\defeq \tilde{t}^{-1}(\{s\})$ for $\sobo\in \rr$. Using  the diffeomorphism $\chi$  we  identify $\Sigma_{s}$ with $\rr^{d}$  to define the Sobolev spaces $H^{\sobo}(\Sigma_{s})$. We introduce the energy spaces:
\[
\cE^{\sobo}(s)\defeq  H^{1+\sobo}(\Sigma_{s})\oplus H^{\sobo}(\Sigma_{s}), \ \sobo\in \rr.
\]
Of course  under $\chi$ all spaces $\cE^{\sobo}(s)$ equal $\cE^{\sobo}(\rr^{d})$ with uniformly equivalent norms. 
We denote by $\cU(t,s): \cE^{\sobo}(s)\to \cE^{\sobo}(t)$ the Cauchy evolution associated to $P$. 
We recall that:
\begin{equation}
\label{e11.10}
\cU(t,s)= Z(t)\cU^{\adg}(t,s)Z^{-1}(s),
\end{equation}
where $Z(t): \cE^{\sobo}(t)\to \cH^{\sobo+\12}$ was defined in Lemma \ref{defdeZ}. We set
\[
P_{\free}\defeq \pe_{t}^{2}- \Delta_{\rx}+ \altm^{2}, 
\]
 and 
 \[
Z_{\outin}\defeq  (\chi_{\outin}^{*})^{-1}T_{\outin}, \ \ \chi_{\outin}(t, \rx)\defeq (t, \ry_{\outin}(\rx)).
\]
 Denoting by $\cU_{\free}(t,s)$ the usual Cauchy evolution for $P_{\free}$ we have by \eqref{e11.3b}:
\[
\cU_{\free}(t,s)= Z_{\outin}\cU^{\adg}_{\outin}(t,s)Z_{\outin}^{-1}.
\]
\begin{proposition}\label{propowavo}
 The limits
 \beq\label{e11.3bb}
W_{\outin}\defeq\lim_{t\to \pm\infty}\cU(0, t)\cU_{\free}(t, 0)
\eeq
exist in $B(\cE^{\sobo}(0))$ with inverses
\beq\label{e11.3c}
W_{\outin}^{-1}=W_{\outin}^{\dag}= \lim_{t\to \pm\infty}\cU_{\free}(0, t)\cU(t, 0).
\eeq
Moreover one has:
\beq\label{e11.4c}
W_{\outin}= Z(0)W^{\adg}_{\outin}Z_{\outin}^{-1},
\eeq
where we recall that $W^{\adg}_{\outin}= \lim_{t\to\pm\infty}\cU^{\adg}(0,t)\cU^{\adg}_{\outin}(t,0)$.
\end{proposition}
\proof   The existence of the limits \eqref{e11.3bb}, \eqref{e11.3c} follows from the Cook argument, using the short range condition $\delta>1$. The  identity \eqref{e11.4c} follows from 
\[
\lim_{t\to \pm \infty}Z(t)Z_{\outin}-\one=0 \hbox{ in }B(\cH^{\sobo}),
\]
by Lemma \ref{l10.3}. \qeds
\subsection{The $\outin$ Hadamard states}
We now consider the \emph{out}/\emph{in} Hadamard states $\omega_{\outin}$ constructed in Thm. \ref{thm.scat1}, whose covariances are denoted by $c^{\pm}_{\outin}$. We denote by $c^{\pm, \vac}_{\free}$ the covariance of the free vacuum state associated to $P_{\free}$.  An easy computation shows that
\beq\label{e20.0}
c^{\pm, \vac}_{\free}= Z_{\out}\pi^{\pm}Z_{\out}^{-1}=Z_{\inn}\pi^{\pm}Z_{\inn}^{-1}.
\eeq
\begin{proposition}
 We have 
 \[
c^{\pm}_{\outin}=  W_{\outin}c^{\pm, \vac}_{\free}W_{\outin}^{-1}= \wlim_{t\to \pm \infty}\cU(0,t)c^{\pm, \vac}_{\free} \cU(t,0)\hbox{ in }\cE^{\sobo}.
\]
\end{proposition}
\proof We have 
\[
\bea
&\cU(0,t)\cU_{\free}(t,0)c^{\pm, \vac}_{\free}\cU_{\free}(0,t)\cU(t,0)\\[2mm]
&=Z(0)\cU^{\adg}(0,t)Z(t)^{-1}Z_{\outin}\cU^{\adg}_{\outin}(t,0)Z_{\outin}^{-1}c^{\pm, \vac}_{\free}Z_{\outin}\\
&\phantom{=} \times\cU^{\adg}_{\outin}(0,t)Z_{\outin}^{-1}Z(t)\cU^{\adg}(t,0)Z(0)^{-1}\\[2mm]
&=Z(0)\cU^{\adg}(0,t)\cU^{\adg}_{\outin}(t,0)Z_{\outin}^{-1}c^{\pm, \vac}_{\free}Z_{\outin}
\cU^{\adg}_{\outin}(0,t)\cU_{\adg}(t,0)Z(0)^{-1}+ o(1)\\[2mm]
&=Z(0)\cU^{\adg}(0,t)\cU^{\adg}_{\outin}(t,0)\pi^{\pm}\cU^{\adg}_{\outin}(0,t)\cU^{\adg}(t,0)Z(0)^{-1}+ o(1)\\[2mm]
&=Z(0)\cU^{\adg}(0,t)\pi^{\pm}\cU^{\adg}_{\outin}(0,t)\cU^{\adg}(t,0)Z(0)^{-1}+ o(1)=c^{\pm}_{\outin}+ o(1),
\eea
\]
using Thm. \ref{thm.scat1}. Letting $t\to \pm \infty$ we obtain the proposition. \qeds

\subsection{Fredholm problems and Feynman pseudo-inverse}\label{ss:fredf}
Following the notation in \eqref{e11.333}, the objects introduced in Sect. \ref{sec:abstract} will be denoted with tildes, like $\tilde{\cX}^{\sobo},\tilde{\cY}^{\sobo}$, etc. 
We define the spaces  
\beq
\bea
\cY^{\sobo}&\defeq (\chi^{-1})^{*}\tilde{\cY}^{\sobo}=(\chi^{-1})^{*}\big( \langle t\rangle^{-\gamma}L^{2}(\rr; H^{\sobo})\big),\\ 
\cX^{\sobo}&\defeq (\chi^{-1})^{*}\tilde{\cX}^{\sobo}=(\chi^{-1})^{*}\big\{ \tilde u \in \big(C^{1}(\rr; H^{\sobo+1})\cap C^{0}(\rr; H^{\sobo})\big): \ \tilde P u \in \tilde\cY^m\big\}.  
\eea
\eeq
In particular $\cX^{\sobo}$ is the space of $u\in \cD'(\rr^{1+d})$ such that $u\circ \chi\in C^{1}(\rr; H^{\sobo+1})\cap C^{0}(\rr; H^{\sobo})$ and $Pu\in \cY^{\sobo}$. We equip $\cY^{m}$ and $\cX^{m}$ with the norms obtained from $\tilde{\cY}^{m}$ and  $\tilde{\cX}^{m}$.
\begin{definition}\label{def:final}
 We set $\varo_{\outin}\defeq \slim_{t\to \pm \infty}\cU_{\free}(0,t)\varo_{t}$ and
 \[
 \bea
 \varo_{\F}\defeq c^{+,\vac}_{\free}\varo_{\out}+ c^{-, \vac}_{\free}\varo_{\inn}, \ \ W_{\F}^{\dag}\defeq c^{+,\vac}_{\free}W^{\dag}_{\out}+ c^{-, \vac}_{\free}W^{\dag}_{\inn},\\[2mm]
 \varo_{\bar{\F}}\defeq c^{-,\vac}_{\free}\varo_{\out}+ c^{+, \vac}_{\free}\varo_{\inn}, \ \ W_{\bar \F}^{\dag}\defeq c^{-,\vac}_{\free}W^{\dag}_{\out}+ c^{+, \vac}_{\free}W^{\dag}_{\inn},
 \eea
 \]
 and for $I\in \{\inn,\out,\F, \aF\}$:
 \[
 \cX^{\sobo}_{I}\defeq\{u\in \cX^{\sobo}: \ \varo_{I^{\rm c}}u=0\}.
 \]
 \end{definition}

\begin{theorem} Assume $(\aM)$ and let $P$, $\cX^\sobo_I$ be as defined in \eqref{e11.1} and Def. \ref{def:final} for $\sobo\in\rr$ and $I\in \{\inn,\out,\F, \aF\}$. Then:
\ben
 \item  $P: \cX^{\sobo}_{I}\to \cY^{\sobo}$ is Fredholm of index $\ind W_{I^{\rm c}}^{\dag}$, and invertible with inverse $G_{\pm}$ if $I= \outin$. Furthermore, $\Ker P|_{\cX^{\sobo}_{I}}\subset \cf(M)$ and the index does not depend on the Sobolev order $m$. 
 \item Let
 \[
 G_{\F}\defeq (\chi^{-1})^{*}(\altch^{1+n/2}\tilde{G}_{\F}\altch^{1-n/2}),
 \]
 where   $\tilde{G}_{\F}$ is the operator defined in \eqref{e100.20} and $\altch$, $\chi$ are defined in Subsect. \ref{s11.1}. Then:
\[
\begin{array}{rl}
i)&G_{\F}\in B(\cY^{\sobo}, \cX^{\sobo}_{\F}), \  PG_{\F}= \one_{\cY^{\sobo}}+ K_{\cY^{\sobo}}, \hbox{where }K_{\cY^{\sobo}}\hbox{ is compact on }\cY^{\sobo},\\[2mm]
ii)&G_{\F}P= \one_{\cX^{\sobo}_{\F}}+ K_{\cX^{\sobo}_{\F}}, \hbox{where }K_{\cX^{\sobo}_{\F}}\hbox{ is compact on }\cX^{\sobo}_{\F},\\[2mm]
iii)&\i^{-1}(G_{\F}- G_{\F}^{*})\geq 0\hbox{ on }\cY^{\sobo}, \hbox{ for }\sobo\geq 0,\\[2mm]
iv)&P G_{\F}-\one, G_{\F}P-\one \hbox{ are smoothing operators},\\[2mm]
v)& \WF(G_{\F})'= (\diag_{T^*M})\cup\textstyle\bigcup_{t\leq 0}(\Phi_t(\diag_{T^*M})\cap \pi^{-1}\cN).
\end{array}
\] 
In particular, $G_{\F}$ is a Feynman parametrix of $P$ in the sense of \cite{DH}.
 \een
 \end{theorem}
\proof  The maps \[
\begin{array}{l}
\cY^{\sobo}\ni f  \mapsto\tilde{f} \defeq \hat{c}^{1-n/2}f\circ \chi\in \tilde{\cY}^{\sobo},\\[2mm]
\cX^{\sobo}\ni u  \mapsto\tilde{u} \defeq \hat{c}^{1+n/2}u\circ \chi\in \tilde{\cY}^{\sobo},\\[2mm]
\end{array}
\]
are boundedly invertible and furthermore, $Pu=f$ iff $\tilde{P}\tilde{u}= \tilde{f}$. Moreover by the computations in Subsect. \ref{cauchycauchy} we obtain that $Z^{-1}\varo u= T^{-1}\tilde{\varo}\tilde{u}$ and hence $u\in \cX_{I}^{\sobo}$ iff $\tilde{u}\in \tilde{\cX}_{I}^{\sobo}$. The theorem follows hence from Thm. \ref{teuheuteuheu}  provided we  check that 
\beq\label{finito}
\ind W_{I}^{\dag}= \ind W^{\adg\dag}_{I}.
\eeq
This is obvious for $I= \outin$ since  the  operators are then bijective. Let us check \eqref{finito} for $I=\aF$ for example. We denote   by $Z_{\free}$ the analog of $Z_{\outin}$ with $\epsilon_{\outin}$ replaced by $\epsilon_{\free}= (-\Delta_{\rx}+\altm^{2})^{\12}$ and $\chi_{\outin}$ replaced by $\one$.  Using \eqref{e11.4c} and \eqref{e20.0} we obtain that
\[
Z_{\free}^{-1}W_{\aF}^{\dag}= (Z_{\free}^{-1}Z_{\out}\pi^{-}W^{\adg\dag}_{\out}+ Z_{\free}^{-1}Z_{\inn}\pi^{+}W^{\adg\dag}_{\inn}) Z(0)^{-1}=  S\circ Z_{\aF}^{\adg\dag}\circ Z(0)^{-1},
\] 
for $S= Z_{\free}^{-1}Z_{\out}\pi^{-}+ Z_{\free}^{-1}Z_{\inn}\pi^{+}$. But since 
$c^{\pm, \vac}_{\free}= Z_{\free}\pi^{\pm}Z_{\free}^{-1}$, $Z_{\free}^{-1}Z_{\outin}$ commutes with $\pi^{+}$ and $\pi^{-}$, using again \eqref{e20.0}. Therefore $S= \pi^{-}Z_{\free}^{-1}Z_{\out}\pi^{-}+ \pi^{+}Z_{\free}^{-1}Z_{\inn}\pi^{+}$ is invertible and hence $\ind W^{\dag}_{\aF}= \ind W^{\adg\dag}_{\aF}$. \qeds

\appendix
\section{}\init\label{secapp1}
\subsection{Proof of Prop. \ref{l5.1}}\label{ssecap1}
To prove Prop. \ref{l5.1} we first need an auxiliary lemma about parameter-dependent pseudo\-differential calculus.

 We start by introducing parameter dependent versions of the spaces  $\Psi^{m}(\Sigma)$, $S^{0}(\rr; \Psi^{m}(\Sigma))$ and $\Psi^{m, 0}_{\std}(\rr; \rr^{d})$.
 
We define the  symbol classes $\widetilde{S}^{m}(T^{*}\Sigma)$  for $m\in \rr$ as the space of functions $c(\rx, \spexi, \lambda)\in \cinf(T^{*}\Sigma\times \rr)$ such that:
\[
\p_{\lambda}^{\gamma}\p_{\rx}^{\alpha}\p_{\spexi}^{\beta}c(\rx, \spexi, \lambda)\in O(\langle \spexi\rangle + \langle \lambda\rangle)^{m-|\beta|- \gamma}, \ \alpha, \beta\in \nn^{d}, \ \gamma\in \nn,
\]
as usual understood after fixing a good chart cover and good chart diffeomorphisms, with uniformity of the constants with respect to the element of the cover.  The standard example of such a symbol is $c(\rx, \spexi, \lambda)= (a(\rx, \spexi)+\langle \lambda\rangle^{m})$, for $a\in S^{m}(T^{*}\Sigma)$ elliptic and positive.
 
The subspaces of symbols poly-homogeneous in $(\spexi, \lambda)$ are denoted by $\widetilde{S}^{m}_{\rm ph}(T^{*}\Sigma)$.
We define  $\widetilde{\cW}^{-\infty}(\Sigma)$ as the set of smooth maps $\rr\ni \lambda\mapsto a(\lambda)\in \cW^{-\infty}(\Sigma)$ such that:
 \[
\| \p_{\lambda}^{\gamma}a(\lambda)\|_{B(H^{-p}(\Sigma), H^{p}(\Sigma))}\in O(\langle \lambda\rangle^{-n}), \ \forall m, n,\gamma\in \nn,
 \]
 and we set
 \[
 \widetilde{\Psi}^{m}(\Sigma)\defeq  \Op(\widetilde{S}_{\rm ph}^{m}(T^{*}\Sigma)) + \widetilde{\cW}^{-\infty}(\Sigma).
 \]
We  also define the time-dependent versions:
\[S^{0}(\rr; \widetilde{S}^{m}_{({\rm ph})}(T^{*}\Sigma)),  \ S^{0}(\rr; \widetilde{\cW}^{-\infty}(\Sigma)), \ S^{0}(\rr; \widetilde{\Psi}^{m}(\Sigma)),
\]
in analogy with Subsect. \ref{symbolo}. For example $c(t, \rx, \spexi, \lambda)\in S^{0}(\rr; \widetilde{S}^{m}(T^{*}\Sigma))$ if
\[
\p_{t}^n\p_{\lambda}^{\gamma}\p_{\rx}^{\alpha}\p_{\spexi}^{\beta}c(t,\rx, \spexi, \lambda)\in O(\langle t\rangle^{-n}(\langle \spexi\rangle + \langle \lambda\rangle)^{m-|\beta|- \gamma}), \ \alpha, \beta\in \nn^{d}, \ \gamma, n\in \nn.
\]

If $\Sigma= \rr^{d}$ we  define similarly $\widetilde{S}_{\std}^{m,0}(\rr; T^{*}\rr^{d})$ to be the space of functions $c(t, \rx, \spexi, \lambda)$ such that:
\[
\p_{t}^{n}\p_{\lambda}^{\gamma}\p_{\rx}^{\alpha}\p_{\spexi}^{\beta}c(t, \rx, \spexi, \lambda)\in O((\langle x\rangle + \langle t \rangle)^{-n}(\langle \spexi\rangle + \langle \lambda\rangle)^{m-|\beta|- \gamma}), \ \alpha, \beta\in \nn^{d}, \ \gamma, n\in \nn.
\]
We define $\widetilde{\cW}^{-\infty}_{\std}(\rr; \rr^{d})$ as the set of smooth maps $\rr\ni \lambda\mapsto a( \lambda)\in \cW^{-\infty}_{\std}(\rr; \rr^{d})$ such that:
\[
\|\p_{t}^{n}\p_{\lambda}^{\gamma}(D_{\rx}^{2}+\rx^{2})^{m}a(t, \lambda)(D_{\rx}^{2}+\rx^{2})^{m}\|_{B(L^{2}(\rr^{d}))}\in O(\langle t\rangle^{-n}\langle \lambda\rangle^{-n}), \ \forall n\in \nn,
\]
and we set
\[
\widetilde{\Psi}^{m, 0}_{\std}(\rr; \rr^{d})= \Op^{\rm w}(\widetilde{S}_{\std, {\rm ph}}^{m,0}(\rr; T^{*}\rr^{d}))+ \widetilde{\cW}^{-\infty}_{\std}(\rr; \rr^{d}).
\]

\begin{lemma}\label{lemomo}
 Let $a(t)\in S^{0}(\rr; \Psi^{2}(\Sigma))$ resp. $\Psi^{2, 0}_{\std}(\rr; \rr^{d})$ such that $a(t)$ is elliptic, selfadjoint on $L^{2}(\Sigma)$ with $a(t)\geq c_{0}\one$,  $c_{0}>0$. Then $(a(t)+ \lambda^{2})^{-1}\in S^{0}(\rr; \widetilde{\Psi}^{-2}(\Sigma))$ resp. $\widetilde{\Psi}^{2, 0}_{\std}(\rr; \rr^{d})$.
 \end{lemma}

\proof  The proof is based on a reduction to the situation without   the parameter $\lambda$. 
We first present the argument in the time-independent case.

Let us denote by $l\in \rr$ the dual variable to $\lambda$. We consider the manifold of bounded geometry $\Sigma_{\rx}\times \rr_{l}$ equipped with the metric $\alth_{ij}(\rx)d\rx^{i}d\rx^{j}+ dl^{2}$. As good chart covering we can take $\widetilde{U}_{i}= U_{i}\times \rr$, $\widetilde{\psi}_{i}(\rx, l)= (\psi_{i}(\rx), l)$ where $\{U_{i}, \psi_{i}\}_{i\in \nn}$ is a good chart covering for $(\Sigma, \alth)$.  A subordinate good partition of unity is $\widetilde{\chi}_{i}(\rx, l)= \chi_{i}(\rx)$.
 
 The classes $S^{m}_{\rm ph}(T^{*}(\Sigma\times \rr))$ are then defined as in Subsect. \ref{symbolo} and one sets as in Subsect. \ref{sobolo}:
 \[
 \cW^{-\infty}(\Sigma\times \rr)= \bigcap_{m\in \nn}B(H^{-m}(\Sigma\times \rr), H^{m}(\Sigma\times\rr)),
 \]
 and $\Psi^{m}(\Sigma\times \rr)= \Op(S^{m}_{\rm ph}(T^{*}(\Sigma\times \rr)))+ \cW^{-\infty}(\Sigma\times \rr)$,
 where $\widetilde{\Op}$ is defined as in Subsect. \ref{pdosec} with $\Sigma$ replaced by $\Sigma\times \rr$. Note that because of our choice of the chart covering $\widetilde{\Op}$ is the usual Weyl quantization w.r.t. the $(l, \lambda)$ variables.
  We note that
 \[
 \widetilde{S}^{m}(T^{*}\Sigma)= \{c\in S^{m}_{\rm ph}(T^{*}(\Sigma\times \rr)): \ \p_{l}c=0\},
 \]
 and denoting by $T_{l}$ the group of translations in $l$ we have
 \[
 [T_{l}, \widetilde{\Op}(c)]=0, \forall\  l\in \rr \ \Leftrightarrow \ c\in \widetilde{S}^{m}(T^{*}\Sigma). 
 \]
Equivalently, if $\mathcal{F}$ is the Fourier transform in $l$ we have
 \beq\label{e.deco1}
 \bea
 &c\in S^{m}_{\rm ph}(T^{*}(\Sigma\times\rr)), \ [T_{l}, \widetilde{\Op}(c)]=0\\[2mm]
  \Leftrightarrow & \ \mathcal{F}\widetilde{\Op}(c)\mathcal{F}^{-1}= \int^{\oplus}_{\rr}\Op(c(\lambda))d\lambda, \hbox{ for } c(\lambda)\in \widetilde{S}^{m}(T^{*}\Sigma).
 \eea
  \eeq
 
 Let now $w\in \cW^{-\infty}(\Sigma\times \rr)$ with $[w, T_{l}]=0$.  We have:
 \beq\label{four}
 \mathcal{F}w\mathcal{F}^{-1}= \int^{\oplus}_{\rr} w(\lambda)d\lambda.
 \eeq
 Since $w\in \bigcap_{m\in \nn}B(H^{-m}(\Sigma\times \rr), H^{m}(\Sigma\times\rr))$ we obtain that:
 \[
 \int_{\rr}\langle \lambda\rangle^{n}\| w(\lambda)u(\lambda)\|^{2}_{H^{p}(\Sigma)}d\lambda\leq C_{n, p}\int_{\rr}\langle \lambda\rangle^{-n}\| u(\lambda)\|^{2}_{H^{-p}(\Sigma)}d\lambda, \ \forall n, p\in \nn,
 \]
 or equivalently
 \[
 \int^{\oplus}_{\rr}\langle \lambda\rangle^{n}(-\Delta_{\alth}+1)^{p/2}w(\lambda)(-\Delta_{\alth}+1)^{p/2}d\lambda\in B(L^{2}(\Sigma\times \rr)).
 \]
 This implies that
 \[
 \|w(\lambda)\|_{B(H^{-p/2}(\Sigma), H^{p/2}(\Sigma))}\in O(\langle \lambda\rangle^{-n})
 \]
 
 The same estimates hold for $\p_{\gamma}w(\lambda)$, which shows that
 \[
 w(\lambda)\in \widetilde{\cW}^{-\infty}(\Sigma).
 \]
 Conversely, if $w(\lambda)\in  \widetilde{\cW}^{-\infty}(\Sigma)$  it is immediate that  $w$ defined by \eqref{four}
 belongs to $\cW^{-\infty}(\Sigma\times \rr)$. Hence we have shown
 \begin{equation}
 \label{e.deco2}
 \bea
 & w\in \cW^{-\infty}(\Sigma\times \rr), \ [w, T_{l}]=0\\[2mm]
 \Leftrightarrow& \ \mathcal{F}w\mathcal{F}^{-1}= \int^{\oplus}_{\rr}w(\lambda)d\lambda, \hbox{ for }  w(\lambda)\in \widetilde{\cW}^{-\infty}(\Sigma). 
 \eea
  \end{equation}
 Let us now consider the time-dependent situation. If we  define the  time-dependent classes  $\cinfb(\rr; \widetilde{S}^{m}(T^{*}\Sigma))$, $\cinfb(\rr; \widetilde{\cW}^{-\infty}(\Sigma))$ and $\cinfb(\rr; \widetilde{\Psi}^{m}(\Sigma))$ in the obvious way, then 
 \begin{equation}
 \label{e.deco3}
 \bea
 &c\in \cinfb(\rr; S^{m}_{\rm ph}(T^{*}(\Sigma\times \rr))), \  [T_{l}, \widetilde{\Op}(c)(t)]=0\\[2mm]
 \Leftrightarrow& \ \mathcal{F}\widetilde{\Op}(c)(t)\mathcal{F}^{-1}= \int^{\oplus}_{\rr}\Op(c(t, \lambda))d\lambda, \ c(t, \lambda)\in \cinfb(\rr; \widetilde{S}^{m}(T^{*}\Sigma)),\\[2mm]
 &w\in \cinfb(\rr; \cW^{-\infty}(\Sigma\times \rr)), \ [w(t), T_{l}]=0\\[2mm]
\Leftrightarrow& \ \mathcal{F}w(t)\mathcal{F}^{-1}= \int^{\oplus}_{\rr}w(t, \lambda)d\lambda, \ w(t, \lambda)\in \cinfb(\rr; \widetilde{\cW}^{-\infty}(\Sigma)). 
 \eea
 \end{equation}
  The same results hold also if we replace $\cinfb(\rr; A)$ by $S^{\delta}(\rr; A)$ for $A= S^{m}_{\rm ph}(T^{*}(\Sigma\times \rr))$, $\widetilde{S}^{m}(T^{*}\Sigma)$ etc. In fact it suffices to note that $c(t)\in S^{\delta}(\rr; A)$ iff $\langle t\rangle^{-\delta+n}\p_{t}^{n}c(t)\in \cinfb(\rr; A)$ for all $n\in \nn$.

Let now $a(t)\in S^{0}(\rr; \Psi^{2}(\Sigma))$  be as in the lemma and let $A(t)= a(t)+ D_{l}^{2}$ acting on $L^{2}(\Sigma\times \rr)$. The operator $A(t)$ is elliptic in $S^{0}(\rr; \Psi^{2}(\Sigma\times \rr))$, selfadjoint on $H^{2}(\Sigma\times \rr)$ with $A(t)\geq c_{0}\one$ for $c_{0}$ as in the lemma. Applying Thm. \ref{seeley}  in the special case $\alpha= -1$ we obtain that $A(t)^{-1}\in S^{0}(\rr; \Psi^{-m}(\Sigma\times \rr))$. We have
\[
\mathcal{F}A(t)^{-1}\mathcal{F}^{-1}= \int^{\oplus}_{\rr}(a(t)+ \lambda^{2})^{-1}d\lambda, 
\]
which by \eqref{e.deco3} implies that $(a(t)+ \lambda^{2})^{-1}\in S^{0}(\rr; \widetilde{\Psi}^{-2}(\Sigma))$. 

If $a(t)\in \Psi^{2, 0}_{\std}(\rr; \rr^{d})$ we  consider the operator $A(t)= a(t)+ D_{l}^{2}$ again. One introduces analogous classes of time-dependent pseudodifferential operators acting on $\rr^{d}_{\rx}\times \rr_{l}$. For example the symbol classes are defined by the conditions
\[
\p_{t}^{n}\p_{\rx}^{\alpha}\p_{\spexi}^{\beta}\p_{l}^{p}\p_{\lambda}^{\gamma}a(t, \rx, \spexi, l, \lambda)\in O\big((\langle \rx\rangle+ \langle t\rangle)^{|\alpha|-n}(\langle \spexi\rangle+ \langle \lambda\rangle)^{m-|\beta|-\gamma}\big), \ \alpha, \beta\in \nn^{d}, \ n, p, \gamma\in \nn.
\]
To obtain a pdo calculus which is stable under composition one has to add an ideal included in  $S^{0}(\rr; \cW^{-\infty}(\rr^{d}_{\rx}\times \rr_{l}))$, consisting of operator-valued functions $\rr\ni t\mapsto a(t)$
such that
\[
\|\p_{t}^{n}(D_{l}^{2}+D_{\rx}^{2}+  \rx^{2})^{n}a(t)(D_{l}^{2}+ D_{\rx}^{2}+ \rx^{2})^{n}\|_{B(L^{2}(\rr^{d}_{\rx}\times \rr_{l}))}\in O(\langle t\rangle^{-n})  \ \forall n\in \nn.
\]
Again Seeley's theorem  and the analog of \eqref{e.deco3} are   valid for this class of pseudodifferential operators. The proof can be completed by exactly the same arguments. \qed

{\bf \noindent Proof of Prop. \ref{l5.1}.} In view of the identity
\[
a_{1}^{1+ \alpha}- a_{2}^{1+ \alpha}= (a_{1}- a_{2})a_{1}^{\alpha}+ a_{2}(a_{1}^{\alpha}- a_{2}^{\alpha}),
\]
 we see that it suffices to prove the proposition for $0<\alpha<1$. We  will use the following formula, valid for example  if $a$ is a selfadjoint operator on a Hilbert space $\cH$  with $a\geq c\one$, $c>0$:
\beq\label{powers}
a^{\alpha}= C_{\alpha}\int_{0}^{+\infty}(a+ s)^{-1}s^{\alpha}ds= C_{\alpha}\int_{\rr}(a+ \lambda^{2})^{-1}\lambda^{2\alpha+1}d\lambda, \ \alpha\in \rr,
\eeq
where the integrals are norm convergent in say, $B(\Dom a^{m}, \cH)$ for $m$ large enough.

We will detail the proof in the first case of Prop. \ref{l5.1}, i.e. $a_{i}\in S^{0}(\rr; \Psi^{2}(\Sigma))$. The second case can be handled similarly.

We have for $r(t)= a_{1}(t)- a_{2}(t)$:
\[
\bea
(a_{1}(t)+ \lambda^{2})^{-1}&= ( a_{2}(t)+ \lambda^{2})^{-1}(\one + (a_{2}(t)+ \lambda^{2})r(t)(a_{1}(t)+ \lambda^{2})^{-1})\\
&= ( a_{2}(t)+ \lambda^{2})^{-1}+  ( a_{2}(t)+ \lambda^{2})^{-2}(a_{2}(t)+ \lambda^{2})r(t)(a_{1}(t)+ \lambda^{2})^{-1}\\
&= (a_{2}(t)+ \lambda^{2})^{-1}+ (a_{2}(t)+ \lambda^{2})^{-2}a_{2}^{k/2}(t)c_{1}(t, \lambda)\\
&= (a_{2}(t)+ \lambda^{2})^{-1}+ a_{2}(t)c_{2}(t, \lambda), 
\eea
\]
where using Lemma \ref{lemomo}, $c_{1}(t,\lambda)\in S^{-\delta}(\rr; \widetilde{\Psi}^{0}(\Sigma))$ and $c_{2}(t, \lambda)\in S^{-\delta}(\rr; \widetilde{\Psi}^{-4}(\Sigma))$. From \eqref{powers} we obtain that:
\beq\label{potopoto}
a_{1}^{\alpha}(t)- a_{2}^{\alpha}(t)= C_{\alpha}a_{2}^{k/2}(t)\int_{\rr}r_{2}(t, \lambda)\lambda^{2\alpha+1}d\lambda.
\eeq
We now write $r_{2}(t, \lambda)$ as $\Op(d_{2}(t, \lambda))+ w_{2}(t, \lambda)$, for $d_{2}\in S^{-\delta}(\rr; \widetilde{S}_{\rm ph}^{-4}(T^{*}\Sigma))$ and $w_{2}(t, \lambda)\in S^{-\delta}(\rr; \widetilde{\cW}^{-\infty}(\Sigma))$.
Using that
\[
\int_{\rr}(\langle \xi\rangle + \langle \lambda \rangle)^{-4-k}\lambda^{2\alpha+1}d\lambda\sim \langle \xi\rangle^{2\alpha-2-k},
\]
we first obtain that
\[
\int_{\rr}d_{2}(t, \lambda)\lambda^{2\alpha+1}d\lambda\in S^{-\delta}(\rr; S_{\rm ph}^{2\alpha-2}(\Sigma)).
\]
Similarly we obtain that $\int_{\rr}w_{2}(t, \lambda)\lambda^{2\alpha+1}d\lambda\in S^{-\delta}(\rr; \cW^{-\infty}(\Sigma))$.  Using \eqref{potopoto} this implies that $a_{1}^{\alpha}(t)- a_{2}^{\alpha}(t)\in S^{-\delta}(\rr; \Psi^{2\alpha}(\Sigma))$, as claimed.
\subsection{Proof of Prop. \ref{p5.1}}\label{to1}

We follow the proof in \cite{GW}. The ${\rm out}$ and ${\rm in}$ cases are treated similarly.
We set $a_{0}= \frac{\i}{2}(\epsilon^{-1}\p_{t}\epsilon+ \epsilon^{-1}r\epsilon)$,
\[
F(c)\defeq  \12 \epsilon^{-1}\left( \p_{t}c+ [\epsilon, c]+ \i rc - c^{2}\right)= G(c)-\12 \epsilon^{-1}c^{2}.
\]
and look for $b(t)$ as $\epsilon(t)+ b_{0}$, where $b_{0}= a_{0}+ F(b_{0})$.
Let us start by studying some properties of  the map $F$. First if $c\in \Psi^{0, -\mu}_{(*)}$ then:
\[
\begin{array}{l}
G(c)\in \Psi_{(*)}^{-1, 0}\times \Psi_{(*)}^{0, -1- \mu}+ \Psi_{(*)}^{-1, 0}\times \Psi_{(*)}^{0, -\mu}+ \Psi_{(*)}^{-1, 0}\Psi_{(*)}^{0, -1- \delta}\times \Psi_{(*)}^{0, -\mu},\\[2mm]
\epsilon^{-1}c^{2}\in \Psi_{(*)}^{-1, -2\mu},
\end{array}
\]
hence 
\begin{equation}
\label{tito}
c\in \Psi_{(*)}^{0, -\mu}\ \Rightarrow \ F(c)\in \Psi_{(*)}^{-1, -\mu}.
\end{equation}
Secondly, if $c_{1}, c_{2}\in \Psi_{(*)}^{0, -\mu}$ and $c_{1}- c_{2}\in \Psi_{(*)}^{-j, -\mu}$ then:
\[
\bea
&G(c_{1})- G(c_{2})= G(c_{1}-c_{2})\\[1mm]
&\in \Psi_{(*)}^{-1, 0}\times \Psi_{(*)}^{-j, -1-\mu}+ \Psi_{(*)}^{-1, 0}\times\Psi_{(*)}^{-j, -\mu}+ \Psi_{(*)}^{-1, 0}\times \Psi_{(*)}^{0, -1- \delta}\times \Psi_{(*)}^{-j, -\mu},\\[1mm]
&\epsilon^{-1}(c_{1}^{2}- c_{2}^{2})
=\epsilon^{-1}c_{1}(c_{1}- c_{2})+ \epsilon^{-1}(c_{1}- c_{2})c_{2}\\[1mm]
&\in \Psi_{(*)}^{-1, 0}\times \Psi_{(*)}^{0, -\mu}\times \Psi_{(*)}^{-j, -\mu}+ \Psi_{(*)}^{-1, 0}\times \Psi_{(*)}^{-j, -\mu}\times \Psi_{(*)}^{0, -\mu},
\eea
\] 
hence
\begin{equation}
\label{titu}
c_{1}, c_{2}\in  \Psi_{(*)}^{0, -\mu}, \ c_{1}- c_{2}\in \Psi_{(*)}^{-j, -\mu}\ \Rightarrow \ F(c_{1})- F(c_{2})\in\Psi_{(*)}^{-j-1, -\mu}.
\end{equation}
 We also have
 \[
 \bea
&a_{0}= \frac{\i}{2}(\epsilon^{-1}\p_{t}\epsilon+ \epsilon^{-1}r\epsilon)\\[1mm]
&\in  \Psi_{(*)}^{-1, 0}\times  \Psi_{(*)}^{1, -1-\delta}+  \Psi_{(*)}^{-1, 0}\times  \Psi_{(*)}^{0, -1- \delta}\times \Psi_{(*)}^{1,0}\in \Psi_{(*)}^{0, -1- \delta}.
\eea
\]
We now follow the proof in \cite[Lemma A.1]{GW}, setting $b_{0}= a_{0}$, $b_{n}= a_{0}+ F(b_{n-1})$, and obtain by induction that 
$b_{n}- b_{n-1}\in \Psi_{(*)}^{-n, -1-\delta}$.
 We set 
 \[
 b_{0}\sim a_{0}+ \sum_{n=1}^{\infty}b_{n}- b_{n-1}\in \Psi_{(*)}^{0, -1-\delta}
 \]
  by Lemma \ref{l5.2}. We obtain that
 \[
\i\p_{t}b- b^{2}+ a + \i rb\in \Psi_{(*)}^{-\infty,-1-\delta}.
\]
 By construction we have $b(t)= \epsilon(t)+ \Psi_{(*)}^{0, -1-\delta}(\rr; \Sigma)$. Applying Prop. \ref{l5.1} we get
 \[
\epsilon(t)= \epsilon_{\outin}+ \Psi_{(*)}^{1, -\delta}(\rr_{\pm}; \Sigma)\hbox{ in }\rr_{\pm}\times \Sigma.
\]
 \qeds 
 \subsection{Proof of Prop. \ref{propoesti}}\label{to2}
 From Prop. \ref{p5.1} we first obtain that $b^{+}- b^{-}= (b+ b^{*})= 2\epsilon+ \Psi^{0, -1- \delta}_{(*)}(\rr; \Sigma)$. It follows first that $(b^{+}- b^{-})^{2}= 4 a+ \Psi_{(*)}^{1, -1- \delta}(\rr; \Sigma)$ and then by Prop. \ref{p5.1} that
\[
(b^{+}- b^{-})^{\alpha}= ((b^{+}- b^{-})^{2})^{\alpha/2}= \begin{cases}
(2 \epsilon)^{\12}+ \Psi^{0, - 1- \delta}_{(*)}(\rr; \Sigma), \ \alpha= \12\\
(2\epsilon)^{-\12}+ \Psi^{-3/2, -1- \delta}_{(*)}(\rr; \Sigma), \ \alpha= -\12.
\end{cases}
\]
We obtain again by Prop. \ref{p5.1} that:
\[
\begin{array}{l}
[(b^{+}- b^{-})^{-\12}, b^{\pm}]= [(2\epsilon)^{-\12}+ \Psi_{(*)}^{-3/2, -1- \delta}(\rr; \Sigma), \pm \epsilon+ \Psi^{0, -1- \delta}_{(*)}(\rr; \Sigma)]\in \Psi^{-3/2, -1- \delta}_{(*)}(\rr; \Sigma),\\[2mm]
\p_{t}(b^{+}- b^{-})^{-\12}(b^{+}- b^{-})^{\12}= (\p_{t}(2\epsilon)^{-\12}+ \Psi^{-3/2, -2- \delta}_{(*)}(\rr; \Sigma))\times \Psi^{\12, 0}_{(*)}(\rr; \Sigma)\\[2mm]
=\p_{t}(2\epsilon)^{-\12}\times\Psi^{\12, 0}_{(*)}(\rr; \Sigma)+ \Psi^{-1, -2-\delta}_{(*)}(\rr; \Sigma).
\end{array}
\]
Since by Prop. \ref{l5.1} $(2\epsilon)^{-\12}= (2 \epsilon_{\outin})^{-\12}+ \Psi^{-3/2, -\delta}_{(*)}(\rr_{\pm}; \Sigma)$, we have 
\[
\p_{t}(2\epsilon)^{-\12}\in \Psi^{-3/2, -1- \delta}_{(*)}(\rr; \Sigma)\ \Rightarrow \ \p_{t}(b^{+}- b^{-})^{-\12}(b^{+}- b^{-})^{\12}\in \Psi^{-1, -1- \delta}_{(*)}(\rr; \Sigma).
\]  Since by hypothesis $(H*)$, $r\in \Psi^{0, -1- \delta}_{(*)}(\rr; \Sigma)$, we obtain that $r_{b}^{\pm}\in \Psi^{0, -1- \delta}_{(*)}(\rr; \Sigma)$. Finally we obtain immediately from Prop. \ref{p5.1}  that $r_{-\infty}^{\pm}= \i\p_{t}b^{\pm}- (b^{\pm})^{2}+ a + \i rb^{\pm}\in \Psi^{-\infty, -1- \delta}_{(*)}(\rr; \Sigma)$. \qed

\subsection{Proof of Lemma. \ref{l10.1}}\label{apoti}
Let us fix two good chart coverings $\{U_{i}, \psi_{i}\}_{i\in \nn}$ and $\{\tilde{U}_{i}, \tilde{\psi}_{i}\}_{i\in \nn}$ with $U_{i}\Subset \tilde{U}_{i}$.
Since $\altb\in \cinfb(\rr; \BT^{1}_{0}(\Sigma, \altk))$, we obtain easily by transporting $\altb$ to $B_{n}(0,1)$ using $\psi_{i}$ that there exists $t_{+,\epsilon}>0$ such that $\ry(t,s, \cdot)$ is a bounded diffeomorphism of $(\Sigma, \altk)$, uniformly  for $|t-s|\leq t_{+,\epsilon}$. By the group property  of the flow we can replace $t_{+,\epsilon}$ by any $\varT>0$, keeping the above uniformity property.

Moreover if $\altb_{i}\defeq  (\psi_{i}^{-1})^{*}b$ we obtain from $(\ast)$ that $\altb_{i}\in S^{-\delta}(\rr; \BT^{1}_{0}(B_{n}(0,1)))$, uniformly in $i\in \nn$. If $\ry_{i}(t, s, \cdot)$ denotes the flow of $\altb_{i}$ we obtain that:
\[
\ry_{i}(t,s, \rx)= \rx+ \int_{s}^{t}\altb_{i}(\sigma, \ry_{i}(\sigma, s, \rx))d\sigma.
\]
From this we obtain that there exists $\varT\gg 1$ such that $\ry_{i}(\pm t, \pm \varT, \cdot): B_{n}(0, \12)\to B_{n}(0,1)$ for all $t\geq \varT$ and moreover 
\[
\lim_{t\to \pm \infty}\ry_{i}(t, \pm \varT, y)= \int_{\pm \varT}^{\pm \infty}\altb_{i}(\sigma, \ry_{i}(\sigma, \pm \varT, \rx))ds\eqdef \ry_{i}(\pm\infty, \varT, \rx).
\]
We can also choose $\varT$ large enough so that if we set
\beq\label{e10.3}
\ry(\pm\infty, \pm \varT,\rx)\defeq  \psi_{i}^{-1}\circ \ry_{i}(\pm\infty, \pm \varT, \cdot)\circ \psi_{i}(y), \ \rx\in U_{i}
\eeq
then $\ry(\pm\infty, \pm \varT, \cdot)$ is well defined, and is a bounded diffeomorphism of $(\Sigma, \altk)$.  We now set:
\[
\ry_{\outin}\defeq  \ry(\pm\infty, \pm \varT, \cdot)\circ \ry(\pm \varT, 0, \cdot),
\]
which is also a bounded diffeomorphism of $(\Sigma, \altk)$. We also obtain from \eqref{e10.4} and the previous estimates on  $\ry(t, s, \cdot)$ for $|t-s|\leq \varT$ that $\{\ry(t,0,  \cdot )\}_{t\in \rr}$ is a bounded family of bounded diffeomorphisms of $(\Sigma, \altk)$.
Moreover from \eqref{e10.3} we obtain that
\beq\label{e10.4}
\ry_{i}(t, 0, \rx)-\ry_{i, \outin}(\rx)\in S^{1- \delta'}(\rr; \cinfb(B_{n}(0,1))), \hbox{ uniformly in }i\in \nn.
\eeq
Let us now consider the metric $\chi^{*}\altg$.

Since $v\cdot dt=0$, $\chi^{*}\altg= {}^{t}\!D\chi(\altg\circ \chi)D\chi= - \altch^{2}(t, \rx)dt^{2}+ \hat \alth(t, \rx)d\rx^{2}$. Using  \eqref{e10.4} we obtain that
\[
\bea
\altch(t, \rx)&= \altc(t, \ry(t, \rx))+ S^{-2\delta'}(\rr; \BT^{0}_{0}(\Sigma))\\[2mm]
&= \altc_{\outin}(\ry(t,\rx))+ S^{-\min(2\delta', \delta)}(\rr_{\pm}; \BT^{0}_{0}(\Sigma))\\[2mm]
&=\altc_{\outin}(\ry_{\outin}(\rx))+ S^{-\min(1- \delta', \delta)}(\rr_{\pm}; \BT^{0}_{0}(\Sigma)).
\eea
\]
Similarly,
\[
\bea
\hat \alth(t, \rx)&= {}^{t}\!D\ry(t, \rx)\alth(t, \ry(t, \rx))D\ry(t, \rx)\\[2mm]
&= {}^{t}\!D\ry(t, \rx)\alth_{\outin}(\ry(t, \rx))D\ry(t, \rx)+ S^{-\delta}(\rr_{\pm}; \BT^{0}_{2}(\Sigma))\\[2mm]
&= {}^{t}\!D\ry_{\outin}(\rx)\alth_{\outin}(\ry_{\outin}(\rx))D\ry_{\outin}(\rx) +S^{-\min(1- \delta', \delta)}(\rr_{\pm}; \BT^{0}_{2}(\Sigma)),\\[2mm]
\chi^{*}\altV&=r(t,\ry(t, \rx))= r_{\outin}(\ry(t, \rx))+ S^{-\delta}(\rr_{\pm}; \BT^{0}_{0}(\Sigma))\\[2mm]
&=\altV_{\outin}(\ry_{\outin}(\rx))+  S^{-\min(1- \delta', \delta)}(\rr_{\pm}; \BT^{0}_{0}(\Sigma)).
\eea
\]
Since by definition
\[
 \hat \alth_{\outin}= \ry_{\outin}^{*}\alth_{\outin},\  \ \altch_{\outin}= \ry_{\outin}^{*}\altc_{\outin}, \ \ \altVh_{\outin}= \ry_{\outin}^{*}\altV_{\outin},
\]
we obtain the assertion. \qeds

\subsection{Proof of Lemma \ref{turlututi}}\label{apota}
\newcommand\epsi[1]{\langle \epsilon(#1)\rangle}
\def\tpx{(\langle \rx\rangle + \langle t \rangle)}
By interpolation, it suffices to prove the lemma for $m, k\in \nn$. Let us set
\[
\begin{array}{l}
T_{m,k}(t)= \epsi{0}^{m}\langle \rx\rangle^{k}\cU^{\adg}(0,t)\tpx^{-k}\epsi{0}^{-m},\\[2mm]
R_{m,k}(t,s)= \cU^{\adg}(t,s) \epsi{s}^{m}\langle \rx\rangle^{k}\cU^{\adg}(s,t)\tpx^{-k}\epsi{0}^{-m}.
\end{array}
\]
Using the  uniform ellipticity of $\epsilon(t)$  it suffices to prove that
\beq\label{e.turo-1}
\sup_{t\geq 0}\| T_{m,k}(t)\|_{B(\cH^{0})}<\infty.
\eeq We claim that
\begin{equation}
\label{e.turo1}
\sup_{0\leq s\leq t}\| R_{m, k}(t,s)\|_{B(\cH^{0})}<\infty,
\end{equation}
This of course implies   \eqref{e.turo-1}  by taking $s=0$ and using that $\cU^{\adg}(t,s)$ is uniformly bounded in $B(\cH^{0})$ by Prop. \ref{l.scat1}. 
To prove \eqref{e.turo1} we compute
 \beq\label{e.turo0}
\bea
&\p_{s}R_{m,k}(t,s)\\
&= \cU^{\adg}(t,s)\left(\p_{s}\epsi{s}^{m}+ [H^{\adg}(s), -\i \epsi{s}^{m}] \right)\langle \rx\rangle^{k}\cU^{\adg}(s,t)\tpx^{-k}\epsi{0}^{-m}\\
&\phantom{=}+ \cU^{\adg}(t,s)\epsi{s}^{m}[H^{\adg}(s), -\i \langle \rx\rangle^{k}]\cU^{\adg}(s,t)\tpx^{-k}\epsi{0}^{m}.
\eea
\eeq
Recall that as in \eqref{e5.5b}:
\[
H^{\adg}(t)= \mat{\epsilon(t)}{0}{0}{-\epsilon(t)}+ \Psi_{\std}^{0, -1-\delta}\otimes B(\cc^{2})
\]
by Prop. \ref{p5.1}, \ref{propoesti}. Hence:
\[
\left(\p_{s}\epsi{s}^{m}+ [H^{\adg}(s), -\i \epsi{s}^{m}]\right) \in \Psi^{m, -1- \delta}_{\std}(\rr; \Sigma)\otimes B(\cc^{2}),
\]
and we can write:
\begin{equation}
\label{e.turo2}
\left(\p_{s}\epsi{s}^{m}+ [H^{\adg}(s), -\i \epsi{s}^{m}]\right)= A_{m}(s)\epsi{s}^{m}\langle \rx\rangle^{-1},
\end{equation}
where 
\beq\label{e.turo3}
\|A_{m}(s)\|_{B(\cH)}\in O(1),
\eeq
since  $\tpx^{-1- \delta}\leq \langle \rx\rangle^{-1}$.

Similarly we have 
\beq\label{e.turo4}
\epsi{s}^{m}[H^{\adg}(s), \i \langle \rx\rangle^{k}]=  C_{m, k}(s)\epsi{s}^{m}\langle \rx\rangle^{k-1}, 
\eeq
where
\begin{equation}
\label{e.turo5}
\|C_{m,k}(s)\|_{B(\cH^{0})}\in O(1).
\end{equation}
We also set
\[
B_{m,k}(t)= \tpx^{-k+1}\epsi{0}^{-m}\tpx^{k-1}\epsi{0}^{m},
\]
and we have by pseudodifferential calculus
\begin{equation}
\label{e.turo6}
\|B_{m,k}(t)\|_{B(\cH^{0})}\in O(1).
\end{equation}
Hence we can rewrite \eqref{e.turo0} as
\beq\label{e.turo7}
\bea
&\p_{s}R_{m,k}(t,s)\\
&=\cU^{\adg}(t,s)D_{m,k}(s)\cU^{\adg}(s,t)\times R_{m, k-1}(t,s)\times B_{m,k}(t)\times \tpx^{-1},
\eea
\eeq
where
\beq\label{e.turo10}
D_{m,k}(s)= A_{m}(s)+ C_{m, k}(s), \ \ \|D_{m, k}(s)\|_{B(\cH^{0})}\in O(1).
\eeq
We can prove now \eqref{e.turo1} by induction for  $k$. First, note that by  Prop. \ref{l.scat1}, 1), \eqref{e.turo1} holds for $k=0$. Assume that \eqref{e.turo1} holds for $k-1$. Integrating  \eqref{e.turo7} from $t$ to $s$ we obtain:
\[
\bea
\|R_{m,k}(t,s)- R_{m, k}(t,t)\|\leq \int_{0}^{t}\|R_{m, k-1}(t, \sigma)\|\langle t\rangle^{-1}dt\in O(1), \hbox{for }0\leq s\leq t 
\eea
\]
by the induction hypothesis. We conclude the proof of \eqref{e.turo1} using that
\[
\|R_{m, k}(t,t)\|=\| \epsi{t}^{m}\langle \rx\rangle^{k}\tpx^{-k}\epsi{0}^{m}\|\in O(1). 
\]
This completes the proof of the lemma. \qeds

\subsection*{Acknowledgments} It is a pleasure to thank Jan Derezi\'nski, Andr\'as Vasy and Jochen Zahn for stimulating discussions. M.\,W. gratefully acknowledges the France-Stanford Center for Interdisciplinary Studies for financial support and the Department of Mathematics of Stanford University, where part of the work was performed. The authors also wish to thank the Erwin Schr\"odinger Institute in Vienna for its hospitality during the program ``Modern theory of wave equations''.

\end{document}